%% file: LT_main.tex
\documentclass[aps, prx, twocolumn, longbibliography, floatfix, nobibnotes, 10pt, superscriptaddress]{revtex4-2}

\input{preamble.tex}

\makeatletter
\renewcommand{\fnum@figure}{FIG.~\thefigure}
\makeatother

\newcommand{\sm}{Supplemental Material~\cite{supmat}}

\usepackage[english]{babel}
\usepackage{pgfplots}
\pgfplotsset{compat=1.18}
\usetikzlibrary{plotmarks}
\usepackage{enumitem}

\usepackage{scalerel}

\usepackage[normalem]{ulem}

\begin{document}

\title{Universal quantum melting of quasiperiodic attractors in driven-dissipative cavities}

\author{Caroline Nowoczyn}
\affiliation{Center for Optical Quantum Technologies and Institute for Quantum Physics, University of Hamburg, Hamburg 22761, Germany}
\affiliation{The Hamburg Center for Ultrafast Imaging, Hamburg 22761, Germany}
\author{Ludwig Mathey}
\affiliation{Center for Optical Quantum Technologies and Institute for Quantum Physics, University of Hamburg, Hamburg 22761, Germany}
\affiliation{The Hamburg Center for Ultrafast Imaging, Hamburg 22761, Germany}
\author{Kilian Seibold}
\email{seibold.kilian@gmail.com}
\affiliation{Department of Physics, University of Konstanz, 78464 Konstanz, Germany}
%
%
\begin{abstract}
Nonlinear classical mechanics has established rich phenomena. These include limit tori defined by toroidal attractors supporting quasiperiodic motion with incommensurate frequencies. We study the fate of such structures in open quantum systems using two coupled driven-dissipative Kerr cavities modeled via the Lindblad master equation.
Combining Liouvillian spectral theory with the truncated Wigner approximation, we characterize the quantum-to-classical crossover.
In the classical limit, two pairs of purely imaginary Liouvillian eigenvalues signal persistent quasiperiodic modes.
Quantum fluctuations induce small negative real parts to these eigenvalues, giving rise to finite lifetimes and leading to the \emph{quantum melting} of the torus.
The associated Liouvillian gaps vanish algebraically in the classical limit, indicating a dynamical critical crossover with spontaneous breaking of time-translational symmetry.
Quantum trajectory analysis reveals that this melting is driven by fluctuation-induced dephasing. Using a circular-variance-based order parameter, we uncover universal scaling in system size and time.
These results establish quantum melting of limit tori as a distinct and robust non-equilibrium critical phenomenon, with clear experimental signatures in trapped ions and superconducting circuits.

\end{abstract}

\maketitle

\section{Introduction}
Nonlinear classical mechanics has provided a framework for understanding phenomena such as bistability, chaos, synchronization, and robust periodic motion~ \cite{Drummond1980, Aldana2013, Bartolo2016, Landa2020, Lyapunov1992, Sharma2009, Laffargue2016, Goto2021, Datseris2022, Lee2013, Mari2013, Walter2014, Loerch2016, DavisTilley2018, Kato2019, Thomas2021, Thomas2022, Lu2023, Waechtler2023, Moreno2024}. 
Such dynamics are analyzed using tools like bifurcation theory~\cite{Luo1997, Kielhoefer2012}, Floquet analysis~\cite{Floquet1883}, and Lyapunov spectra~\cite{Eckmann1985}, which reveal the stability and structure of underlying attractors.
Among these attractors are limit tori, which emerge through Neimark-Sacker bifurcations~\cite{Strogatz2024, Kolmogorov1954, Moeser1962, Arnold2009, Yusipov2019, Cosme2025}.
They exhibit quasiperiodic motion: Trajectories densely fill a toroidal manifold due to two or more incommensurate frequencies.
The quasiperiodicity and the associated geometric structure make limit tori intrinsically more complex than limit cycles, where a single fundamental frequency yields purely periodic dynamics and a closed orbit in phase space.
Quasiperiodic attractors appear across disciplines, including fluid dynamics, nonlinear optics, chemistry, and biology \cite{Garfinkel1997,Suzuki2016,Bick2020,Voskresensky2020,Kpomahou2022,Zou2024}. Although limit tori are well understood in classical systems, their quantum analogs have received little theoretical study, with only a few recent works~\cite{Yusipov2019}.

Recent advances in driven-dissipative quantum platforms, such as superconducting circuits~\cite{Ferrari2025}, ultracold atomic platforms~\cite{Gross2017,Cosme2025}, and trapped-ion platforms, have opened new possibilities for investigating nonlinear dissipative quantum dynamics.
Examples including quantum synchronization~\cite{Lee2013, Mari2013, Walter2014, Kato2019, Thomas2021, Lu2023, Waechtler2023, Moreno2024} and time-crystalline behavior~\cite{Wilczek2012, Else2016, Kessler_2019a, Seibold2020, Kongkhambut_2022} show that many-body quantum systems can maintain long-lived coherence despite strong dissipation. 

In this work, we address the question of how phenomena established in classical nonlinear mechanics translate into quantum dissipative dynamics.
As an intriguing representative example we focus on limit tori.
We note that while other phenomena such as limit cycles~\cite{Seibold2020, BenArosh2021, Thomas2021, Thomas2022, Dutta2025, Skulte2024} and emergent signatures of quantum chaos~\cite{Zanardi2021, Yoshimura2024, Richter2024, Solanki2024, Ferrari2025} have been explored in the quantum domain, the quantum description, and generally the fate of limit tori and complex attractors, are essentially unexplored.
Quantum coherence, fluctuations, and dissipation can drastically reshape or destabilize such structures, raising fundamental questions about their persistence in open quantum systems.
We consider the following questions:
\begin{enumerate}[label=(\arabic*)]
    \item Can quasiperiodic attractors survive quantum noise and dissipation?
    \item What are the spectral signatures of quasiperiodic attractors in the Liouvillian?
    \item Do universal laws govern the dephasing and disappearance of quasiperiodic attractors in the quantum regime?
\end{enumerate}
Answering these questions is crucial for bridging classical nonlinear dynamics with the physics of open quantum systems.
To address these questions on a concrete example, we develop a quantum description of limit tori and their associated Liouvillian spectrum in a minimal model of two coupled driven-dissipative Kerr cavities.
By combining mean-field analysis, Liouvillian spectral theory~\cite{Albert2014, Minganti2018, Macieszczak2016}, and quantum trajectory methods within the truncated Wigner approximation~\cite{Vogel1989, Polkovnikov2010, Gardiner2014, Drummond2017, Yoneya2024}, we investigate the emergence, stability, and quantum melting, that is, the quantum-fluctuation-induced degradation, of limit tori across the quantum-to-classical crossover.
We identify quasiperiodic motion via pairs of complex-conjugate Liouvillian eigenvalues whose real parts vanish in the classical limit, with Liouvillian gaps that close algebraically. 
This spectral signature reveals universal scaling laws governing the quantum melting and supports its interpretation as a dynamical critical crossover characterized by algebraic softening of Liouvillian phase modes.

The paper is organized as follows.
In Sec.~\ref{sec: Theoretical framework}, we introduce the model and theoretical framework.
In Sec.~\ref{sec: Results}, we present our results on the emergence and melting of quantum limit tori and identify associated universal scaling laws.
In Sec.~\ref{sec: experimental realization}, we discuss prospects for experimental realization, outlining a concrete implementation strategy in a trapped-ion platform.
Finally, Sec.~\ref{sec: Conclusions} summarizes our findings and discusses future directions.

%
\section{Theoretical framework}
\label{sec: Theoretical framework}
\subsection{Coupled Kerr cavities}
\begin{figure}
\centering
\includegraphics[width=\columnwidth]{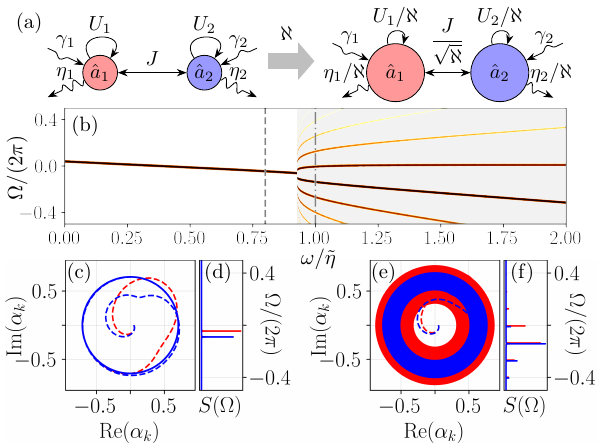}
\caption{
Coupled driven-dissipative Kerr cavities and emergence of limit tori.
(a) Schematic representation of the coupled Kerr cavities, described by Eqs.~\eqref{eq: Hamiltonian} and \eqref{eq: Master equation}, and the quantum-to-classical rescaling of the system. 
The parameter $\aleph$ is a dimensionless scaling parameter that controls the quantum-to-classical crossover; see Eq.~\eqref{eq: scaling relation}.
(b) Heatmap of the Fourier spectrum $S_k(\Omega)$ of mode $k=1$, 
as defined in Eq.~\eqref{eq: spectrum_mode_1}, obtained from the Gross-Pitaevskii (GP) dynamics, Eq.~\eqref{eq: GPE}, as a function of $\omega$. 
The region with a white background corresponds to limit-cycle dynamics, dominated by a single frequency, while the gray background indicates the limit-torus regime, dominated by two fundamental frequencies (and higher harmonics). 
The gray dashed line marks $\omega = 0.8$ for the limit cycle, shown in (c) and (d), and the dashed-dotted line marks $\omega = 1.0$ for the limit torus, shown in (e) and (f).
[(c) and (e)] Phase-space trajectories of the mean fields $\alpha_1$ (red) and $\alpha_2$ (blue), obtained from GP equations, Eq.~\eqref{eq: GPE}. 
[(d) and (f)] Corresponding Fourier spectra $S_k(\Omega)$ of modes $k=1$ and $k=2$, respectively.  The trajectories are obtained by numerically integrating the GP equations up to $\eta t = 10^4$,  with initial conditions $\alpha_1(0) = 0.05$ and $\alpha_2(0) = 0.05$.
Unless stated otherwise, the remaining parameters are $U_1=U_2=0.1$, $J=0.4$, $\gamma_1=\gamma_2=\gamma=1$, and $\eta_1=\eta_2=\eta=1$.
}
\label{fig: figure 1}
\end{figure}

We consider a minimal theoretical model of two coupled Kerr cavities, subject to incoherent driving and two-photon loss processes, as illustrated in Fig.~\ref{fig: figure 1}(a).
The Hamiltonian of the system reads
\begin{equation}
    \dfrac{\hat{\mathcal{H}}}{\hbar} = 
    \sum_{k=1,2} \omega_k \hat{a}_k^\dagger \hat{a}_k
    + \frac{U_k}{2} \hat{a}_k^{\dagger 2} \hat{a}_k^2 
    - J ( \hat{a}_1^{\dagger 2} \hat{a}_2 + \hat{a}_2^\dagger \hat{a}_1^2 )\;,
    \label{eq: Hamiltonian}
\end{equation}
where $\hat{a}_k$ ($\hat{a}_k^\dagger$) are the bosonic annihilation (creation) operators for cavity $k$, satisfying the commutation relation $\commut{\hat{a}_j}{\hat{a}_k^\dagger}=\delta_{jk}$, 
$\omega_k$ denotes the bare cavity frequencies,
$U_k$ the strength of the on-site Kerr nonlinearities, and
$J$ the amplitude of the nonlinear tunneling term.
Throughout this work, we refer to the excitations as \q{photons} for simplicity, though in specific implementations, such as the trapped-ion realization, discussed in Sec.~\ref{sec: experimental realization}, they may represent other bosonic modes, such as phonons.
The incoherent pump provides energy to the system, while the two-photon loss is a nonlinear process that damps oscillations at high amplitudes. This combination of dissipative processes is known to stabilize self-sustained limit cycles (LCs)~\cite{Lee2013}.

The dissipative dynamics of the system is given by the Lindblad master equation for the density matrix $\hat{\rho}$~\cite{Breuer2007,Gardiner2000}:
\begin{equation}
    \dfrac{d\hat{\rho}}{dt} = \mathcal{L}\hat{\rho} = \dfrac{1}{i\hbar}\comm{\hat{\mathcal{H}}}{\hat{\rho}}+ \sum_{k=1,2}
    \gamma_k\mathcal{D}[\hat{a}_k^\dagger]\hat{\rho}+
    \eta_k\mathcal{D}[{\ao}_k^2]\hat{\rho}\;,
    \label{eq: Master equation}
\end{equation}
where
$
\mathcal{D}[\hat{L}]\hat{\rho} = \hat{L} \hat{\rho} \hat{L}^\dagger 
- \frac{1}{2} ( \hat{L}^\dagger \hat{L} \hat{\rho} + \hat{\rho} \hat{L}^\dagger \hat{L})
$ 
denotes the dissipator in Lindblad form associated with jump operator $\hat{L}$. Here $\gamma_k$ and $\eta_k$ represent the incoherent pumping and two-photon loss rates, respectively.

\subsection{Spectral analysis of the Liouvillian}
The dynamics of an open quantum system is governed by the Liouvillian superoperator $\mathcal{L}$, defined in Eq.~\eqref{eq: Master equation}, which acts on the system's density matrix. 
Its spectrum is defined by the eigenvalue equation $\mathcal{L}\op{\rho}_j=\lambda_j\op{\rho}_j$, where $\lambda_j$ are generally complex-valued. According to Spohn's theorem~\cite{Spohn1976,Spohn1977,Nigro2019}, any finite-dimensional open quantum system admits a unique nonequilibrium steady state (NESS) associated with the eigenvalue $\lambda_0=0$, while all other eigenvalues have strictly negative real parts. 
Consequently, any initial state asymptotically relaxes to the NESS.
The timescale of this relaxation is set by the Liouvillian gap, defined as the nonzero eigenvalue with the smallest magnitude of the real part~\cite{Minganti2018}.
The closing of the Liouvillian gap signals a qualitative change in the long-time dynamics, marking the onset of critical phenomena such as dissipative phase transitions or the emergence of persistent oscillations in the stationary state~\cite{Seibold2020}.


\subsection{Quantum-to-classical rescaling}
In the limit of large photon occupation, the dynamics of a driven-dissipative system can be described by the Gross-Pitaevskii equation (GPE), a mean-field approach that neglects quantum fluctuations. The GPE is obtained from the adjoint master equation of Eq.~\eqref{eq: Master equation} by applying the mean-field approximation $\langle\ao[A]\ao[B]\rangle=\langle\ao[A]\rangle\langle\ao[B]\rangle$ and taking the classical limit $\hbar\rightarrow 0$. It reads
\begin{equation}
\begin{aligned}
    i\dfrac{d\alpha_1}{dt}
		 &= \left(\omega_1 +i\dfrac{\gamma_1}{2} + (U_1-i\eta_1)\abs{\alpha_1}^2\right)\alpha_1 -2J\alpha_1^*\alpha_2\\
    i\dfrac{d\alpha_2}{dt}
		 &= \left(\omega_2 +i\dfrac{\gamma_2}{2} + (U_2-i\eta_2)\abs{\alpha_2}^2\right)\alpha_2 -J\alpha_1^2\;,\\
\end{aligned}
        \label{eq: GPE}
\end{equation}
where $\alpha_k = \mathrm{Tr}(\hat{a}_k\hat{\rho})$ denotes the expectation value of the annihilation operator.
To systematically control the crossover between quantum and classical behavior, we introduce a scaling parameter $\aleph$, which rescales the system parameters according to
\begin{equation}
    U_k = \dfrac{\tilde{U}_k}{\aleph}\;,\;\;
	J   = \dfrac{\tilde{J}}{\sqrt{\aleph}}\;,\;\;
	\eta_k = \dfrac{\tilde{\eta}_k}{\aleph} \;,
\label{eq: scaling relation}
\end{equation}
as illustrated in Fig.~\ref{fig: figure 1}(a). 
Under this transformation, the complex fields scale as $\alpha_k = \tilde{\alpha}_k\sqrt{\aleph}$, leading to mean-field populations $n_k = |\alpha_k|^2 \propto \aleph$.
Notably, Eq.~\eqref{eq: GPE} remains invariant under these scaling relations, demonstrating that the limit $\aleph \rightarrow\infty$ defines a well-controlled classical limit characterized by an infinite number of photons. This scaling method is commonly used in the study of driven-dissipative quantum systems to systematically explore quantum-to-classical transitions~\cite{Hwang2016,Casteels2017,Casteels2017a,Puebla2017,Hwang2018,Lledo2019}.
Unless otherwise specified, we consider the symmetric parameter choice $\omega_1=\omega_2= \omega = 1$, $\tilde{U}_1=\tilde{U}_2=0.1$, $\tilde{J}=0.4$, $\gamma_1=\gamma_2=\gamma=1$, and $\tilde{\eta}_1=\tilde{\eta}_2=\tilde{\eta}=1$.
These values enable the emergence of complex dynamics, including LCs and limit tori (LT).

\subsection{Truncated Wigner approximation}
To capture quantum fluctuations and the mixed character of the density matrix beyond mean-field theory, we employ the truncated Wigner approximation (TWA)~\cite{Vogel1989, Polkovnikov2010, Gardiner2014, Drummond2017, Yoneya2024, Dagvadorj2015, Deuar2021, Dujardin2015, Orso2025, Vicentini2018, Vicentini2019}. In the TWA, the density operator $\op{\rho}$ is mapped onto the Wigner quasiprobability distribution $W(\{\alpha_k,\alpha_k^*\})$ over the real and imaginary parts of the complex mode amplitudes $\alpha_k$.
Truncating the dynamical equations for the Wigner distribution at order $\mathcal{O}(\hbar^2)$ yields a Fokker-Planck equation, see \sm, which is equivalent to the ensemble dynamics of stochastic Langevin trajectories.

For the two-cavity model, we initialize the Wigner function with the distribution corresponding to a two-mode coherent state $\ket{\alpha_1(0),\alpha_2(0)}$. The Langevin equations for the rescaled fields $\tilde{\alpha}_k$ take the form:
\begin{equation}
	\begin{aligned}
		i \frac{d\tilde{\alpha}_{1}}{dt} &= \left[\omega_1 + i\frac{\gamma_1}{2} +\left(\tilde{U}_1-i \tilde{\eta}_1\right)\left(\abs{\tilde{\alpha}_1}^2-\frac{1}{\aleph}\right)\right]\tilde{\alpha}_1
        \\
        &-2\tilde{J}\tilde{\alpha}_1^*\tilde{\alpha}_2
        +\sqrt{\dfrac{\gamma_1}{2\aleph}}\chi_{1}(t)
        +\sqrt{\dfrac{2\tilde{\eta}_1\abs{\tilde{\alpha}_1}^2}{\aleph}}\xi_{1}(t)
		\\
		i\frac{d\tilde{\alpha}_2}{dt} &= \left[\omega_2 + i\frac{\gamma_2}{2} +\left(\tilde{U}_2 - i \tilde{\eta}_2\right)\left(\abs{\tilde{\alpha}_2}^2-\frac{1}{\aleph}\right)\right]\tilde{\alpha}_2
        \\
        &-\tilde{J}\tilde{\alpha}_1^2
        +\sqrt{\dfrac{\gamma_2}{2\aleph}}\chi_{2}(t)
        +\sqrt{\dfrac{2\tilde{\eta}_2\abs{\tilde{\alpha}_2}^2}{\aleph}}\xi_{2}(t)\;,
	\end{aligned}
    \label{eq: TWA}
\end{equation}
where $\chi_{k}$ and $\xi_{k}$ are independent, zero-mean and unit-variance complex Gaussian stochastic variables, characterized by correlation functions 
$\langle\chi_{k}(t)\chi_{k'} (t')\rangle = \langle\xi_{k}(t)\xi_{k'} (t')\rangle =0$,
$\langle\chi_{k}^*(t)\chi_{k'}(t')\rangle=\delta_{k,k'}\delta(t-t')$, and
$\langle\xi_{k}^*(t)\xi_{k'}(t')\rangle=\delta_{k,k'}\delta(t-t')$.
The noise terms $\chi_{k}(t)$ and $\xi_{k}(t)$ account for quantum fluctuations.
The expectation value of an observable $\hat{O}$ is approximated by averaging its Weyl symbols $O_{W}\bigl(\{\alpha_{k}, \alpha^*_{k}\}\bigr)$ over $N_{\rm traj}$ stochastic trajectories:
\begin{equation}
  \langle \hat{O} \rangle(t) \approx \frac{1}{N_{\rm traj}} \sum_{\mu=1}^{N_{\rm traj}} O_{W}\bigl(\{\alpha_{k,\mu}(t), \alpha^*_{k,\mu}(t)\}\bigr)\;,
\end{equation}
where the individual stochastic realizations are labeled by $\mu$.
In particular, for symmetrically, i.e., Weyl-ordered products of operators, such as $\mathcal{S}[(\hat{a}_i^\dagger)^n \hat{a}_j^m]$, the expectation value reads:
\begin{equation}
  \bigl\langle \mathcal{S}[(\hat{a}_i^\dagger)^n \hat{a}_j^m] \bigr\rangle(t) \approx \frac{1}{N_{\rm traj}} \sum_{\mu=1}^{N_{\rm traj}} (\alpha_{i,\mu}^*(t))^n (\alpha_{j,\mu}(t))^m\;.
\end{equation}

The system is initialized with each mode prepared in a coherent state of amplitude $\tilde{\alpha}_0 = 0.05$. The initial conditions for the stochastic trajectories are sampled from the corresponding Gaussian Wigner distribution, 
\begin{equation} 
\tilde{\alpha}_k(t=0) = \tilde{\alpha}_0 + \frac{1}{\sqrt{2\aleph}}\zeta_k, \label{eq: initial condition TWA} 
\end{equation} 
where $\zeta_k$ is a unit-variance complex random variable with zero mean, satisfying \mbox{$\langle \zeta_k\zeta_{k'} \rangle = 0$} and \mbox{$\langle \zeta_k^*\zeta_{k'} \rangle = \delta_{k,k'}$}.
Each realization corresponds to a possible measurement outcome, providing a semi-classical picture of the system's quantum dynamics.
%

%
\section{Quantum melting of limit tori and its universality}
\label{sec: Results}
\subsection{Quantum melting of LT dynamics}
\label{subsec: Quantum melting of LT dynamics}
\subsubsection{Semiclassical analysis}
We begin by analyzing the mean-field dynamics to identify parameter regimes where LC and LT attractors emerge.
Figure~\ref{fig: figure 1}(b) shows the Fourier spectrum of mode~$k=1$,
\begin{equation}
S_k(\Omega) = \left|\int_0^{\infty} d\tau 
\alpha_k(\tau) e^{i\Omega \tau}\right|\;,
\label{eq: spectrum_mode_1}
\end{equation}
computed from the GPE, as a function of $\omega$.
On the left side of the plot, with white background, a single dominant frequency is visible, signaling LC behavior, while on the right, with gray background, the presence of two fundamental incommensurate frequencies signals the LT phase, characterized by quasiperiodic dynamics.
A Lyapunov exponent analysis confirms this distinction. The LC phase shows a single zero Lyapunov exponent, while the LT phase is marked by two zero exponents, see \sm.
The phase-space trajectories further illustrate the nature of the attractors. For $\omega = 0.8$, the system exhibits periodic oscillations with well-defined amplitude and frequency; see Fig.~\ref{fig: figure 1}(c). For $\omega = 1$, the amplitude modulation over multiple periods reveals a toroidal structure; see Fig.~\ref{fig: figure 1}(e).
The ratio of the dominant frequencies is analyzed in the \sm.
%

\begin{figure}
\centering
\includegraphics[width=\columnwidth]{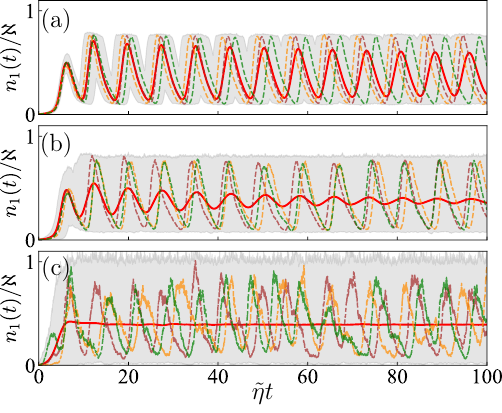}
\caption{
    Dynamics of the rescaled occupation $n_1(t)/\aleph$ of mode 1, in the TWA.
    Panels (a)--(c) correspond to decreasing values of the scaling parameter: (a) $\aleph=30\,000$, (b) $\aleph=4000$, and (c) $\aleph=200$.
    Dashed lines show the evolution of individual TWA trajectories, while the solid red lines indicate the expectation value of $n_1(t)/\aleph$.
    The gray area represents the range between the minimal and maximal occupation values across all trajectories.
    Both modes are initialized in a coherent state centered at
    $\tilde{\alpha}_0 = 0.05$. The expectation values are averaged over $3\times10^4$ TWA trajectories. 
    Mode 2 shows qualitatively similar behavior.
    }
    \label{fig: figure 2}
\end{figure}
\subsubsection{Quantum transient}
Individual stochastic TWA trajectories show that the LT dynamics observed in the mean-field regime persist in the quantum regime.
Figure~\ref{fig: figure 2} shows the time evolution of single-trajectory dynamics, represented by dashed lines.
After a rapid transient regime ($\tilde{\eta} t \lesssim 10$), each trajectory settles into the LT attractor and explores its full amplitude range without damping.

The gray-shaded area in Fig.~\ref{fig: figure 2} delineates the envelope of minimal and maximal occupations across all trajectories. 
Although quantum fluctuations are present, each individual trajectory stays within the envelope, maintaining dynamics that is confined to the LT attractor. They remain quasiperiodic, with persistent undamped oscillations.
Over time, stochastic noise induces gradual dephasing between trajectories, leading to a decay of phase coherence across the ensemble of single trajectories.
This results in an effective damping of ensemble-averaged TWA expectation values, as shown in Fig.~\ref{fig: figure 2}, where the solid red lines depict the rescaled occupation $n_1(t)/\aleph$ approaching a non-oscillatory steady state.
The progressive temporal broadening of the envelope (gray-shaded area) in Fig.~\ref{fig: figure 2}(a) further reflects the dephasing dynamics.
This dephasing-driven melting differs fundamentally from bistable Kerr dynamics, where stochastic quantum trajectories exhibit telegraphic switching between metastable attractors~\cite{Dykman2007,Chan2007}; here trajectories remain confined to the LT manifold and relaxation occurs via diffusive phase decoherence.
This process constitutes the dominant relaxation mechanism and is governed by the Liouvillian gap: Quantum noise induces random phase drift, effectively melting the LT.

%
\subsubsection{Liouvillian spectral signature of LT}
As discussed above, at the mean-field level, the two coupled driven-dissipative Kerr cavities in our model undergo a Neimark-Sacker bifurcation, giving rise to a two-dimensional LT attractor embedded in four-dimensional phase space.
This attractor is characterized by exactly two vanishing Lyapunov exponents, one for each neutrally stable direction of the torus, while all other exponents are strictly negative; see \sm.
The resulting quasiperiodic dynamics are governed by two incommensurate frequencies, a hallmark of toroidal motion.
These classical features leave a clear imprint on the quantum dynamics. Specifically, the emergence of LTs is encoded in the spectral structure of the Liouvillian superoperator $\mathcal{L}$.
The low-lying spectrum encodes two key ingredients: the quasiperiodic motion inherited from mean-field dynamics and the slow dephasing induced by quantum fluctuations.
\begin{figure}
\centering
\includegraphics[width=\columnwidth]{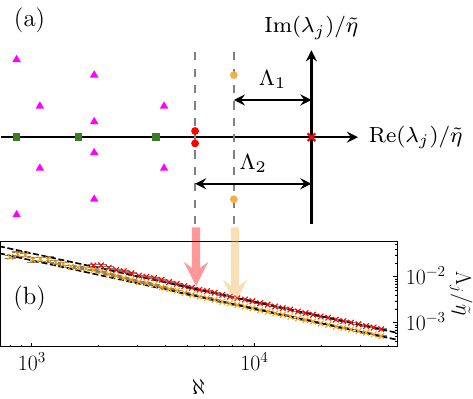}
\caption{
Liouvillian spectrum analysis.
(a) Schematic representation of the spectrum of the Liouvillian. A zero eigenvalue, indicated with a red cross, corresponds to the unique steady state. 
In the LT regime, two pairs of complex conjugate eigenvalues (red and gold) emerge, each with small real parts that vanish in the classical limit and incommensurate imaginary parts.
(b)  Scaling of the Liouvillian gaps with $\aleph$. The power-law fits of the two Liouvillian gaps for $5\times10^{3} \leq \aleph \leq 4\times10^{4}$ is shown by the dashed and dash-dotted lines. The coefficients of the fit are $\Lambda_j/\tilde{\eta} = \aleph^{-a_j}\times b_j$, with $a_1=1.054\pm0.01$ and $b_1 = 47.144 \pm 5.724$ and $a_2=1.052\pm0.01$ and $b_2 = 32.922 \pm 3.452$.
}
\label{fig: figure 3}
\end{figure}

Figure~\ref{fig: figure 3}(a) schematically illustrates this structure of the low-lying Liouvillian spectrum. The steady state corresponds to a unique zero eigenvalue $\lambda_0 = 0$, denoted as a red cross, as required by Lindblad formalism.
In a LC regime, the Liouvillian spectrum is dominated by a single pair of complex conjugate eigenvalues. 
The imaginary parts of these eigenvalues determine the oscillation frequency. The small negative real parts set the decay rate, which vanishes in the classical limit where quantum fluctuations are negligible~\cite{Seibold2020, Dutta2025}.
By contrast, the low-lying Liouvillian spectrum of a LT comprises two sets of complex conjugate eigenvalues
\begin{equation}
	\lambda_{j}^\pm = -\Lambda_{j} \pm i\nu_{j}
	\quad\text{for}\quad {j}=1,2\;,
\end{equation}
with incommensurate frequencies $\nu_{1,2}$ associated with motion along the two angular directions of the torus.
Each direction is subject to quantum-fluctuation-induced dephasing, quantified by a distinct Liouvillian gap $\Lambda_j$.
In the classical limit, where $\Lambda_j\to0$, this leads to two sets of purely imaginary eigenvalues, recovering the undamped quasiperiodic motion.

To probe the resulting quasiperiodicity in cavity 1, we compute its time-resolved emission spectrum via a windowed Fourier transform of the first-order correlation function \footnote{The time-resolved spectral analysis enables a high-precision extraction of the individual frequency components and their lifetimes, directly corresponding to the imaginary and real parts of Liouvillian eigenvalues, respectively. It provides a robust alternative to linewidth-based estimates, particularly in the presence of noise and non-stationarity.},
\begin{equation}
    S_1(t,\Omega) = 2 \mathrm{Re} \int_{0}^{\infty}\hspace{-0.3cm}d\tau w(t-\tau) \langle \ao_1^\dagger(t_0+\tau )\ao_1(t_0)\rangle e^{i\Omega\tau}  \;,
\label{eq: time-resolved emission spectrum}
\end{equation}
with window function $w(s)$.
This quantity is numerically evaluated using a discrete short-time Fourier transform of the correlation function sampled from $N_{\text{traj}} = 10^5$ TWA trajectories; see \sm.
The resulting spectrogram displays two dominant peaks at constant incommensurate frequencies $\nu_{1,2}$, whose magnitudes decay over time. We numerically extract the corresponding lifetimes; see \sm. The frequencies $\nu_{1,2}$ coincide with the imaginary parts of the low-lying Liouvillian eigenvalues, while the decay rates are set by the corresponding Liouvillian gaps $\Lambda_{1,2}$. The spectrogram Eq.~\eqref{eq: time-resolved emission spectrum} therefore provides a direct probe of the Liouvillian spectrum.
Since the observable considered here is sensitive only to cavity 1, spectral features associated with cavity 2 are not directly resolved. Nevertheless, cavity 2 contributes an additional pair of complex-conjugate eigenvalues to the Liouvillian spectrum, corresponding to its own quasiperiodic motion.
Figure~\ref{fig: figure 3}(b) shows the scaling of $\Lambda_j$ with scaling parameter $\aleph$. We find a \textit{power-law} behavior,
\begin{equation}
  \Lambda_j \;\propto\; \aleph^{-a_j}, 
  \quad a_j \approx 1\;.
  \label{eq: power-law liouvillian gap}
\end{equation}
The exponent $a_j$ therefore characterizes the algebraic softening of Liouvillian phase modes and can be interpreted as a dynamical critical exponent governed by diffusive phase decoherence along the two neutrally stable directions of the torus, rather than an equilibrium critical exponent associated with spatial correlations.
In particular, the inverse gaps $1/\Lambda_j$ set the slowest timescales over which quasiperiodic oscillations in single-trajectory observables lose phase coherence, and are directly linked to the relaxation of population oscillations in Fig.~\ref{fig: figure 2}.
The algebraic gap closing reflects diffusive phase decoherence along each angular coordinate, with a diffusion constant scaling as $D\propto \aleph^{-1}$.
Accordingly, each Liouvillian gap behaves as $\Lambda_j\approx D\propto \aleph^{-1}$. When probing an observable sensitive to \textit{both} angular directions [e.g. $S_1(\omega)$], the total dephasing rate becomes $\Lambda_1 + \Lambda_2 \approx 2D$,  preserving the $\aleph^{-1}$ scaling.
This power-law scaling can also be understood from signal-to-noise ratio (SNR) considerations: The coherent amplitude scales as $|\alpha|\propto\sqrt{\aleph}$, while quantum fluctuations remain $\aleph$ independent, leading to $\mathrm{SNR}\sim \aleph$.
Since phase diffusion is driven by quantum noise, dephasing rates scale inversely with SNR, yielding $\Lambda_j \propto \aleph^{-1}$.
This inverse scaling of the dephasing rate with the SNR can be seen in analogy to the scaling of the Schawlow-Townes laser linewidth in laser theory, where spontaneous emission drives dephasing of the coherent field, yielding a Lorentzian spectrum with width proportional to the inverse intracavity SNR~\cite{Schawlow1958,Chia2019}.

Crucially, the frequencies $\nu_j$ remain invariant with increasing $\aleph$, reflecting their mean-field origin.
Indeed, they match the frequencies obtained from the spectrum of the GPE, confirming that the Liouvillian spectrum inherits the angular frequencies of the underlying quasiperiodic mean-field dynamics.

Importantly, the extracted scaling exponents $a_j\approx 1$ for the Liouvillian gaps $\Lambda_j$ are found to be robust across different parameter regimes that support limit tori.
This indicates that the exponent $a_j \approx 1$ is a robust and universal feature of the quantum-to-classical crossover governed by diffusive phase decoherence in quasiperiodic open systems.

\begin{widetext}
\onecolumngrid
\begin{figure}[H]
\centering
\includegraphics[width=1\textwidth]{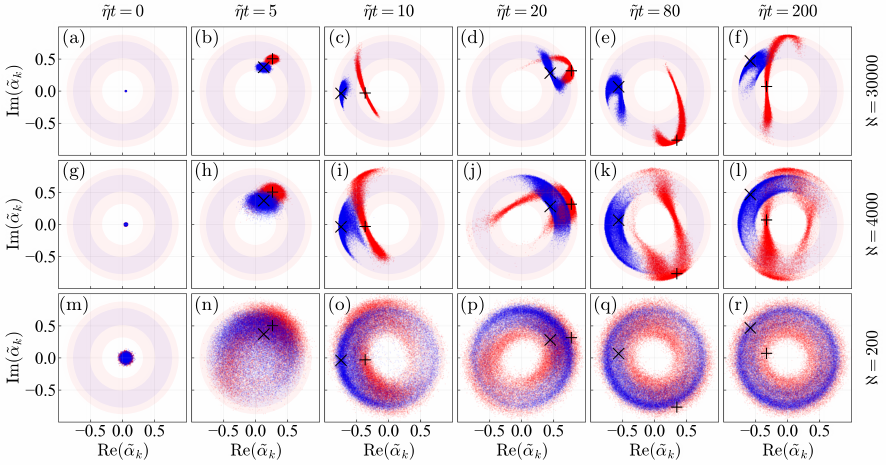}
\caption{
    Wigner function reconstructions based on the TWA fields $\alpha_1$ (red) and $\alpha_2$ (blue), shown for different times $t$ (columns) and scaling parameter $\aleph$ (rows).
    Classical GPE solutions for mode 1 marked with $+$, and mode 2 marked with $\times$ are shown at corresponding times.
    The background light red and light blue  regions indicate the phase-space area traced by the classical LT attractor, as obtained from GPE simulations.
    The three values of $\aleph$ match those in Fig.~\ref{fig: figure 2}, decreasing from the top to the bottom row.
    Each distribution is obtained from $3\times 10^4$ TWA trajectory realizations.
    }
\label{fig: figure 4}
\end{figure}
\twocolumngrid
\end{widetext}

%
\subsubsection{Quantum melting}
The quantum melting of the LT structure is further illustrated through Wigner function reconstructions presented in Fig.~\ref{fig: figure 4}. The reconstructions are obtained from TWA fields $\alpha_1$ (red) and $\alpha_2$ (blue) for different times (columns) and scaling parameters $\aleph$ (rows).
For reference, the phase-space area of the classical LT obtained from the GPE, and also shown in Fig.~\ref{fig: figure 1}(e), is overlaid as light red and light blue regions.
Initially, the system is prepared according to Eq.~\eqref{eq: initial condition TWA} for both modes shown in the first column of Fig.~\ref{fig: figure 4}. 
During time evolution, the Wigner distribution migrates towards the phase-space region of the LT, see second and further columns, circulating while remaining confined within the toroidal attractor.
Remarkably, the distribution is well-confined to the phase-space region of the classical LT.
Single trajectories remain confined in this structure, consistent with Fig.~\ref{fig: figure 2}.
The noise manifests itself along both temporal and scaling dimensions, depicted horizontally and vertically, respectively.
At later times, dephasing at the level of single trajectories induces broadening of the Wigner function
over the entire toroidal structure in both modes.
Similarly, decreasing $\aleph$ results in a broader distribution, indicating that quantum fluctuations speed up the diffusion process.
A visualization of the limit torus dynamics in the full four-dimensional phase space of the system is presented in the \sm.
We also provide an alternative representation of the Wigner function reconstructions shown in Fig.~\ref{fig: figure 4} based on two-dimensional Gaussian kernel density estimates; see \sm. 

%
\subsection{Universal scaling}
\label{subsec: Universal scaling}
\begin{figure}
\centering
\includegraphics[width=\columnwidth]{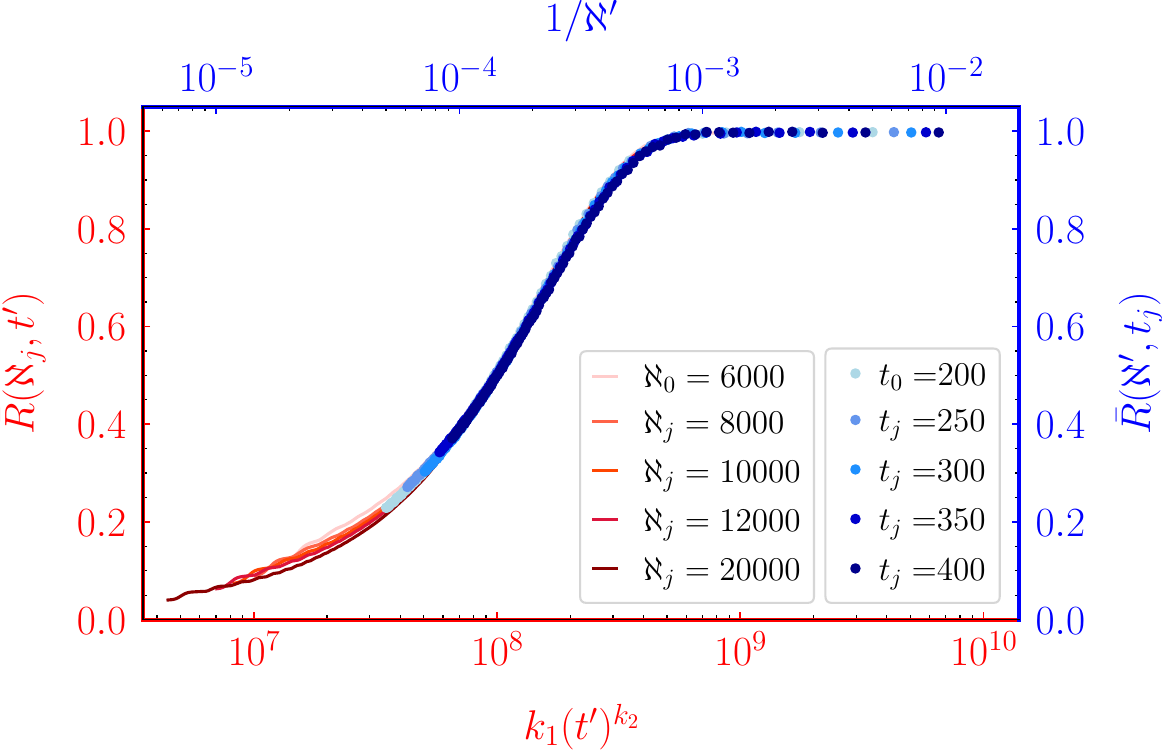}
\caption{
Universal scaling behavior of the quantum-to-classical crossover.
Data collapse of the period-averaged circular variance is achieved by plotting $\bar{R}$ against
(i) the rescaled time $t^{\prime}$ (red curves and axes), $\bar{R}(\aleph_j,t')$, and
(ii) the rescaled scaling parameter $\aleph^{\prime}$ (blue curves and axes), $\bar{R}(\aleph',t_j)$.
The time (red) and inverse scaling parameter (blue) axes are rescaled according to Eq.~\eqref{eq: scaling relations}, revealing a power-law dependence of the relaxation dynamic along both dimensions.
Further rescaling the time axis according to Eq.~\eqref{eq: time rescaling 2} achieves an overall data collapse, highlighting the universality of the scaling behavior.
}
\label{fig: figure 5}
\end{figure}

Having established the emergence of LT and their melting via both trajectory dynamics and spectral signatures, we now investigate how coherence degrades as a function of 
time $t$ and  scaling parameter $\aleph$. In the following, we demonstrate that this melting follows universal scaling laws.
To this end, we quantify phase diffusion in phase space via the circular variance,
\begin{equation}
		R(t) = 1 - \abs{\frac{1}{N_{\rm traj}} \sum_{n=1}^{N_{\rm traj}} e^{i\theta_n(t)}}\;,
        \label{eq: circular variance 2}
\end{equation}
where $\theta_n$ denotes the phase-space angle of the $n$th trajectory, and $N_{\rm traj}$ is the total number of trajectories. We note that the scaling parameter $\aleph$ is closely related to the number of photons in the cavities.
The metric in Eq.~\eqref{eq: circular variance 2} exploits the intrinsic circular symmetry of the system dynamics in phase space, and serves as a sensitive measure of phase diffusion~\cite{Ferrari2025_2, Kruglikov2025}.
For an ideally circular-invariant state, $R$ solely and faithfully captures the dephasing dynamics.
However, when the Wigner distribution deviates from perfect circular symmetry, see Fig.~\ref{fig: figure 4}, oscillations in the mean phase-space angle emerge, thereby introducing an additional time-dependent modulation in $R$.
As the deviation from perfect circular symmetry, and thus the oscillation in $R$, stem solely from the initialization, and are not associated with the melting of the LT structure, we mitigate their influence by introducing the period-averaged circular variance
\begin{equation}
\label{eq: period-averaged circular variance}
    \bar{R}(t) = \frac{1}{T}\int_{t-T/2}^{t+T/2} d\tau R(\tau)\;,
\end{equation}
where $T$ is the oscillation period of the mean phase-space angle. 
We compute $\bar{R}(\aleph,t)$ in cavity 1 over increasing values of the scaling parameter $\aleph$ and time $t$, fitting the resulting curves to an exponential relaxation form,
$\bar{R}(x) = 1 - \exp(-\delta_{\bar{R}} x)$, with $x$ representing either time or the inverse scaling parameter $1/\aleph$. 
The relaxation rate $\delta_{\bar{R}}(x)$ thus plays an analogous role to the Liouvillian gap extracted earlier from the emission spectrum: Both quantify the characteristic decay rate associated with the dephasing of coherent dynamics due to quantum noise~\footnote{We note that the same scaling analysis can equivalently be performed using the time-resolved spectrum Eq.~\eqref{eq: spectrum_mode_1}, which resolves both temporal decay at fixed $\aleph$ and its complementary dependence on $\aleph$ at fixed time.}.
The relaxation rate $\delta_{\bar{R}}(x)$ is found to exhibit a power-law dependence, $\delta_{\bar{R}}(x) = c x^d$, with $c$ and $d$ positive constants.
Figure~\ref{fig: figure 5} illustrates the scaling property of $\delta_{\bar{R}}(x)$. The time dependence of $\bar{R}(\aleph,t)$ is displayed for five fixed values $\aleph_j$, while the dependence on the inverse scaling parameter $1/\aleph$ is shown for five fixed time points $t_j$.
Rescaling the time and the $1/\aleph$ axis using the coefficients that characterize the power-law dependencies of $\delta_{\bar{R}}(t)$ and $\delta_{\bar{R}}(1/\aleph)$,
\begin{equation}
    \begin{aligned}
        t \to t^{\prime}
        &= (\aleph_j/\aleph_0)^{d_1} t
        \\
        1/\aleph \to 1/\aleph^{\prime}
        &= (t_j/t_0)^{d_2} 1/\aleph
    \end{aligned}
\label{eq: scaling relations}
\end{equation}
leads to a scaling collapse of the curves, with $d_1\approx 1.08$ and $d_2\approx0.90$.
We find that the two scaling dimensions, time and $\aleph$, are intrinsically linked by the relation
\begin{equation}
t = \alpha\aleph^\beta\;,
\label{eq: universality relation}
\end{equation}
with $\alpha = (c_1/c_2)^{1/d_2} = 1.04\pm0.01$ and $\beta = d_1/d_2 = -1.196\pm0.001$.
In Fig.~\ref{fig: figure 5}, this universality is further demonstrated by a collapse of the data when Eq.~\eqref{eq: universality relation} is used to rescale the time axis as
\begin{equation}
    t^{\prime} \to k_1 (t^{\prime})^{k_2}\;,
\label{eq: time rescaling 2}
\end{equation}
with
$k_1 = \left(c_1/c_2\right)^{d_2/d_1-1} t_0^{d_2}/\aleph_0 ^{d_1}$,
$k_2 = -d_2/d_1$.
This clearly shows that the rate of phase diffusion is inherently tied to the scaling parameter and quantum fluctuations. 
The scaling relation demonstrates a universal mechanism for the emergence of classical behavior from quantum dynamics, highlighting the universal nature of dephasing in quantum systems exhibiting LT dynamics.

\subsection{Consistent indicators of quantum melting}
We now demonstrate that the observables analyzed in Secs.~\ref{subsec: Quantum melting of LT dynamics} and \ref{subsec: Universal scaling} reflect the same underlying mechanism: quantum-fluctuation-induced dephasing of coherent torus dynamics.
While conceptually distinct, the Liouvillian spectral gap and the phase diffusion measured by circular variance yield consistent scaling behavior.
Specifically, the Liouvillian gap $\Lambda_j$ extracted from the emission spectrum defined in Eq.~\eqref{eq: time-resolved emission spectrum} scales as $\aleph^{-a_j}$, while the relaxation rate $\delta_{\bar{R}}$ of the period-averaged circular variance scales as $\aleph^{-d_1}$.
Both exponents are approximately unity, with $a_j \approx 1.05$ and $d_1 \approx 1.08$, as seen in Eq.~\eqref{eq: power-law liouvillian gap}.
This agreement confirms that both observables capture the same fluctuation-driven relaxation process.
Importantly, the spectral approach probes dissipation along each angular direction of the torus individually, while the circular variance captures global phase decoherence.
The agreement between their scaling exponents therefore implies that quantum dephasing is isotropic across the torus manifold and that the underlying mechanism is universal.

This dual characterization not only strengthens the interpretation of LT melting as a universal, fluctuation-induced crossover but also provides two experimentally accessible probes for identifying this regime.
Furthermore, it demonstrates that the universal scaling law, revealed along one angular dimension of the LT via circular variance, applies uniformly across all toroidal directions.
This reveals a deep structural invariance of the scaling, highlighting a new universality class for topologically nontrivial attractors in open quantum systems.

\subsection{Quantum melting of LT as an emergent dynamical critical crossover}
The above analysis of the Liouvillian spectrum shows that the gradual melting of LT dynamics with increasing quantum fluctuations is governed by the softening of toroidal phase modes.
We observe an algebraic Liouvillian gap closing $\Lambda_j\propto\aleph^{-1}$, while the steady state is unique and remains qualitatively unchanged for finite $\aleph$.
The algebraic gap closing thus does not indicate a dissipative phase transition, but a universal critical-like crossover governed by symmetry breaking in the classical limit and the algebraic restoration of these symmetries at finite $\aleph$.
The algebraic scaling is also distinct from exponential gap closing in metastable systems, as for example observed in dissipative bistable Kerr resonators, where slow dynamics originate from noise-activated switching between metastable attractors.
In our case, the slow dynamics stems from diffusive phase motion along the torus manifold, rather than activation-driven switching.
Below we clarify the microscopic mechanism underlying the universal critical-like crossover.

\subsubsection{Symmetry structure and soft modes}
The Hamiltonian Eq.~\eqref{eq: Hamiltonian} and the Liouvillian Eq.~\eqref{eq: Master equation} are both invariant under a global continuous $\mathrm{U}(1)$ symmetry
\begin{equation}
    \ao_1 \rightarrow e^{i\phi} \ao_1\;,
    \qquad
    \ao_2 \rightarrow e^{2i\phi} \ao_2\;.
\end{equation}
Additionally, $\mathcal{L}$ is invariant under continuous time translations, as it is explicitly time-independent.
In the classical limit, the LT breaks the global $\mathrm{U}(1)$ symmetry and the continuous time-translation symmetry, giving rise to two independent phase directions associated with the incommensurate frequencies $\nu_{1,2}$.
Applying the open-system Nambu-Goldstone theorem~\cite{Hidaka2020} yields two diffusive Goldstone modes.
One of these modes is a \q{mixed} mode, resulting from a nonzero commutator between the $\mathrm{U}(1)$ generator and the time-translation generator. The second mode corresponds to phase diffusion along the remaining torus angle.
For any finite system parameters and $\aleph$, the Liouvillian generically possesses a unique, time-translationally invariant steady state~\cite{Albert2014, Nigro2019, Minganti2018}, indicating that all spontaneously broken symmetries are eventually restored at long times.
The Goldstone modes therefore acquire finite damping rates.

\subsubsection{Softening of toroidal Goldstone modes}
In the classical limit $\aleph\to\infty$, each incommensurate frequency parametrizes an independent phase direction on the torus arising from the breaking of time-translation symmetry~\cite{Strogatz2019, Goldenfeld2018}.
A two-frequency torus therefore supports two independent Goldstone modes, each associated with phase motion along one of the torus angles.
Quantum fluctuations restore these symmetries, softening the Goldstone modes into damped modes with finite coherence times.
This softening manifests in the Liouvillian spectrum: in the classical limit, each gapless Goldstone mode appears as a purely imaginary pair of eigenvalues,
\begin{equation}
\lambda_j^\pm = \pm i \nu_j\;,
\end{equation}
characterizing undamped phase motion.
At finite $\aleph$, quantum corrections introduce small negative real parts,
\begin{equation}
    \lambda_j^\pm = -\Lambda_j\pm i \nu_j\;,
\end{equation}
where the Liouvillian gaps $\Lambda_j$ quantify phase diffusion along the corresponding angular directions and define the finite lifetimes of the soft modes.
This generalizes previous results for limit cycles~\cite{Alaeian2021, Dutta2025}, where the breaking of a single continuous time-translation symmetry yields one soft mode.
For LTs, the presence of two incommensurate frequencies, and thus two independent phase directions generated by the single broken time-translation symmetry on the LT, gives rise to two distinct soft modes, reflecting the richer quasiperiodic symmetry-breaking structure of the torus.

\subsubsection{Critical-like behavior and algebraic softening}
Dissipative critical phenomena in open quantum systems are commonly classified according to the closing of the Liouvillian gap:
(i) exponential closing, typically associated with bistability, metastability, and first-order-like transitions and
(ii) algebraic closing, which is linked to diffusive soft modes and continuous (second-order) dynamical criticality~\cite{Casteels2017a, Minganti2018, Ptaszynski2024}.
In our system, $\Lambda_j \propto \aleph^{-1}$, implying diverging coherence times $\tau_j \sim \Lambda_j^{-1}$ and dynamical critical slowing down of quasiperiodic phase modes as the semiclassical limit $\aleph \to \infty$ is approached~\cite{Beaulieu2025}.
Importantly, this algebraic gap softening does not signal a dissipative phase transition at finite parameters: The Liouvillian retains a unique steady state for all finite $\aleph$~\cite{Minganti2018,Macieszczak2016}. Moreover, the low-lying spectrum does not exhibit quasidegenerate real eigenvalues indicative of metastable sectors or noise-activated switching between competing attractors~\cite{Dykman2007,Chan2007} as, for example, observed in bistable Kerr resonators. 
Instead, the slow dynamics originates from diffusive phase motion along the torus manifold.
More precisely, the algebraic closing reflects the gradual softening of diffusive phase (Goldstone-like) modes associated with motion along the angular directions of the torus, which acquire finite damping at finite $\aleph$ due to quantum fluctuations.
The system therefore undergoes a continuous, fluctuation-induced deformation from coherent quasiperiodic attractor dynamics to an incoherent, noise-dominated steady state.
We thus interpret the observed algebraic scaling as a universal quantum-to-classical critical crossover governed by diffusive phase decoherence rather than a genuine dissipative phase transition.
\subsubsection{Two-dimensional scaling structure}
Despite the system being $0+1$-dimensional, i.e., zero spatial dimensions plus time, we uncover a two-dimensional scaling structure in $(\aleph, t)$.
In the absence of spatial degrees of freedom, conventional finite-size scaling does not apply. Instead, the scaling parameter $\aleph$, which controls photon occupation and noise strength, acts as an effective system size governing the dynamical scaling of the Liouvillian soft modes.
A genuine dynamical critical point emerges in the limit $\aleph \to \infty$, even in the absence of spatial structure~\cite{Beaulieu2025}.
Simultaneously, the nontrivial scaling relation $t\sim\aleph^\beta$ shows that time itself becomes a critical scaling dimension.
The resulting dynamical scaling collapse signals the emergence of a universal behavior, echoing space-time scaling phenomena seen in Berezinskii-Kosterlitz-Thouless transitions and quasi-long-range order in 2D condensates such as superfluids, where a genuine order arises only asymptotically in a singular limit (here, $\aleph\to\infty$).

Taken together, these results show that the \q{melting} of LT dynamics is not a dissipative phase transition in the conventional sense but a universal quantum-to-classical crossover governed by the algebraic restoration of broken symmetries.
The softening of toroidal Goldstone modes provides a unified framework connecting the spectral signatures, trajectory-level phase diffusion, and the emergent scaling structure across ($\aleph,t$).
This identifies a novel class of dynamical universality in open quantum systems, combining topological attractors, symmetry-protected soft modes, and fluctuation-driven decoherence in low-dimensional systems.

\subsection{Robustness against single-photon loss and thermal noise}
To ensure experimental relevance, we evaluated the robustness of our results against unavoidable single-photon loss and thermal fluctuations. 
These effects are incorporated into the master equation and TWA formalism, as detailed in the Supplemental Material~\cite{supmat}. 
We systematically analyzed how key quantities, such as the Liouvillian gap and the circular variance, are affected by these perturbations.
These findings demonstrate that our conclusions remain valid under experimentally realistic levels of dissipation and thermal noise.

%
\section{Experimental Realization}
\label{sec: experimental realization}

\subsection{Trapped ions}
The driven-dissipative coupled Kerr oscillator system described by Eqs.~\eqref{eq: Hamiltonian} and \eqref{eq: Master equation} can be experimentally realized with trapped ions~\cite{Thomas2022,Lee2013}. In this realization, two motional modes of a single ion serve as the bosonic modes~\cite{Steinbach_1997}, offering precise control over system parameters and facilitating the study of quantum-to-classical crossovers.

\subsubsection{Motional modes and trapping potential}
In a linear Paul trap, an ion experiences harmonic confinement along the orthogonal directions. By adjusting the trap potentials, one can achieve nearly degenerate motional frequencies, $\omega_1 \approx \omega_2$, as required for the model. The noninteracting Hamiltonian, including internal electronic states, is given by
\begin{equation}
    \dfrac{\hat{H}_0}{\hbar} = \omega_1 \hat{a}^{\dagger}_1\hat{a}_1 +\omega_2 \hat{a}^{\dagger}_2\hat{a}_2 +\sum_j \omega_{eg}^{(j)} \ket{e_j}\bra{e_j} \;,
\end{equation}
where $\omega_{eg}^{(j)}$ denotes the energy difference between the $j$th electronic excited state $\ket{e_j}$ and the electronic ground state $\ket{g}$.

\subsubsection{Engineering Kerr nonlinearity}
Anharmonicities in the trapping potential can be introduced with multipole trap designs or by applying tailored static electric fields~\cite{Kajita_2022_BOOK, Agarwal_1998, Walz_1994}, leading to a Kerr-type nonlinearity in the motional modes. In the interaction picture and under the rotating wave approximation (RWA), the quartic potential terms $\propto (\hat{a}_k+\hat{a}^{\dagger}_k)^4$ yield an effective Kerr interaction:
\begin{equation}
\dfrac{\hat{V}^{\text{RWA}}_I}{\hbar}\approx \frac{U_k}{2}\hat{a}^{\dagger}_k\hat{a}^{\dagger}_k\hat{a}_k\hat{a}_k \;,
\end{equation}
where $U_k$ is the Kerr coefficient, tunable via the trap's electric field configuration.

\subsubsection{Nonlinear mode coupling}
Nonlinear coupling between the two motional modes can be achieved through laser driving that induces two-phonon and single-phonon transitions via an intermediate electronic state $\ket{e_j}$~\cite{Thomas2022}.
This is achieved by applying two laser fields that off-resonantly drive the second red sideband of mode 1 and the first red sideband of mode 2.
The laser addressing mode 1 has frequency $\omega_{d,1}$ with detuning from the electronic transition defined as $\delta_1^{(j)} \equiv \omega_{d,1} - \omega_{eg}^{(j)} = -2\omega_1 + \Delta$.
The laser addressing mode 2 has frequency $\omega_{d,2}$, with detuning $\delta_2^{(j)} \equiv \omega_{d,2} - \omega_{eg}^{(j)} =  -\omega_2 + \Delta$. Here $\Delta$ is the common detuning of the drives from the respective sidebands. Under these conditions and within the RWA, the resulting interaction Hamiltonian takes the form:
\begin{equation}
\begin{aligned}
\dfrac{\hat{V}^{\text{RWA}}_I}{\hbar} \approx
        &
        -\frac{\eta_{\text{LD},1}^2\Omega_1}{4}
        \hat{\sigma}_{+}^{(j)}
        \hat{a}_1\hat{a}_1
        e^{-i\Delta t +i\phi_1}
        + \mathrm{H.c.}
        \\
        &
        +
        i \frac{\eta_{\text{LD},2}\Omega_2}{2}
        \hat{\sigma}_{+}^{(j)}
        \hat{a}_2
         e^{-i\Delta t +i\phi_2}
        + \mathrm{H.c.} \;,
\end{aligned}
\end{equation}
where $\hat{\sigma}_{+}^{(j)}$ and $\hat{\sigma}_{-}^{(j)}$ are the raising and lowering operators between $\ket{g}$ and $\ket{e_j}$,
$\eta_{\text{LD},k}$ are the Lamb-Dicke parameters,
$\Omega_k$ the Rabi frequencies, and 
$\phi_k$ the phases of the driving fields.

Under the condition $\abs{\Delta} \gg \Omega_k$, the excited state can be adiabatically eliminated~\cite{Cirac1992, Reiter2012} leading to an effective nonlinear coupling term:
\begin{equation} 
\dfrac{\hat{H}_{\text{int}}}{\hbar} = J \left( \hat{a}_1^{\dagger} \hat{a}_1^{\dagger} \hat{a}_2 + \hat{a}_2^{\dagger} \hat{a}_1 \hat{a}_1 \right), \end{equation}
with coupling strength $J = \frac{\eta_{\text{LD},1}^2 \eta_{\text{LD},2} \Omega_1 \Omega_2}{8\abs{\Delta}}$, adjustable via laser parameters. This effective coupling arises at second order through virtual excitation of the electronic state; see \sm.

\subsubsection{Dissipative processes}
Dissipation, as described in Eq.~\eqref{eq: Master equation}, can be engineered using sideband cooling and heating techniques~\cite{Lee2013}. For each mode individually, incoherent pumping is implemented by driving the first blue sideband, i.e. $+\omega_k$ detuning, resulting in the following:
\begin{equation}
\dfrac{\hat{V}^{\text{RWA}}_I}{\hbar} \approx
    \sum_{k=1,2}\frac{\eta_{\text{LD},k}\Omega_k}{2}
    \left( i\hat{\sigma}_{+}^{(j)}\hat{a}^{\dagger}_k e^{i\phi_k}
    +\mathrm{H.c.} \right)\;,
\label{eq: V_int incoherent pump}
\end{equation}
while two-phonon loss is achieved by driving the second red sideband, i.e. $-2\omega_k$ detuning:
\begin{equation}
\dfrac{\hat{V}^{\text{RWA}}_I}{\hbar}
        \approx
        -\sum_{k=1,2}
        \frac{\eta_{\text{LD},k}^2\Omega_k}{4}
        \left(
              \hat{\sigma}_{+}^{(j)} \hat{a}_k\hat{a}_k e^{i\phi_k}
            + \mathrm{H.c.}
        \right) \;.
\label{eq: V_int two-phonon loss}
\end{equation}
Note that the two processes described in Eqs.~\eqref{eq: V_int incoherent pump} and~\eqref{eq: V_int two-phonon loss} can be independently implemented using different internal states $\ket{e_j}$.
Under the assumption of rapid decay of the excited state, i.e., decay rate $\Gamma_j\gg\eta_{\text{LD},k}\Omega_k$, adiabatic elimination leads to effective dissipators in the master equation:
\begin{equation}
\gamma_{k,\text{eff}}\mathcal{D}[\hat{a}^{\dagger}_k]\hat{\rho}
\quad \text{and}\quad 
\eta_{k,\text{eff}}\mathcal{D}[\hat{a}_k^2]\hat{\rho}
\;,
\end{equation}
with rates 
\begin{equation}
\gamma_{k,\text{eff}} = \eta_{\text{LD},k}^2 \Omega_k^2 / \Gamma_j
\;\;\text{and}\;\;
\eta_{k,\text{eff}} = \eta_{\text{LD},k}^4 \Omega_k^2 / (4\Gamma_j)
\;,
\end{equation}
controllable via laser intensities; see~\sm.

\subsection{Alternative platforms}
Alternatively, the model can be realized in circuit QED using two microwave cavities, each coupled to Josephson junction elements to produce on-site Kerr nonlinearities $U_k$~\cite{Hillmann2022}. 
The nonlinear tunneling $J$ is induced by a parametric pump at frequency $ \omega_p = 2\omega_1 - \omega_2$, enabling three-wave mixing between the cavities~\cite{Lescanne2020}. 
Two-photon loss $\eta_k$ is engineered by coupling each cavity to a lossy auxiliary mode via similar parametric drives, while incoherent pumping $\gamma_k$ can be emulated either by coupling each cavity to a driven, lossy transmon qubit or by injecting broadband microwave noise into the input line~\cite{Narla2016}.

Cavity optomechanical systems offer another platform, where the two dissipative modes are realized as mechanical oscillators in a \q{membrane-in-the-middle} setup. Here nonlinear coupling is achieved by coupling both mechanical modes to an additional cavity mode~\cite{Thomas2022, Walter2014}.

\subsection{Quantum-to-classical scaling}
The crossover from quantum to classical behavior, characterized by the scaling parameter $\aleph$, can be probed across a variety of experimental platforms by tuning $U_k$, $J$, and the effective dissipation rates according to Eq.~\eqref{eq: scaling relation}.
As apparent from the rescaling in Eq.~\eqref{eq: scaling relation}, the classical regime corresponds to large occupations of the oscillator modes, whereas the quantum regime corresponds to low occupations.

An alternative approach to accessing the quantum-to-classical scaling involves a dual-scaling procedure of system parameters, similarly to the procedure proposed in Ref.~\cite{Beaulieu2025}.
In this case, the scaling relation in Eq.~\eqref{eq: scaling relation} is replaced by the modified relation
\begin{equation}
    \omega_k = \tilde{\omega}_k \aleph \, ,\;
    J = \tilde{J}\sqrt{\aleph}\,, \;\; \text{and}\;\;
    \gamma_k = \tilde{\gamma}_k \aleph  \;.
\label{eq: dual_scaling}
\end{equation}

\subsection{State characterization}
Both, trapped ion and circuit QED systems permit direct measurement of the bosonic states through techniques such as Wigner function tomography~\cite{Ding_2017, Lutterbach_1997, Hofheinz2009, He2024}, enabling reconstruction of phase-space distributions described in Sec.~\ref{subsec: Universal scaling}. While single quantum trajectories can be challenging, ensemble measurements provide access to averaged quantities, enabling comparison with the predictions discussed above.

\section{Conclusions}
\label{sec: Conclusions}
In this work, we have developed a quantum theory of limit tori (LTs), establishing a dynamical crossover that bridges open quantum dynamics and nonlinear classical dynamics.
By focusing on a minimal model of two coupled driven-dissipative Kerr cavities, we systematically explored the quantum-to-classical crossover using Liouvillian spectral theory and stochastic unraveling of the density matrix. Our results reveal a universal structure underlying this crossover, suggesting a broader connection between classical attractor dynamics and dissipative quantum systems.

In the classical limit, we identified robust LTs characterized by quasiperiodic motion on a toroidal phase-space manifold. 
Each incommensurate frequency of the torus is encoded in the Liouvillian spectrum as a pair of purely imaginary eigenvalues, corresponding to Goldstone-like modes arising from the spontaneous breaking of continuous time-translation symmetry.
As quantum fluctuations increase, these eigenvalues acquire finite negative real parts, indicating the onset of quantum phase diffusion and the restoration of time-translation symmetry, a process we term quantum melting.


This quantum melting manifests as a breakdown of persistent quasiperiodicity. Through analysis of individual quantum trajectories, we identified quantum-fluctuation-induced dephasing as the microscopic mechanism driving this transition. 
Remarkably, even in the presence of strong noise, these trajectories remain confined to a topologically nontrivial toroidal manifold, demonstrating the robustness of the underlying structure.


The crossover is characterized by matching power-law scaling in both the Liouvillian spectral gap and a circular-variance-based order parameter, revealing a two-dimensional dynamical scaling structure in system size and time. 
Time itself emerges as an effective critical dimension, a hallmark of dynamical criticality reminiscent of Berezinskii-Kosterlitz-Thouless transitions, superfluid crossovers, and scaling phenomena in low-dimensional dissipative systems. 
This establishes quantum melting as an emergent dynamical critical crossover governed by universal scaling laws.


We further propose concrete implementations of LTs and their quantum melting in trapped ion systems and superconducting circuits, offering clear dynamical signatures and observables for detection.

Our findings contribute to a broader theoretical framework that connects topologically nontrivial attractors in classical nonlinear systems with their quantum counterparts. 
They also enrich current efforts to understand quantum chaos in open quantum systems~\cite{Zanardi2021, Yoshimura2024, Richter2024, Ferrari2025}, where the complex structure of the Liouvillian spectrum plays a central role in signaling chaotic dynamics and their relation to classical chaos.

Together, these advances offer a path toward classifying, stabilizing, and realizing coherent quasiperiodic behavior in engineered platforms operating at the frontier between classical and quantum nonlinear dynamics.
Understanding how quantum fluctuations modify classical attractors will not only deepen our theoretical understanding but also provide insights into the design of robust quantum states for sensing, simulation, and information processing.
%

\begin{acknowledgments}
We thank D. K. J. Bone{\ss} and L. P. Peyruchat for constructive feedback on the manuscript. C.N. and L.M. acknowledge funding by the Cluster of Excellence \q{Advanced Imaging of Matter} (EXC 2056), Project No. 390715994. The project is co-financed by ERDF of the European Union and by \q{Fonds of the Hamburg Ministry of Science, Research, Equalities and Districts (BWFGB).} K.S. acknowledges funding from the Deutsche Forschungsgemeinschaft (DFG) via Project No. 449653034.
\end{acknowledgments}

\bibliographystyle{KilianStyle}
\bibliography{biblio_Q_LT_paper.bib}

\clearpage
\newpage

\onecolumngrid
\appendix

\renewcommand{\appendixname}{}

\setcounter{section}{0}
\renewcommand{\thesection}{S\arabic{section}}
\renewcommand{\thesubsection}{\Alph{subsection}}

\renewcommand{\thefigure}{\thesection\arabic{figure}}
\counterwithin{figure}{section}

\renewcommand{\theequation}{\thesection\arabic{equation}}
\counterwithin{equation}{section}

\renewcommand{\thetable}{\thesection\arabic{table}}
\counterwithin{table}{section}

\newgeometry{
	top=1.5cm,
	bottom=1.4cm,
	left=1.65cm,
	right=1.65cm
}

\section*{Supplemental Material for \q{Universal quantum melting of quasiperiodic attractors in driven-dissipative cavities}}

\begin{center}
\begin{minipage}{0.75\textwidth}
	\centering
	Caroline Nowoczyn,\textsuperscript{1,2}
	Ludwig Mathey,\textsuperscript{1,2}
	and Kilian Seibold\textsuperscript{3}
	
	\medskip
	
	\textsuperscript{1}\textit{Center for Optical Quantum Technologies and Institute for Quantum Physics,
	University of Hamburg, Hamburg 22761, Germany}
	
	\textsuperscript{2}\textit{The Hamburg Center for Ultrafast Imaging, Hamburg 22761, Germany}
	
	\textsuperscript{3}\textit{Department of Physics, University of Konstanz, 78464 Konstanz, Germany}
	%
	%
\end{minipage}
\end{center}

In this Supplemental Material, we begin by deriving the classical Gross-Pitaevskii equations (GPE) from the quantum master equation.
In the second section, we present phase diagrams that map out the classical dynamical regimes of the system---such as limit cycles, limit tori, and chaos---based on the Lyapunov spectra.
The third section analyzes the GPE frequency spectrum, highlighting the presence of two incommensurate fundamental frequencies and their harmonic structure.
In the fourth section, we derive the truncated Wigner approximation (TWA) used to describe the quantum dynamics of the system.
In the fifth section, we provide an alternative visual representation of the Wigner function reconstructions using a two-dimensional Gaussian kernel density estimate (KDE).
In the sixth section, we discuss the full 4D visualization of the TWA dynamics on the classical torus attractor that is provided in the form of mp4 movies (files: \texttt{Movie\_S1\_aleph\_2000.mp4} and \texttt{Movie\_S2\_aleph\_6000.mp4}).
In the seventh section, we detail the procedure used to extract the universal scaling behavior of the Liouvillian gap and the circular variance.
Next, we investigate the robustness of our results against quantum and thermal fluctuations, including the effects of single-photon loss and finite temperature on toroidal attractors.
Finally, we provide a comprehensive outline of a trapped-ion implementation capable of realizing the required nonlinear interactions and dissipation channels in an experimental setting.
Interested readers are welcome to contact the authors for additional details or complementary results.

\section*{Contents}
\startcontents[supplement]

\titlecontents{section}
[2.5em]                  
{}
{\contentslabel{2.5em}}  
{}
{\titlerule*[0.5pc]{.}\contentspage}

\titlecontents{subsection}
[5.0em]
{}
{\contentslabel{3.0em}}
{}
{\titlerule*[0.5pc]{.}\contentspage}

\titlecontents{subsubsection}
[5.0em]
{}
{\contentslabel{3.0em}}
{}
{\titlerule*[0.5pc]{.}\contentspage}

\printcontents[supplement]{}{1}[2]{}
\clearpage


\section{Derivation of the Gross-Pitaevskii Equation}

In this section, we derive the GPE, which describes the dynamics of the system in the classical limit.
We start from the Lindblad master equation for the density matrix $\hat{\rho}$, as defined in Eq.~(2) in the main text:
\begin{equation}
	\frac{d\hat{\rho}}{dt} = \mathcal{L}\hat{\rho} = \frac{1}{i\hbar}\comm*{\hat{\mathcal{H}}}{ \hat{\rho}}
	+ \sum_{k=1,2} \gamma_k \mathcal{D}[\hat{a}_k^\dagger] \hat{\rho} + \eta_k \mathcal{D}[\hat{a}_k^2] \hat{\rho},
	\label{eq: supp: master_equation}
\end{equation}
where
$\mathcal{D}[\hat{L}]\hat{\rho} = \hat{L} \hat{\rho} \hat{L}^\dagger - \frac{1}{2} \{\hat{L}^\dagger \hat{L}, \hat{\rho}\}$
denotes the Lindblad dissipator.
In this quantum description, the time evolution of the expectation values of $\hat{a}_k$, $\hat{a}_k^{\dagger}$ can be computed via
\begin{equation}
	\frac{d}{dt}\expval*{\hat{a}_k}
	=
	\expval{\hat{a}_k \frac{d\hat{\rho}}{dt} }
	=
	\frac{i}{\hbar}\expval*{\comm*{\hat{\mathcal{H}}}{\hat{a}_k}}
	+\sum_{k=1,2} \gamma_k \expval*{\bar{\mathcal{D}}[\hat{a}_k^{\dagger}]\hat{a}_k}
	+\sum_{k=1,2} \eta_k \expval*{\bar{\mathcal{D}}[\hat{a}_k^2]\hat{a}_k}\;,
	\label{eq: supp: adjoint_master_equation}
\end{equation}
with adjoint Lindblad dissipator defined as
$\bar{D}[\hat{L}]\hat{O}=\hat{L}^{\dagger}\hat{O}\hat{L}-\frac{1}{2}\{\hat{L}^{\dagger}\hat{L},\hat{O}\}$.
With the Hamiltonian $\hat{\mathcal{H}}$ defined by Eq.~(1) in the main text, this leads to
\begin{equation}
	\begin{cases}
		\frac{d}{dt}\expval*{\hat{a}_1}
		=
		-i\left(\omega_1\expval*{\hat{a}_1} + U_1\expval*{\hat{a}_1^{\dagger}\hat{a}_1\hat{a}_1} -J\expval*{2\hat{a}_1^{\dagger}\hat{a}_2 }\right)
		-\frac{\gamma_1}{2}\expval*{\hat{a}_1}
		-\eta_1\expval*{\hat{a}_1^{\dagger}\hat{a}_1^2}
		\\
		\frac{d}{dt}\expval*{\hat{a}_2}
		=
		-i\left(\omega_2\expval*{\hat{a}_2} + U_2\expval*{\hat{a}_2^{\dagger}\hat{a}_2\hat{a}_2} -J\expval*{\hat{a}_1^2}\right)
		-\frac{\gamma_2}{2}\expval*{\hat{a}_2}
		-\eta_2\expval*{\hat{a}_2^{\dagger}\hat{a}_2^2} \;.
	\end{cases}
	\label{eq: supp: adjoint_master_equation_explicit}
\end{equation}
We consider the mean-field approximation $\alpha_k(t) = \langle \hat{a}_k \rangle$ and factorize higher-order moments:
\[
\langle \hat{a}_j^\dagger \hat{a}_k \rangle \approx \alpha_j^* \alpha_k, \qquad \langle \hat{a}_j \hat{a}_k \rangle \approx \alpha_j \alpha_k,
\]
which yields the classical Gross-Pitaevskii equations, as given in Eq.(3) in the main text:
\begin{equation}
	\begin{cases}
		i\frac{d\alpha_1}{dt} =
		\omega_1 \alpha_1 + U_1 \abs{\alpha_1}^2\alpha_1 - 2J\alpha^*_1\alpha_2 -\frac{\gamma_1}{2}\alpha_1 - \eta_1 \abs{\alpha_1}^2\alpha_1
		\\
		i\frac{d\alpha_2}{dt} =
		\omega_2 \alpha_2 + U_2 \abs{\alpha_2}^2\alpha_2 - J\alpha_1^2 -\frac{\gamma_2}{2}\alpha_2 -\eta_2 \abs{\alpha_2}^2\alpha_2 \;.
	\end{cases}
\end{equation}  

\section{Lyapunov Phase Diagrams}
To characterize the classical dynamical behavior of the system and identify possible parameter regions of periodic, quasiperiodic and chaotic behavior, we perform a stability analysis based on the computation of the Lyapunov exponents $\{\lambda_i\}$, which quantify the average rates of exponential divergence (or convergence) of initially infinitesimal close trajectories along different directions~\cite{Escot2020, Benettin1980}. Specifically, $\lambda_i$ are defined via the long-time limit
\begin{equation}
	\lambda_i=\lim_{t\to\infty}\frac{1}{t}\ln\left( \frac{\lVert \delta\boldsymbol{\alpha}_i(t)\rVert}{\lVert \delta\boldsymbol{\alpha}_i(0)\rVert} \right)
	\;,
\end{equation}
with $\lVert\cdot\rVert$ denoting the vector norm, $\boldsymbol{\alpha}$ the state vector, and $\delta\boldsymbol{\alpha}_i(t)$ is the evolution of a small initial perturbation $\delta\boldsymbol{\alpha}_i(0)$ along the direction $i$.
This implies that in the long-time limit, the norm of the perturbation behaves like
\begin{equation}
	\lVert\delta\boldsymbol{\alpha}_i(t)\rVert \approx e^{\lambda_i t} \lVert\delta\boldsymbol{\alpha}_i(0)\rVert\;.
\end{equation}
As can be seen from this equation, a negative Lyapunov exponent $\lambda_i$ means that trajectories are converging along the corresponding direction. A positive exponents means trajectories are diverging, indicating chaotic behavior. Lyapunov exponents equal to zero indicate neutrally stable directions of the system and can thus support the classification of (quasi-)periodic dynamics.
The number of zero exponents determines the dimensionality of the invariant manifold~\cite{Escot2020}.

Given the dynamical equations
\begin{equation}
	\frac{d\boldsymbol{\alpha}}{dt} = \boldsymbol{F}(\boldsymbol{\alpha}),
\end{equation}
the linearized dynamics for a perturbation $\delta \boldsymbol{\alpha}$ is:
\begin{equation}
	\frac{d}{dt} \delta \boldsymbol{\alpha} = \mathbf{J}(\boldsymbol{\alpha}(t)) \cdot \delta \boldsymbol{\alpha}\;,
\end{equation}
where $\mathbf{J}$ is the Jacobian matrix of $\boldsymbol{F}$. This forms the basis for numerical computation of Lyapunov exponents.
Following the method originally stated in~\cite{Benettin1980}, we evolve a set of orthonormal perturbation vectors along a trajectory and use the QR decomposition or Gram-Schmidt orthonormalization to extract the full Lyapunov spectrum. The numerical simulations were performed using the \textit{DynamicalSystems.jl} library~\cite{Datseris2018}.

The number of Lyapunov exponents in the spectrum is given by the number of dimension of the system.
The maximal Lyapunov exponent $\lambda_{\max}$, is used to classify the system’s dynamics:
\begin{itemize}[noitemsep, topsep=0pt]
	\item $\lambda_{\max} > 0$: chaotic motion (sensitive dependence on initial conditions)
	\item $\lambda_{\max} = 0$: periodic (limit cycle) or quasiperiodic behavior
	\item $\lambda_{\max} < 0$: convergence to a fixed point (stable equilibrium) 
\end{itemize}
More refined dynamics classification can be achieved via the full Lyapunov spectrum~\cite{Baier_1991_BOOK}:
\begin{itemize}[noitemsep, topsep=0pt]
	\item A limit cycle (1D periodic orbit) exhibits one zero exponent and all other exponents negative.
	\item A 2-torus (quasiperiodic motion with two incommensurate frequencies) gives two zero exponents and remaining exponents negative. 
	\item Higher-dimensional tori correspondingly feature more zero exponents and remaining exponents negative.
\end{itemize}
However, we note this criterion is necessary but not always sufficient, as Lyapunov exponents equal to zero can also indicate conserved quantities or continuous symmetries of the system.
In fact, our system has a global continuous $U(1)$ symmetry, see Eq.(18) in the main text. As a consequence, there is always at least one Lyapunov exponent equal to zero, associated with the direction of the $U(1)$ symmetry.
This occurs because the presence of the continuous symmetry prohibits the existence of an isolated fixed point, instead it only allows for a continuous $U(1)$ orbit of fixed points, and this orbit defines a neutrally stable direction of motion.

\begin{figure}[h]
	\centering
	\includegraphics[width=0.75\linewidth]{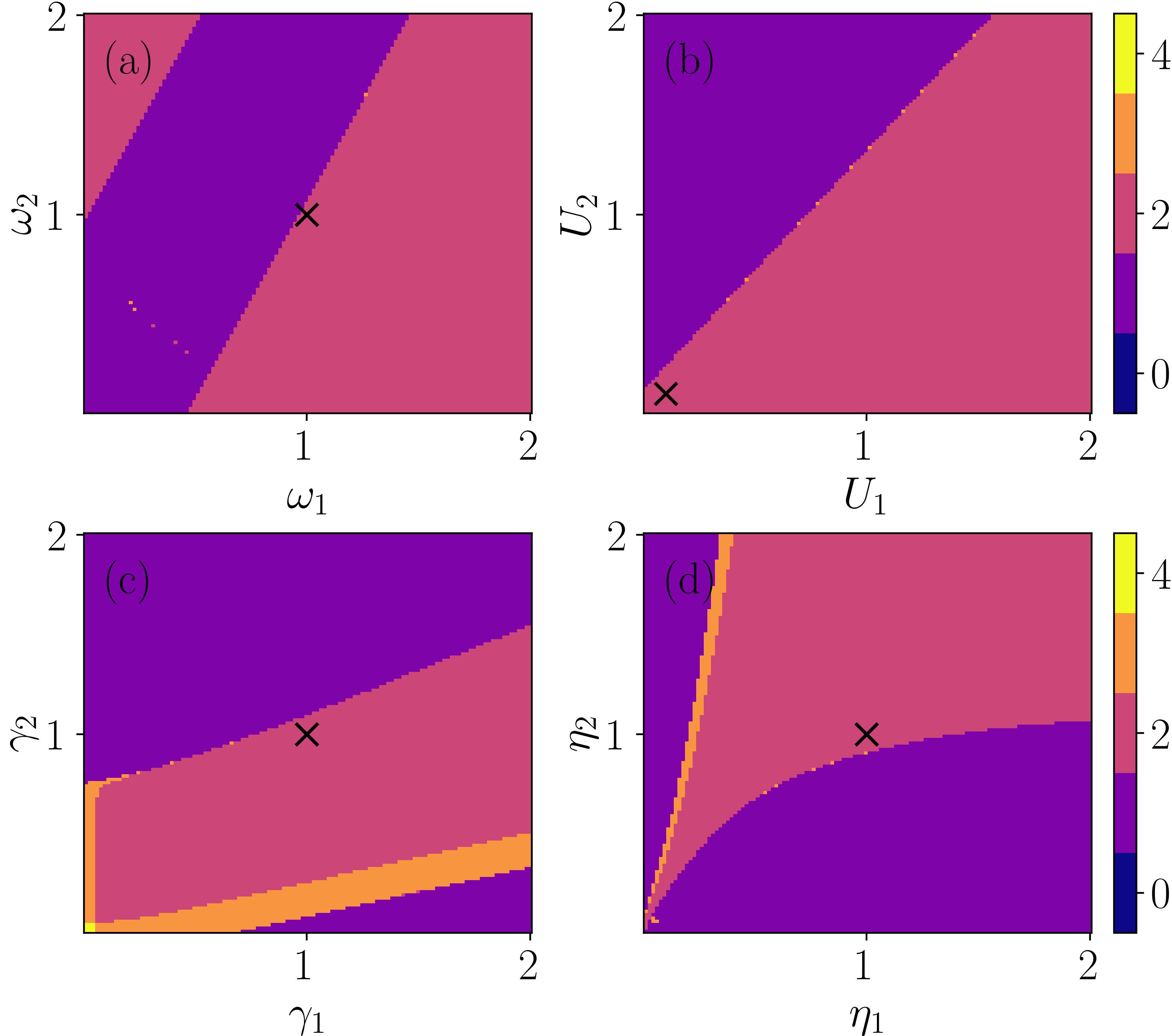}
	\caption{
		Phase diagrams showing the number of zero (marginal) Lyapunov exponents computed from the GPE. The number of zero exponents indicates the dimension of the underlying neutral manifold. 
		The number of marginal Lyapunov exponents is computed numerically using a threshold $5\times10^{-2}$, where exponents with absolute value less than this are considered approximately zero.
		Here we study the asymmetry of parameters in cavity 1 and cavity 2.
		Parameters used in the main text are marked by \q{$\times$}.
	}
	\label{fig: supp: Lyapunov 1}
\end{figure}

\begin{figure}[h]
	\centering
	\includegraphics[width=0.75\linewidth]{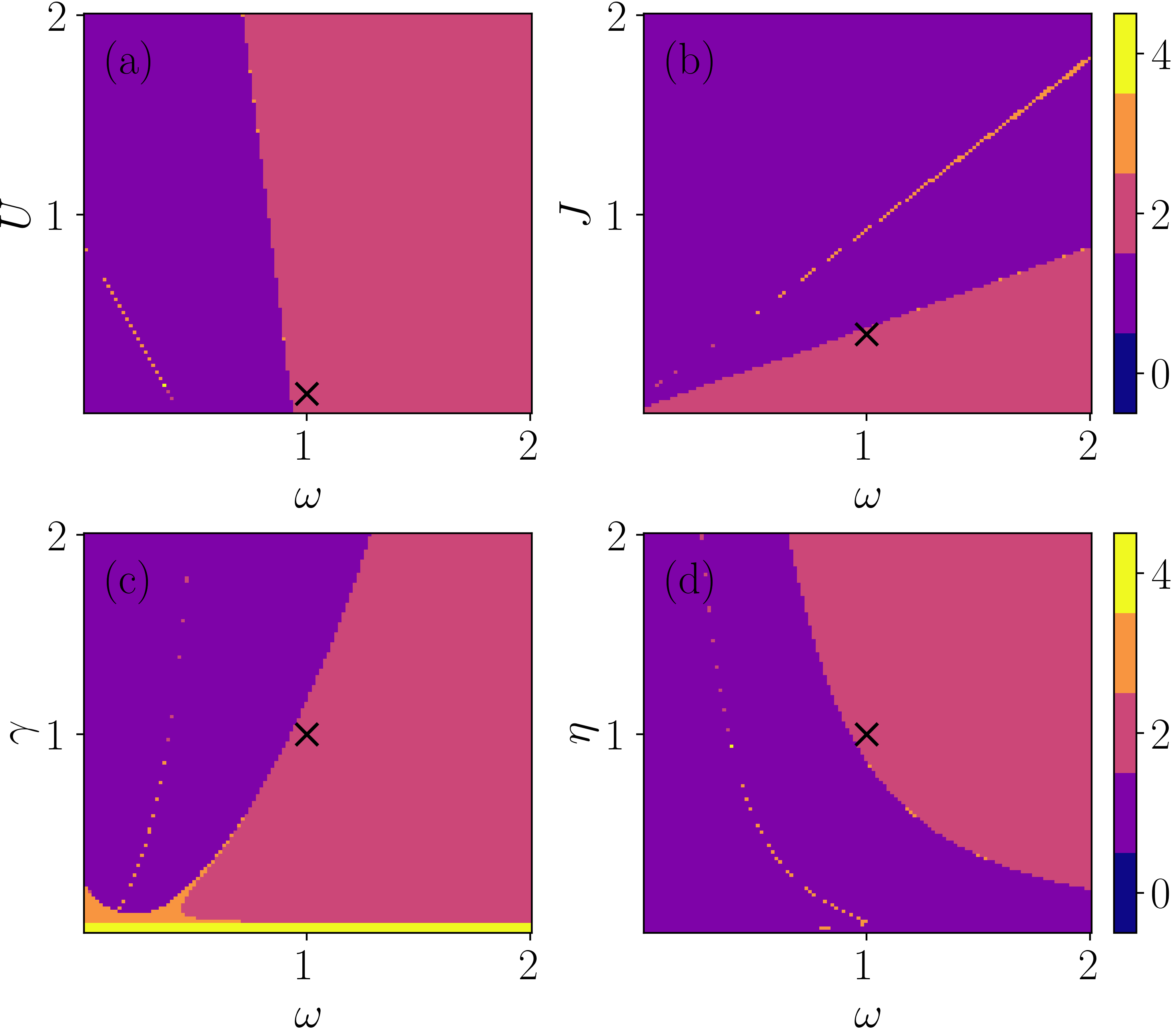}
	\caption{
		Phase diagrams showing the number of zero (marginal) Lyapunov exponents computed from the GPE.
		Here, the varied parameter is applied identically to both cavity 1 and cavity 2, while the cavity frequencies are systematically tuned.
		Parameters used in the main text are marked by \q{$\times$}.
	}
	\label{fig: supp: Lyapunov 2}
\end{figure}

Figures \ref{fig: supp: Lyapunov 1} and \ref{fig: supp: Lyapunov 2} show the number of eigenvalues in the Lyapunov spectra that are zero for different system parameters.
We first note that all other (non-zero) exponents are negative, indicating the absence of chaotic dynamics in the explored parameter regime.
As explained above, due to the continuous $U(1)$ symmetry there is always at least one Lyapunov exponent equal to zero.
The parameters used in the main text lie in the region of two zero Lyapunov coefficients, supporting the classification of the studied system as a 2-torus.
In the next section, this classification is confirmed by analyzing the spectrum of the GPE, confirming the presence of two incommensurate frequencies that lead to the observed quasiperiodic motion.


\section{GPE spectrum}
\begin{figure}[h]
	\centering
	\includegraphics[width=0.85\linewidth]{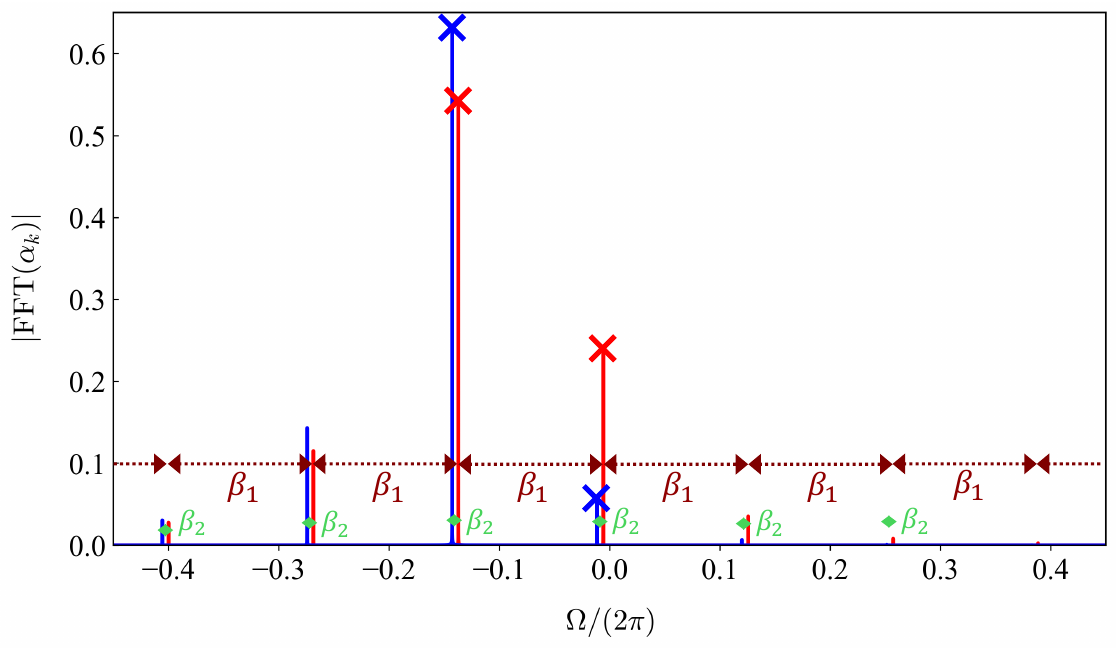}
	\caption{
		Frequency spectrum of the limit torus, obtained from the classical GPE.
		The spectrum of cavity 1 is shown in red, the spectrum of cavity 2 is shown in blue.
		The spectrum shows two fundamental peaks at incommensurate frequencies and additional harmonics in each of the cavities. The fundamental peaks are marked by red and blue markers \q{$\times$}, respectively.
		Further the spectrum shows periodic structures with peaks that are equidistant in frequency, with period spacing $\Delta f = \beta_1$ and $\Delta f = \beta_2$, indicated by brown and green dotted arrows, respectively.
		The system parameters are the same used throughout the LT analysis in the main text: $\omega_1=\omega_2=1$, $U_1=U_2=0.1$, $J=0.4$, $\gamma_1=\gamma_2=1$, $\eta_1=\eta_2=1$.
		The discrete FFT is performed for a signal of length $T=10^6$, corresponding to a frequency resolution of $\Delta \Omega/2\pi = 10^{-6}$.
	}
	\label{fig: supp: GPE spectrum}
\end{figure}

In this section, we analyze the frequency spectrum defined in Eq.(9) of the main text, obtained from the classical GPE.
The spectrum is shown in Figure~\ref{fig: supp: GPE spectrum}, for
the same LT supporting parameters as we use in the main text, $\omega_1=\omega_2=1$, $U_1=U_2=0.1$, $J=0.4$, $\gamma_1=\gamma_2=1$, $\eta_1=\eta_2=1$, thus corresponding to the cut at $\omega=1.0$ marked by the dashed-dotted line in Fig.1(b) of the main text.
The quasiperiodic motion of the limit torus is reflected by two fundamental, incommensurate frequencies within each of the cavities.
These fundamental frequencies, $\nu_{1,2}^{\text{cav}1}$ and $\nu_{1,2}^{\text{cav}2}$, are highlighted in Fig.~\ref{fig: supp: GPE spectrum} by red and blue \q{$\times$} markers, respectively.
Using a numerical discrete Fourier transform with frequency resolution $\Delta \Omega/(2\pi)=10^{-6}$, we find
\begin{equation}
	\begin{aligned}
		\nu_{1}^{\text{cav}1}/(2\pi) &= -0.005662\\
		\nu_{2}^{\text{cav}1}/(2\pi) &= -0.137070\\
		\nu_{1}^{\text{cav}2}/(2\pi) &= -0.011325\\
		\nu_{2}^{\text{cav}2}/(2\pi) &= -0.142732 \;.
	\end{aligned}
\end{equation}
The frequencies $\nu_{1,2}^{\text{cav}1}$ are the LT modes of cavity 1 that are analyzed across the quantum-to-classical crossover in the main text.
The ratios of the fundamental torus frequencies are given by:
\begin{equation}
	\begin{aligned}
		\nu^{\text{cav1}}_2/\nu^{\text{cav1}}_1 &= 24.225874867444325
		\\
		\nu^{\text{cav2}}_2/\nu^{\text{cav2}}_1 &= 12.608833922261484 \;,
	\end{aligned}
	\label{eq: supp: ratios of fundamental frequencies}
\end{equation}
confirming that the frequencies are incommensurate.

In addition to the fundamental frequencies, the spectrum displays harmonics at linear combinations of $\nu_{1,2}^{\text{cav}1,2}$. A complete list of all peak frequencies in the interval $\Omega/(2\pi)\in [-0.5, 0.5]$ is given in Table~\ref{tab: supp: peak_frequencies_gpe_spectrum}.
Notably, the spectrum exhibits periodic frequency structures.
Equal frequency spacings are highlighted in Fig.~\ref{fig: supp: GPE spectrum} by dotted arrows.
We observe two different period spacings,
\begin{equation}
	\begin{aligned}
		\beta_1 &= 0.131408
		\\
		\beta_2 &= 0.005662 \;.
	\end{aligned}
\end{equation}

\begin{table}[h]
	\renewcommand{\arraystretch}{1.4}
	\centering
	\begin{tabular}{ |c|c|c| }
		\hline
		peak index $i$ &
		$\nu^{\text{cav1}}_i/(2\pi)$ &
		$\nu^{\text{cav2}}_i/(2\pi)$ \\
		\hline
		1 & -0.005662 & -0.011325\\
		\hline
		2 & -0.137070 & -0.142732\\
		\hline
		3 & -0.268477 & -0.274139\\
		\hline
		4 & -0.399884 & -0.405546\\
		\hline
		5 & 0.125745 & 0.120082 \\
		\hline
		6 & 0.257152 & 0.251490 \\
		\hline
		7 & 0.388559 & 0.382897 \\
		\hline
	\end{tabular}
	\caption{
		Frequencies observed in the GPE spectrum, listing only values $\Omega/(2\pi) \in [-0.5,0.5]$.
		The frequency resolution of the numerically performed discrete FFT is $\Delta \Omega/(2\pi)=10^{-6}$.
	}
	\label{tab: supp: peak_frequencies_gpe_spectrum}
\end{table}


\section{Truncated Wigner Approximation}
\label{sec: appendix TWA}
We study how quantum fluctuations affect system dynamics using the truncated Wigner approximation (TWA).
The TWA is a phase-space method based on the Wigner-Weyl quantization, which maps bosonic operators $\hat{O}(\ao,\co)$ to their corresponding Weyl symbols $O_W(\alpha,\alpha^*)$~\cite{Polkovnikov2010},
\begin{equation}
	\label{eq: weyl symbol}
	O_W(\alpha,\alpha^*) = \iint d\lambda d\lambda^* \bra{\alpha-\lambda/2}\hat{O}(\hat{a},\hat{a}^{\dagger})\ket{\alpha+\lambda/2}e^{\frac{1}{2}(\lambda^*\hat{a} -\lambda\hat{a}^{\dagger})} \;.
\end{equation}
The Weyl symbol of the density matrix, referred to as \textit{Wigner distribution} $W(\alpha,\alpha^*)$, represents quantum states as quasiprobability distributions in complex phase space.
Operators acting on the density matrix are corresponding to differential operators---also referred to as \textit{Bopp operators}---acting on the Wigner distribution \cite{Gardiner2000}:
\begin{equation}
	\label{eq: operator mapping TWA}
	\begin{aligned}
		\ao \hat{\rho} &\to \left(\alpha+\frac{1}{2}\pdv{}{\alpha^*}\right)W\\
		\co \hat{\rho} &\to \left(\alpha^*-\frac{1}{2}\pdv{}{\alpha}\right)W\\
		\hat{\rho}\ao   &\to \left(\alpha-\frac{1}{2}\pdv{}{\alpha^*}\right)W\\
		\hat{\rho}\co   &\to \left(\alpha^*+\frac{1}{2}\pdv{}{\alpha}\right)W
	\end{aligned}
\end{equation}
Applying this mapping to the rescaled master equation defined by Eqs.~(1, 2) and scaling relation Eq.~(4) in the main text, we obtain the following equation for the time-evolution of the Wigner distribution in terms of rescaled system parameters and rescaled complex phase-space variables $\tilde{\alpha}_k(t)$:
\begin{equation}
	\begin{aligned}
		\pdv{W}{t}
		=
		-\bigg\{
		&\sum_{k=1,2}
		(-i)\pdv{}{\tilde{\alpha}_k} \tilde{\alpha}_k
		\left[
		\omega_k
		+i\frac{\gamma_k}{2}
		+(\tilde{U}_k -i\tilde{\eta}_k) \left(\abs{\tilde{\alpha}_k}^2-\frac{1}{\aleph}\right)
		\right]
		+2i \tilde{J} \pdv{}{\tilde{\alpha}_1}  \tilde{\alpha}_1^*\tilde{\alpha}_2
		+i \tilde{J}\frac{\partial}{\partial\tilde{\alpha}_2}  (\tilde{\alpha}_1)^2 
		+c.c.
		\bigg\} W
		\\
		+\bigg\{
		&\frac{1}{2\aleph}\sum_{k=1,2}
		\left[
		\gamma_k
		\frac{\partial^2}{\partial\tilde{\alpha}_k \partial\tilde{\alpha}_k^*}
		+2\tilde{\eta}_k
		\frac{\partial^2}{\partial\tilde{\alpha}_k\partial\tilde{\alpha}_k^*}(2\abs{\tilde{\alpha}_k}^2-\frac{1}{\aleph})
		\right]
		\bigg\} W
		\\
		+ \bigg\{
		&\frac{1}{4\aleph^2}\sum_{k=1,2}
		(-i)\frac{\partial^3}{\partial\tilde{\alpha}_k \partial\tilde{\alpha}_k \partial\tilde{\alpha}_k^*}
		\left( \tilde{U}_k +i\tilde{\eta}_k \right) \tilde{\alpha}_k
		-i\tilde{J}\frac{\partial^3}{\partial\tilde{\alpha}_1^* \partial\tilde{\alpha}_1^* \partial\tilde{\alpha}_2}
		+c.c.
		\bigg\} W
	\end{aligned}
\end{equation}
Within a semi-classical approximation ($\aleph \gg 1$) we can neglect contributions $\mathcal{O}( (1/\aleph)^2)$ and higher.
Under this approximation, the above equation of motion takes the form of a Fokker-Planck equation
\begin{equation}
	\frac{\partial}{\partial t}W(\tilde{\boldsymbol{\alpha}},\tilde{\boldsymbol{\alpha}}^*) = \left[-\frac{\partial}{\partial \tilde{\alpha}_i}A_i(\tilde{\boldsymbol{\alpha}})+\frac{1}{2}\frac{\partial^2}{\partial \tilde{\alpha}_i\partial \tilde{\alpha}_j^*}D_{ij}(\tilde{\boldsymbol{\alpha}})\right]W(\tilde{\boldsymbol{\alpha}},\tilde{\boldsymbol{\alpha}}^*) 
\end{equation}
where $\boldsymbol{A}$ is the drift matrix and $\boldsymbol{D}$ is the diffusion matrix which can be factorized as $\boldsymbol{D} = \boldsymbol{B}\boldsymbol{B}^T$.
This Fokker-Planck equation describing the evolution of the quasiprobability density $W(\tilde{\boldsymbol{\alpha}},\tilde{\boldsymbol{\alpha}}^*)$ corresponds to the Langevin equations for the variables $\tilde{\alpha}_k$, $\tilde{\alpha}_k^*$, as given in Eq.~(5) in the main text.
In the classical limit $\aleph\to\infty$, the Langevin equations reduce to the GPE equations, given in Eq.~(3) in the main text.


\section{Wigner function reconstructions via Gaussian Kernel Density Estimate (KDE)}
\label{sec: appendix KDE Fig4}

We provide an alternative visual representation of the Wigner function reconstructions shown in Fig.~4 of the main text.
Instead of directly scattering the outcomes of all TWA trajectories onto the subspaces spanned by real- and imaginary parts of the respective modes, here we apply a two-dimensional Gaussian kernel density estimate (KDE) in these subspaces.

The KDE is evaluated on a uniform grid covering the sample support, with a small margin added to prevent edge effects.
Bandwidth selection is based on Scott’s rule of thumb, producing a smooth estimate of the marginal Wigner distribution $W(\mathrm{Re}(\alpha_k),\mathrm{Im}(\alpha_k))$ of mode k, obtained by projecting the TWA ensemble onto the $\alpha_k$ subspace.
The resulting density provides a clear, non-parametric visualization of each mode’s phase-space structure while preserving the statistics encoded by the TWA ensemble.

\begin{figure}
	\centering
	\includegraphics[width=\linewidth]{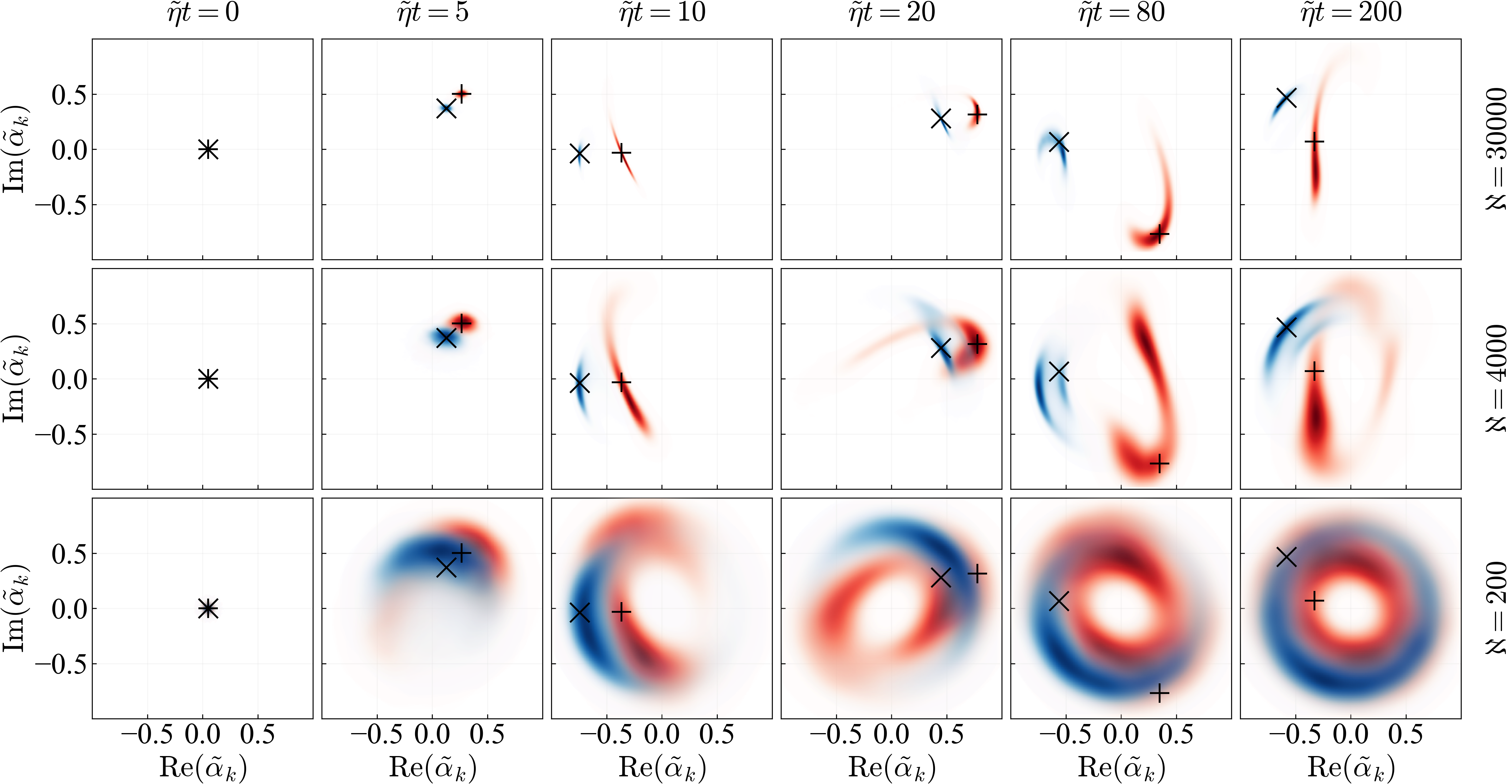}
	\caption{
		Wigner function reconstructions shown as marginal phase-space distributions of modes 1 (red) and 2 (blue), constructed from the same TWA ensemble as Fig.~4 of the main text. The densities are obtained using a two-dimensional Gaussian kernel density estimate (KDE).
	}
	\label{fig: wigner_function_reconstructions_KDE}
\end{figure}

The resulting plot is shown in Fig~\ref{fig: wigner_function_reconstructions_KDE}. It can be seen as a complementary representation of Fig.~4 of the main text, constructed from the same TWA ensemble. The KDE suppresses marker-overlap effects inherent to scatter plots and resolves internal density gradients and fine structure of the phase-space distribution more clearly, in particular making the angular spreading along the toroidal manifold directly visible.

\section{Full 4D visualization of the TWA dynamics on the classical torus attractor}
\label{sec: 4D Wigner reconstructions}
We provide two supplementary movies that visualize the full four-dimensional phase space dynamics of the system of two coupled Kerr cavities (quantum model given by Eqs.~(1) and (2), classical dynamics by the GPE, Eq.~(3); and the TWA equations given in Eq.~(5) in the main text). 
Because the system consists of two bosonic modes, the total phase space is 4D; the movies show the attractor structure without tracing out one subsystem. 
We visualize the 4D dynamics via a 3D projection with axes $Re(\tilde\alpha_1)/\sqrt{\aleph}$ and $Im(\tilde \alpha_1)/\sqrt{\aleph}$, and $|\tilde\alpha_2|^2/\aleph$. 
The classical stationary attractor (GPE stationary solution; transients removed) is a static torus and rendered in light-gray.
The time-dependent TWA Wigner reconstruction is rendered on top of this torus.
The color of the TWA cloud encodes the phase $arg(\tilde \alpha_2)$ using a cyclic colormap with wrap at $\pm\pi$.
Movies S1 and S2 start from the coherent initial condition $\tilde\alpha_k = 0.05$ as also used for all simulations presented in the main text.
Movie S1 corresponds to $\aleph=2000$ (file: \texttt{Movie\_S1\_aleph\_2000.mp4}) and Movie S2 to $\aleph=6000$ (file: \texttt{Movie\_S2\_aleph\_6000.mp4}). 

This illustrates how the TWA ensemble dynamics follows and eventually melts away from the classical LT. 


\section{Scaling and Universality}

\subsection{Extraction of the Liouvillian gap}
\begin{figure}[t]
	\centering
	\includegraphics[width=0.63\linewidth]{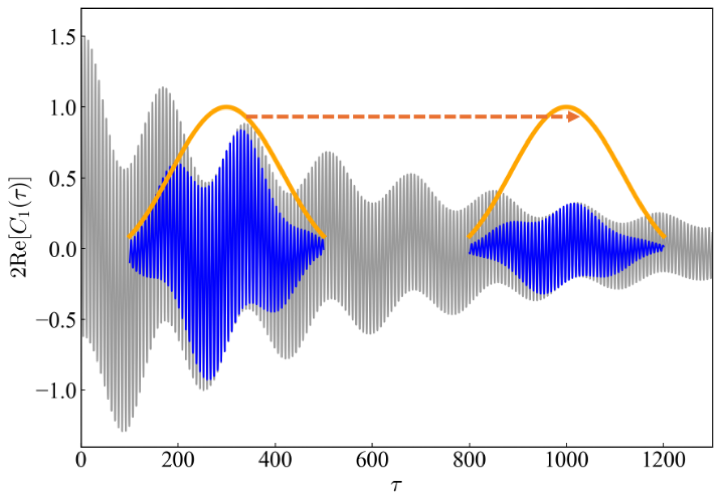}
	\caption{
		Illustration of the windowing process which is used for the STFT.
		The signal $2Re[C_{1}(\tau)]$ is shown in gray, the windowed signal at two different window positions $t_{w,1}=300.0$ and $t_{w,2}=1000.0$ is shown in blue.
		The corresponding window functions $w(t_{w,1})$ and $w(t_{w,2})$ are shown in orange.
		The dashed arrow indicates the shifting of the window function.
		We use a Kaiser window of width $T_{\text{window}} = 400.0$ and with beta-parameter $\beta=4.0$.
		The scaling parameter for the signal shown in this plot is $\aleph=14573.27$.
		The signal is dominated by the two incommensurate torus frequencies:
		$\frac{\nu_1}{2\pi} \approx 0.005662$, which corresponds to a period length of $T_1 \approx 176.72 $, and 
		$\frac{\nu_2}{2\pi} \approx 0.137070$, which corresponds to a period length $T_2 \approx 7.30$.
		Both frequency components decay over time, see also Fig.~\ref{fig: supp: STFT 3D example}.
	}
	\label{fig: supp: STFT scheme}
\end{figure}

In the following, we explain how we numerically extract the Liouvillian gaps
from the time-resolved emission spectrum of cavity 1.
The time-resolved spectrum is obtained by a windowed Fourier transform of the first-order correlation function $C_{1}(\tau)=\langle \ao_1^\dagger(t_0+\tau )\ao_1(t_0)\rangle$, see Eq.~(11) in the main text:
\begin{equation}
	S_1(t,\Omega) = 2 Re \int_{0}^{\infty}\hspace{-0.3cm}d\tau w(t-\tau) \langle \ao_1^\dagger(t_0+\tau )\ao_1(t_0)\rangle e^{i\Omega\tau}  \;,
	\label{eq: supp: first-order correlation function}
\end{equation}
with window function $w(s)$.
Using the truncated Wigner approximation (TWA), we compute the correlation function $C_{1}(\tau)$ for 50 values of the scaling parameter $\aleph \in [800,40000]$. We are therefore in the regime $\aleph \gg 1$, where the approximations made in Sec.~\ref{sec: appendix TWA} are well justified.
For each $\aleph$, we numerically compute the corresponding time-resolved emission spectrum $S_1(t_w,\Omega)$ by performing a discrete short-time Fourier transform (STFT) of the signal $C_{1}(t)$~\cite{2020SciPy-NMeth}.
In this procedure, the window function $w(t-\tau)$ is shifted across the time interval of the signal, see Fig.~\ref{fig: supp: STFT scheme}. At each window position $t=t_{w}$, a single discrete FFT of the windowed signal is computed. In this way, we get access to a time-resolved analysis of the emission spectrum.

\begin{figure}[h!]
	\centering
	\includegraphics[width=0.8\linewidth]{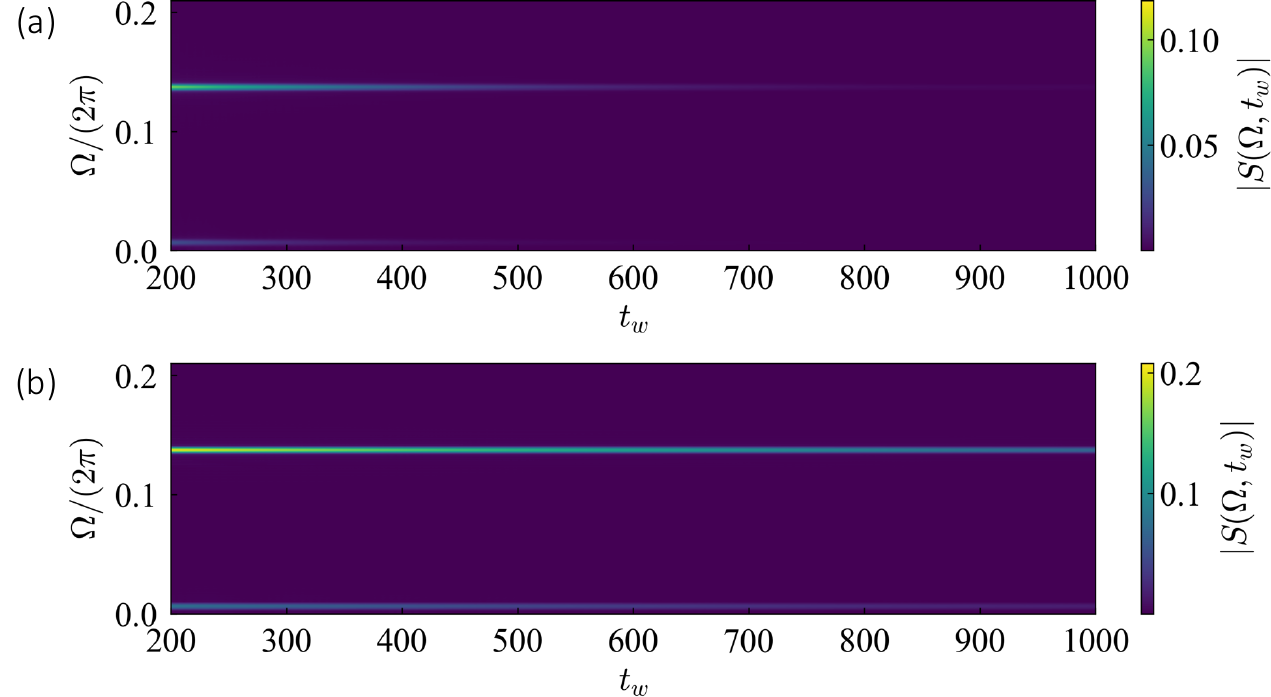}
	\caption{
		Density plot of the time-resolved emission spectrum $S(t_w,\Omega)$ of cavity 1 for
		(a) $\aleph = 4154.23$,
		(b) $\aleph = 14573.27$,
		obtained by a discrete STFT of the first-order correlation function $C_1(\tau)$.
		The peaks are located at the incommensurate torus frequencies $\nu_{j=1,2}$, peak magnitudes decay exponentially over time.
		The decay rates of the peak magnitudes increase for decreasing $\aleph$, as can be seen by comparing (a) and (b): For $\aleph=14573.27$, the lifetime of the two modes is longer than for $\aleph = 4154.23$, with visibly more sustained peaks in the spectrogram.
	}
	\label{fig: supp: STFT 3D example}
\end{figure}
To achieve sufficient resolution in frequency and time, we choose a symmetric Kaiser window of width $T_{\text{window}} = 400.0$, and with beta-parameter $\beta=4.0$.
After initialization, the system undergoes a short transient phase during which the trajectories relax toward the LT attractor. To isolate the stationary LT dynamics, we choose $t_0=20$, thereby excluding early-time transients from the signal.
In Figure~\ref{fig: supp: STFT 3D example}, we show the resulting spectrogram $S_1(t_w,\Omega)$ for two exemplary values of $\aleph$.

The connection between the spectral peaks and the Liouvillian eigenvalues follows from the quantum regression theorem for Markovian Lindblad dynamics. 
For a time-independent Liouvillian superoperator $\mathcal{L}$, two-time correlation functions can be written as 
$\langle \ao_1^\dagger(t)\ao_1(0)\rangle = \mathrm{Tr}\!\left[\ao_1^\dagger e^{\mathcal{L}t}(\ao_1 \rho_{\mathrm{ss}})\right]$. 
Expanding the propagated operator in Liouvillian eigenmodes yields a sum of terms proportional to $e^{\lambda_j t}$, such that the oscillation frequencies are given by $\mathrm{Im}\,\lambda_j$ and the exponential decay rates by $-\mathrm{Re}\,\lambda_j$. 
Consequently, when a spectral peak is dominated by a single isolated Liouvillian mode, its temporal envelope decays exponentially with rate $\Lambda_j = -\mathrm{Re}\,\lambda_j$. 
In the present work, the correlation functions are computed within the truncated Wigner approximation (TWA), which corresponds to an effective Fokker-Planck description valid up to order $\mathcal{O}(1/\aleph)$. 
The extracted decay constants therefore represent effective semiclassical Liouvillian gaps in the regime $\aleph \gg 1$.

The spectrum exhibits two dominant peaks at the incommensurate torus frequencies $\nu_{1}$ and $\nu_{2}$, which correspond to the imaginary parts of the Liouvillian eigenvalues $\lambda_j$ governing the long-time dynamics.
Due to the dephasing of individual TWA trajectories, the magnitudes of these peaks decay exponentially over time.
The decay rates correspond to the small real parts of the Liouvillian eigenvalues $\lambda_j$, and thus directly quantify the Liouvillian gaps $\Lambda_j(\aleph)$.
Since the STFT allows for a time-resolved spectral analysis, we not only get access to the peak positions and their lifetimes, but can also confirm that the peak positions stay constant over time, see Fig.~\ref{fig: supp: STFT 3D example}.
%

%
\begin{figure}[h]
	\centering
	\includegraphics[width=0.7\linewidth]{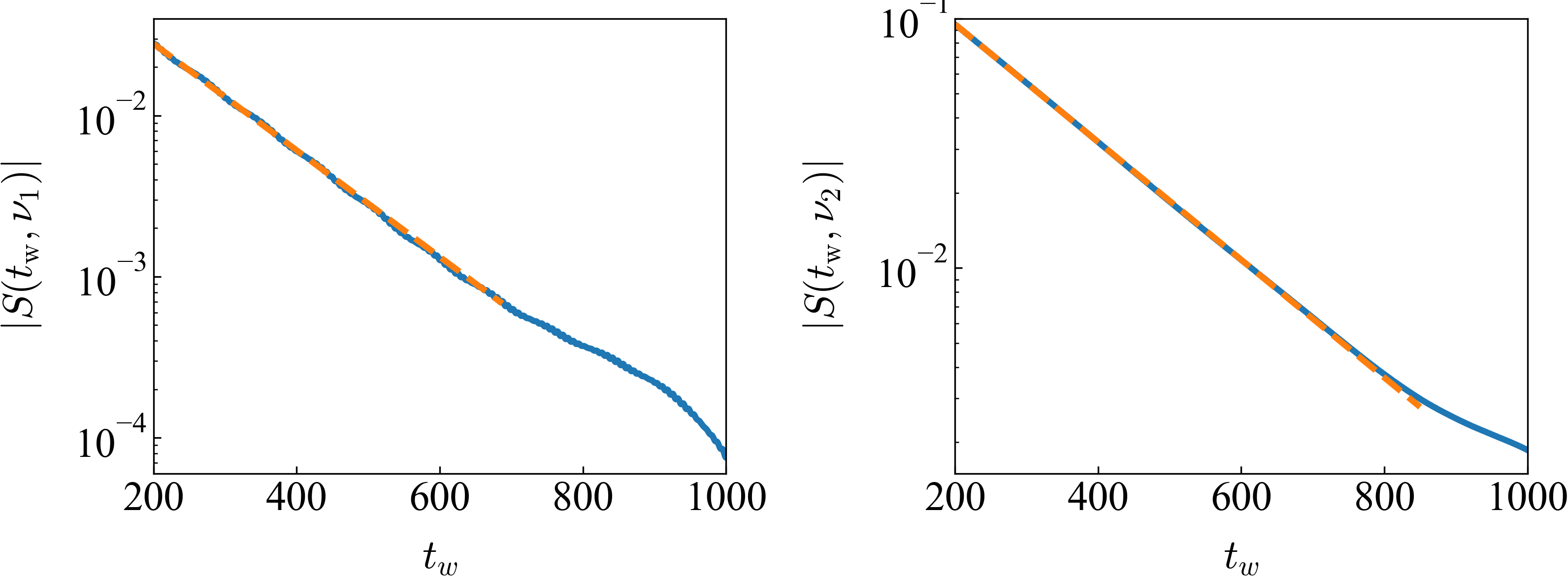}
	\caption{
		Slices of the emission spectrogram $\abs{S(t_w,\Omega)}$ along the the time axis, at the incommensurate torus frequencies $\Omega=\nu_1$ (left) and $\Omega=\nu_2$ (right), for scaling parameter $\aleph=4154.23$
		The fitted exponential functions are shown as orange dashed lines.
		At larger times, $t_w > 700$ for the peak at $\nu_1$, and $t_w > 800$ for the peak at $\nu_2$, the magnitude of the peaks runs into a noise floor. To decrease errors induced by this effect we limit the range of the exponential fit depending on the scaling parameter $\aleph$. Here, the fitting ranges are $t_w\in [200,700]$ for the peak at $\nu_1$, and $t_w\in [200,900]$ for the peak at $\nu_2$, as indicated by the plotting range of the orange dashed lines.
	}
	\label{fig: supp: exp_fits_spectrum_slices_aleph=4154.23}
\end{figure}
\begin{figure}[H]
	\centering
	\includegraphics[width=0.73\linewidth]{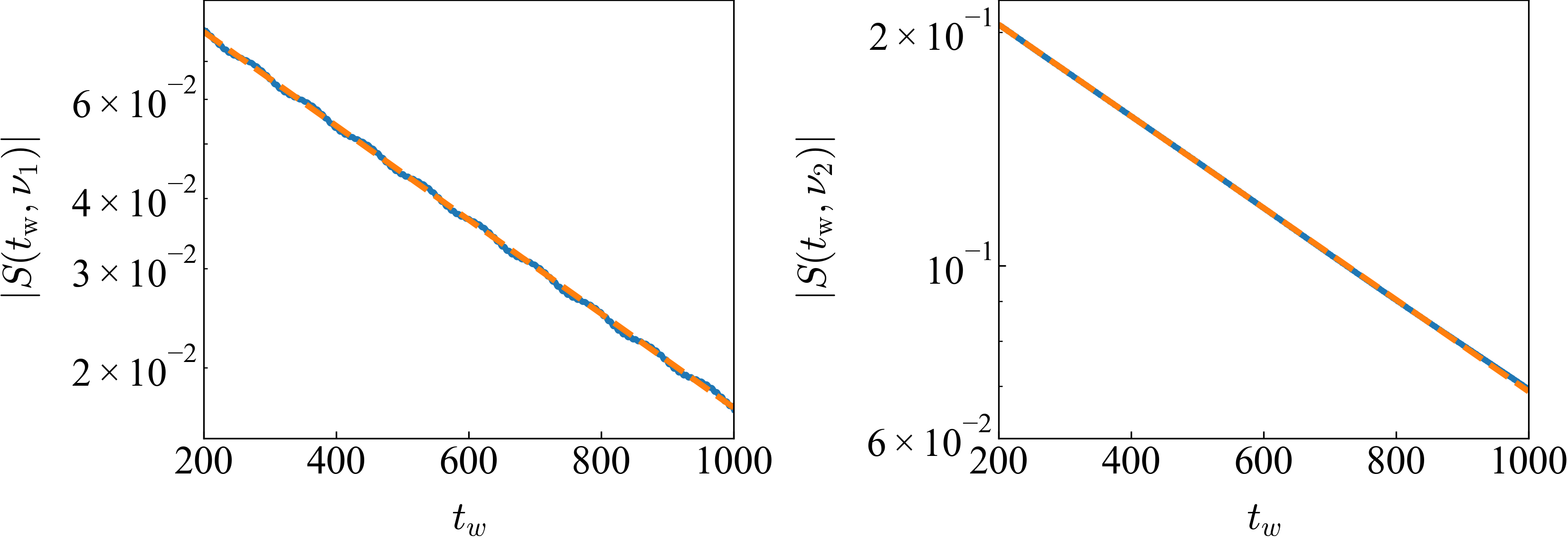}
	\caption{
		Slices of the emission spectrogram $\abs{S(t_w,\Omega)}$ along the the time axis, at the incommensurate torus frequencies $\Omega=\nu_1$ (left) and $\Omega=\nu_2$ (right), for scaling parameter $\aleph=14573.27$.
		The fitted exponential functions are shown as orange dashed lines.
		%
		The detected magnitude of the first peak at $\nu_1$ (left plot) visibly oscillates when shifting the window function. This can be explained by spectral leakage, as the width of the window function does not perfectly fit the frequency of the peak. Because the fitting interval is much larger than the frequency of this oscillation, the impact on the precision of the exponential fit is small.
	}
	\label{fig: supp: exp_fits_spectrum_slices_aleph=14573.27}
\end{figure}
To extract the decay rate $\Lambda_j$ of each of the two peaks, we take two horizontal cuts along the time-axis of the spectrogram $\abs{S(t_w,\Omega)}$ at frequencies $\nu_{1,2}$ and fit the exponential decay of the peak magnitudes along these slices, see Figs.~\ref{fig: supp: exp_fits_spectrum_slices_aleph=4154.23} and~\ref{fig: supp: exp_fits_spectrum_slices_aleph=14573.27}.
In this way, we numerically extract the Liouvillian gaps $\Lambda_j(\aleph)$ for the different values of the scaling parameter $\aleph$. The results, together with the fitted power-law scaling, are shown in Fig.~(3) in the main text.

As the lifetime of the peaks is decreasing for decreasing scaling parameter $\aleph$, this procedure does not allow for fitting the gaps for arbitrarily small $\aleph$.
Accurate and robust exponential fits are obtained if the lifetime of the peak, given by the inverse gap $1/\Lambda_j$, is of the order of the period length of the signal, given by the inverse torus frequency, $T_j = 2\pi/\nu_j$.
This limits the range of good accuracy for the first peak at frequency $\nu_1$ to $\aleph\gtrsim 5000$, and for the second peak at frequency $\nu_2$ to $\aleph\gtrsim 2000$.
For the first peak, the chosen window function only allows to extract the gap $\Lambda_1$ for $\aleph \gtrsim 1800$, which is why in Fig.~(3) of the main text there are no data points for $\Lambda_1$ below this value.
%

%
The results are subject to statistical errors due to the finite number of trajectories in the TWA simulation. To estimate these errors, we use a bootstrapping-resampling procedure, with samples of size $N_{\text{traj}}=10^5$ TWA trajectories, drawn from a dataset of 20 subsets of 5000 trajectories each. We compute the Liouvillian gap as described above for each of the replicates, and estimate the errors via standard deviation of the results. We note that statistical errors increase for small $\aleph$, as stochastic fluctuations are stronger in this regime.
As discussed previously, the TWA is accurate up to order $\mathcal{O}(1/\aleph)$, limiting its range of validity to $\aleph \gg 1$.

We note that the time-resolved spectral analysis used in this section is a powerful tool for analyzing the scaling behavior of multi-frequency structures like LTs.
The method not only allows for an analysis of the scaling with system size $\aleph$, but the time-resolved spectra would also allow for an analysis of the scaling behavior with time $t$ for each individual frequency component. In this work, the scaling with time was studied via the circular-variance metric, see Sec.~\ref{subsec: appendix universality}.

\subsection{Power-law scaling of the Liouvillian gap}
As explained in the previous section, we compute the time-resolved emission spectrum $S(t_w,\Omega)$ for $50$ values of $\aleph \in [800,40000]$ and numerically extract the Liouvillian gaps $\Lambda_j(\aleph)$.
We then numerically fit the data to the power-law
\begin{equation}
	\Lambda_j(\aleph) = \aleph^{-a_j} \times b_j \;.
\end{equation}
Here, we propagate the uncertainties of each data point obtained from the bootstrapping-resampling procedure described in the previous section. This leads to the fitted coefficients and errors as given in the main text.

\begin{figure}[H]
	\centering
	\includegraphics[width=0.65\linewidth]{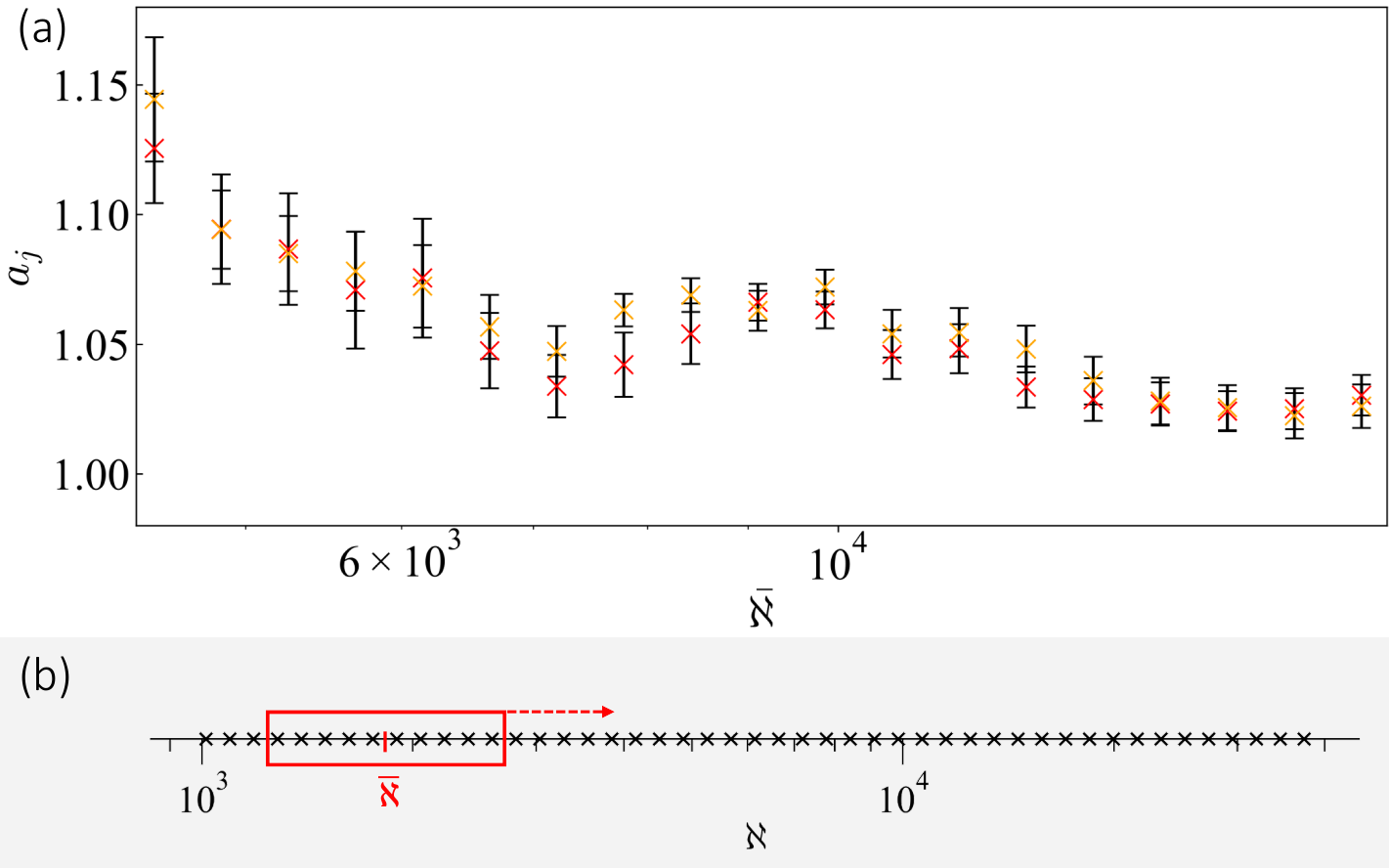}
	\caption{
		Power-law fit of the scaling of the Liouvillian gaps for different positions of the boxcar window $\bar{\aleph}$.
		(a) Fitted power-law coefficients $a_j$ for each of the boxcar-window positions $\bar{\aleph}$.
		(b) Boxcar window used to select the fitting range. The window selects $10$ data points and is shifted from small to large $\aleph$.
	}
	\label{fig: supp: boxcar-windowed power-law fits}
\end{figure}

To get further inside into the numerically extracted power-law coefficient and errors across different regimes of $\aleph$, we perform the power-law fit for different subsets of the available data.
More specifically, we use a boxcar window to select only $10$ points within the range of $\aleph$, see Fig.~\ref{fig: supp: boxcar-windowed power-law fits}(b), and shift this window from small $\aleph$ to large $\aleph$.
For each position of the boxcar window $\bar{\aleph}$, we perform a power-law fit of the data within this window. The resulting power-law coefficients $a_j$ are shown in Fig.~\ref{fig: supp: boxcar-windowed power-law fits}(a).
For higher $\aleph$, stochastic errors of the TWA and fitting errors are decreasing. The fitted power-law coefficients $a_j$ are slightly decreasing and getting closer to $1$ for increasing $\bar{\aleph}$.
As described above, the coefficients $a_1=1.054\pm 0.1$ and $a_2=1.052\pm0.1$ given in the main text are obtained by fitting the power-law for the data within the range $\aleph \in [5000,40000]$.

\newpage
\subsection{Universality}
\label{subsec: appendix universality}
In Sec.~III B of the main text, we show that the relaxation rate of the period-averaged circular variance $\delta_{\bar{R}}$ exhibits power-law scaling with both system size $\aleph$ and time $t$.
We show that the scaling along this two individual dimensions is linked via the relation $t \propto \aleph^{-\beta}$, and demonstrate this universality in Fig.~5 by data collapse of $\bar{R}(\aleph,t)$ under corresponding rescaling transformations.
In the following, we provide a detailed description of the individual steps that lead to the data collapse as presented in Fig.~5.

The circular variance quantifies phase diffusion in phase space along one of the angular directions of the LT. As defined in Eq.~(13) in the main text, it is given by
\begin{equation}
	R(t) = 1 - \abs{\frac{1}{N_{\rm traj}} \sum_{n=1}^{N_{\rm traj}} e^{i\theta_n(t)}}\;,
	\label{eq: supp: circular variance}
\end{equation}
where $\theta_n(t)$ denotes the phase-space angle of the $n$th trajectory, and $N_{\rm traj}$ is the total number of TWA trajectories.
The circular variance can take values between $0$ and $1$, where $R=0$ corresponds to minimal phase diffusion with all $\theta_n$ being equal, and $R=1$ corresponds to maximal phase diffusion with $\theta_n$ being uniformly distributed.

\begin{figure}[b]
	\centering
	\includegraphics[width=0.5\linewidth]{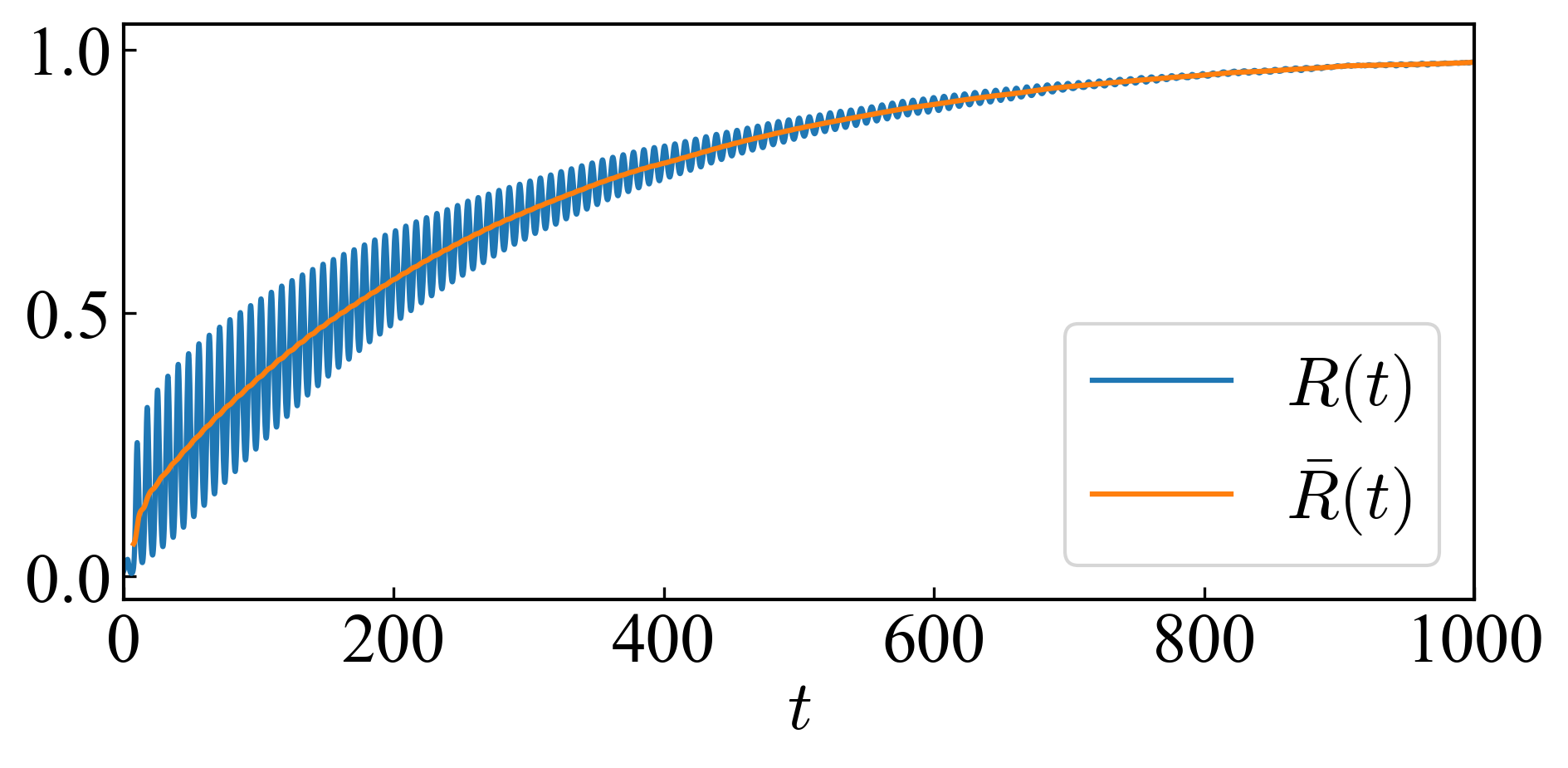}
	\caption{Example of the circular variance $R(t)$ (blue) compared to the period-averaged circular variance $\bar{R}(t)$ (orange), here shown for $\aleph=6000$.}
	\label{fig: supp: example_period_avg_R}
\end{figure}

As explained in the main text, we remove oscillations in $R(t)$ that are caused by deviation of the Wigner distribution from perfect circular symmetry---and therefore do not encode information about melting of the LT structure along the angular direction of interest---by taking the period-averaged circular variance, see also Eq.~(14) in the main text,
\begin{equation}
	\bar{R}(t) = \int_{t-T/2}^{t+T/2} d\tau R(\tau)\;,
	\label{eq: supp: period-averaged circular variance}
\end{equation}
where $T\approx 7.61$ is the oscillation period of the mean phase-space angle.
In Figure~\ref{fig: supp: example_period_avg_R} we show an example comparing the circular variance $R(t)$ to the period-averaged circular variance $\bar{R}(t)$.
%
%
\begin{figure}[t]
	\centering
	\includegraphics[width=0.65\linewidth]{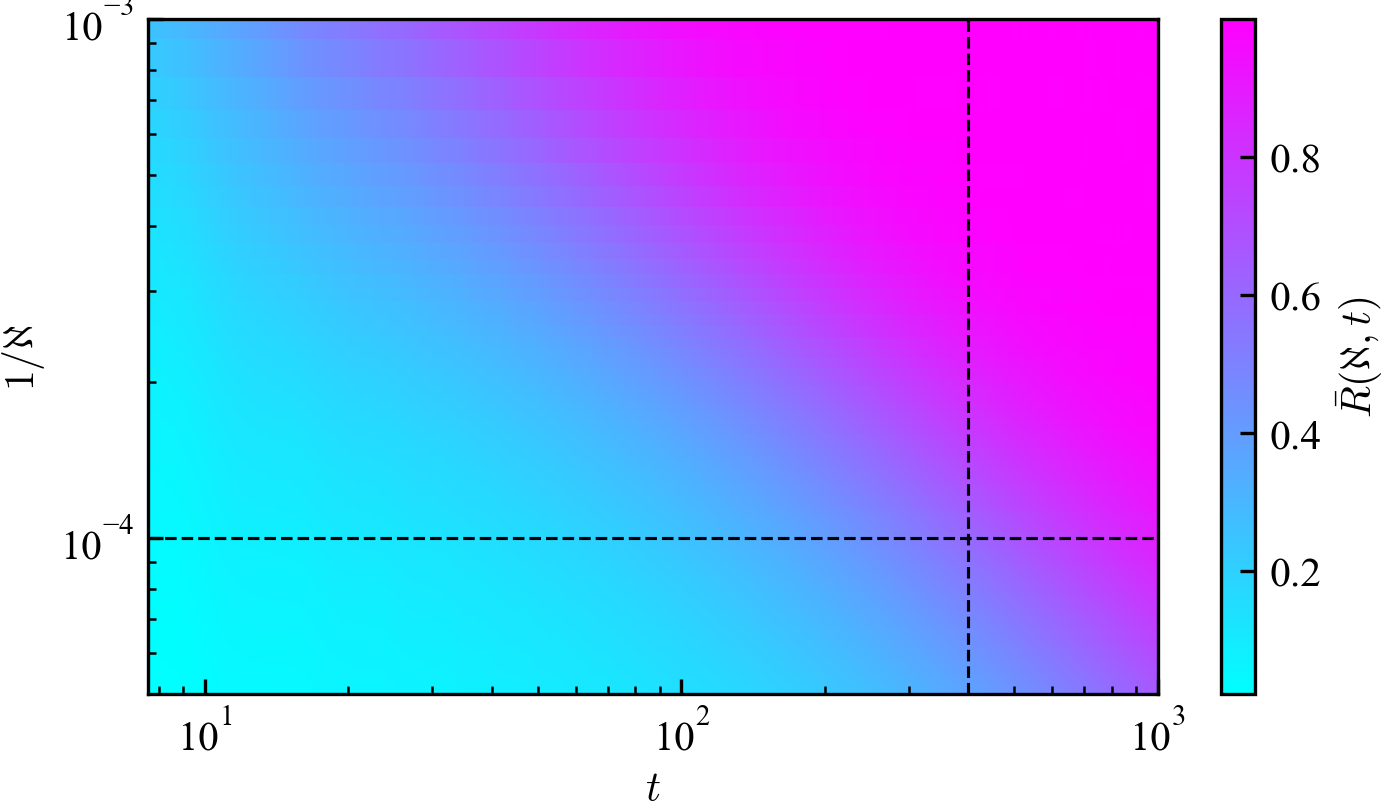}
	\caption{
		Period-averaged circular variance $\bar{R}(\aleph,t)$ as a function of time $t$ and inverse scaling parameter $1/\aleph$.
		Dashed lines show two examples of cuts along the $t$- and $(1/\aleph)$-axis which are used to extract the exponential relaxation rates $\delta_{\bar{R}}(t)$ and $\delta_{\bar{R}}(1/\aleph)$, respectively, see Eq.~\eqref{eq: supp: exponential_relaxation_of_R}.
	}
	\label{fig: supp: density_plot_period_avg_R}
\end{figure}
We compute $\bar{R}(t)$ for different values of $\aleph$, averaging over $N_{\text{traj}}=10^5$ TWA realizations each.
In this way, we obtain the period-averaged circular variance as a function of time \textit{and} scaling parameter,
\begin{equation}
	\bar{R} = \bar{R}(\aleph,t) \;.
\end{equation}
In Figure~\ref{fig: supp: density_plot_period_avg_R}, we show $\bar{R}(\aleph,t)$ for $\aleph\in[1000,20000]$ and up to $t=1000$.
Visibly, the circular variance increases with both, time $t$ and inverse scaling parameter $1/\aleph$.
More precisely, the circular variance exhibits exponential relaxation towards its steady state $\bar{R}=1$ along both directions.
To extract the corresponding relaxation rates $\delta_{\bar{R}}(t)$ and $\delta_{\bar{R}}(1/\aleph)$, we keep either time or scaling parameter fixed at $t=t^{\ast}$ or $\aleph=\aleph^{\ast}$, respectively, see for example dashed lines in Fig.~\ref{fig: supp: density_plot_period_avg_R}, and fit the relaxation rates along the free dimension using the exponential model
\begin{equation}
	\begin{aligned}
		\bar{R}(\aleph^{\ast},t) &= 1- m_1 e^{-\delta_{\bar{R}}(1/\aleph^{\ast}) t} \;,
		\\
		\bar{R}(\aleph, t^{\ast}) &= 1- m_2 e^{-\delta_{\bar{R}}(t^{\ast})\aleph^{-1}} \;.
	\end{aligned}
	\label{eq: supp: exponential_relaxation_of_R}
\end{equation}
In Figures~\ref{fig: supp: exponential_relaxation_and_fit_1} and~\ref{fig: supp: exponential_relaxation_and_fit_2}, we show two examples of this exponential relaxation and corresponding numerical fits for $\bar{R}(\aleph^{\ast}=10000,t)$ and for $\bar{R}(\aleph,t^{\ast}=400)$, respectively.
By repeating this procedure for multiple values of $\aleph^{\ast}$ and $t^{\ast}$, we obtain the numerically fitted relaxation rates $\delta_{\bar{R}}$ as functions of $\aleph$ and functions of $t$, respectively.
We find that the relaxation rates exhibit power-law scaling,
\begin{equation}
	\begin{aligned}
		\delta_{\bar{R}}(1/\aleph) &= c_1 (1/\aleph)^{d_1} \;,
		\\
		\delta_{\bar{R}}(t) &= c_2 t^{d_2} \;.
	\end{aligned}
	\label{eq: supp: power_law_scaling_relaxation_rates}
\end{equation}
\begin{figure}[h!]
	\centering
	\includegraphics[width=0.5\textwidth]{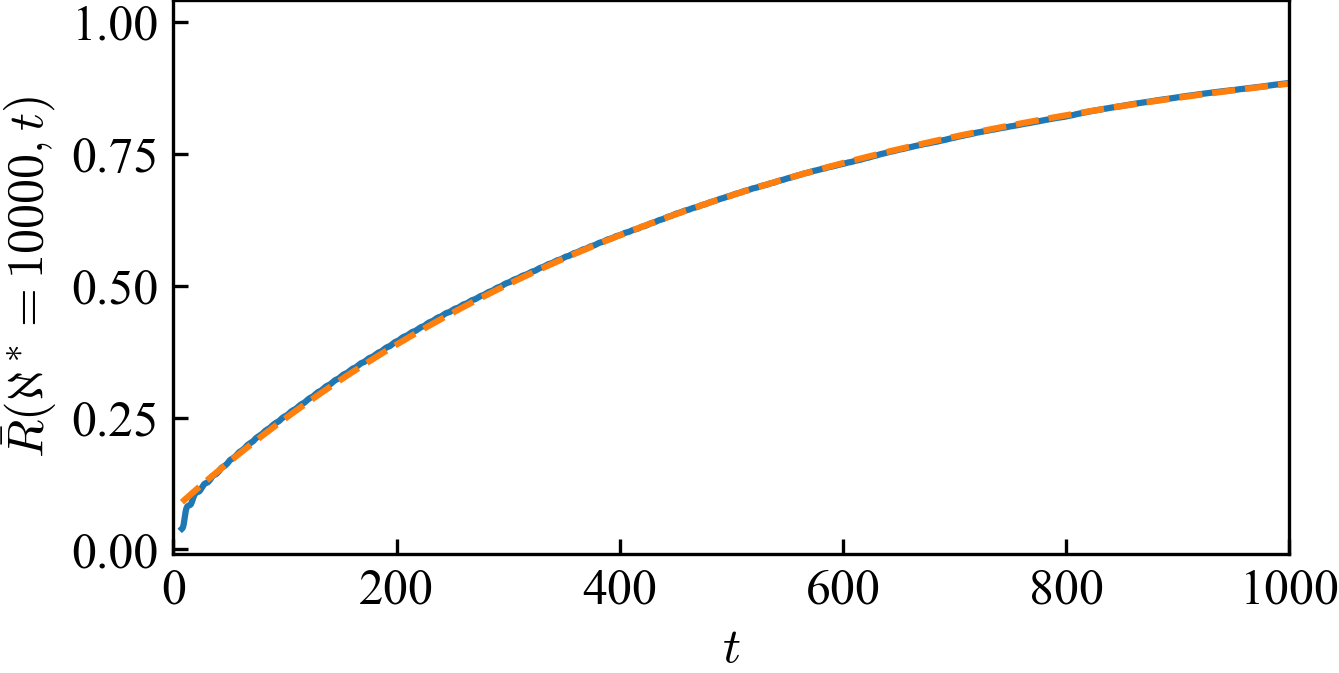}
	\caption{
		Period-averaged circular variance $\bar{R}(\aleph^{\ast},t)$ for $\aleph^{\ast}=10000$ (blue, solid), as indicated by the black dashed horizontal line in Fig.~(\ref{fig: supp: density_plot_period_avg_R}).
		The corresponding exponential fit according to the model given in Eq.~\eqref{eq: supp: exponential_relaxation_of_R} is shown as a orange dashed line.
		To exclude transient motion after initialization, where the distributions still moves to the LT attractor, we exclude early times $t<100$ from the fitting interval. Here, this initial deviation from exponential behavior is visible for $t\lesssim20$.
	}
	\label{fig: supp: exponential_relaxation_and_fit_1}
\end{figure}
\begin{figure}[h!]
	\includegraphics[width=0.5\textwidth]{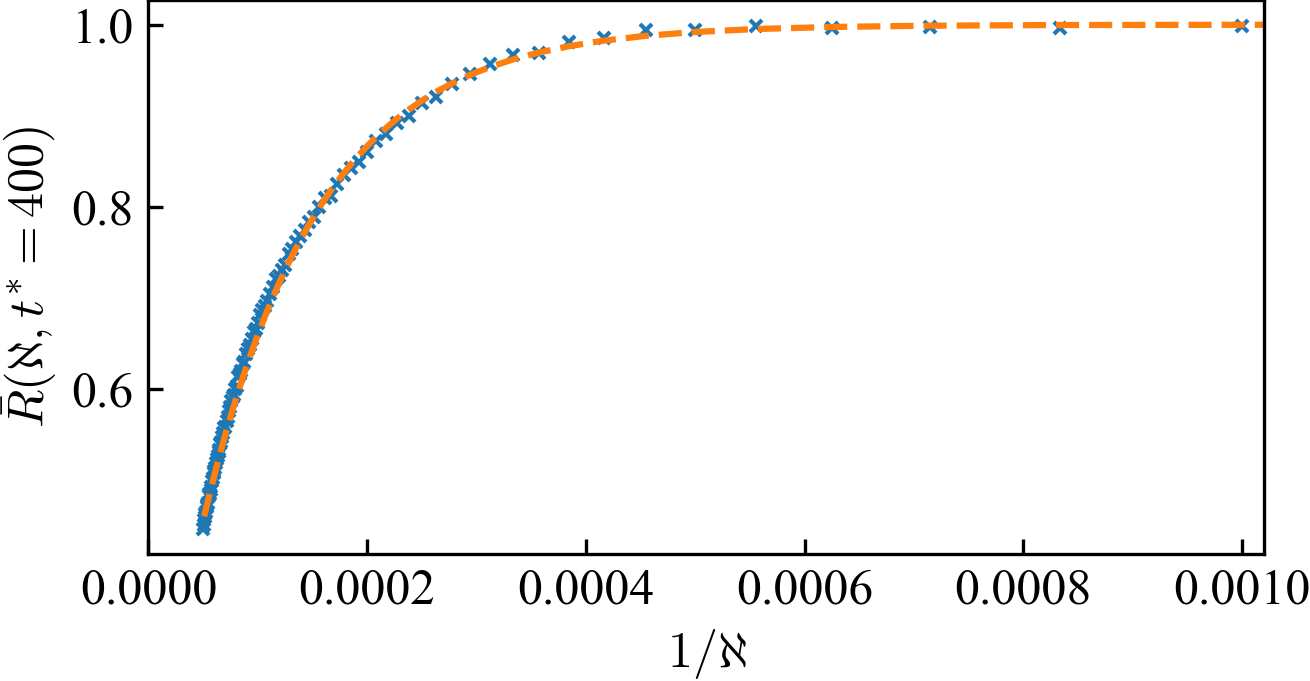}
	\caption{
		Period-averaged circular variance $\bar{R}(\aleph, t^{\ast})$ for $t^{\ast}=400$ (blue), as indicated by the black dashed vertical line in Fig.~\ref{fig: supp: density_plot_period_avg_R}.
		The corresponding exponential fit according to the model given in Eq.~\eqref{eq: supp: exponential_relaxation_of_R} is shown as a orange dashed line.
	}
	\label{fig: supp: exponential_relaxation_and_fit_2}
\end{figure}
The coefficients $c_{1}$, $c_2$ and $d_{1}$, $d_2$ that characterize this power-law behavior are obtained by numerically fitting the data to the model in Eq.~\eqref{eq: supp: power_law_scaling_relaxation_rates}.
The values obtained are
\begin{equation}
	\begin{aligned}
		c_1 &= 44.20 \pm 1.64\\
		c_2 &= 39.56 \pm 0.09\\
		d_1 &= 1.082 \pm 0.004\\
		d_2 &= 0.907 \pm 0.0004 \;,
	\end{aligned}
\end{equation}
where errors are the numerical fitting errors.
In Figure~\ref{fig: supp: power_law_scaling_relaxation_rates_and_fit_1} and Figure~\ref{fig: supp: power_law_scaling_relaxation_rates_and_fit_2}, we show the relaxation rates $\delta_{\bar{R}}(1/\aleph)$ and $\delta_{\bar{R}}(t)$ and the corresponding numerically fitted power-law models.
\begin{figure}[t]
	\centering
	\includegraphics[width=0.5\textwidth]{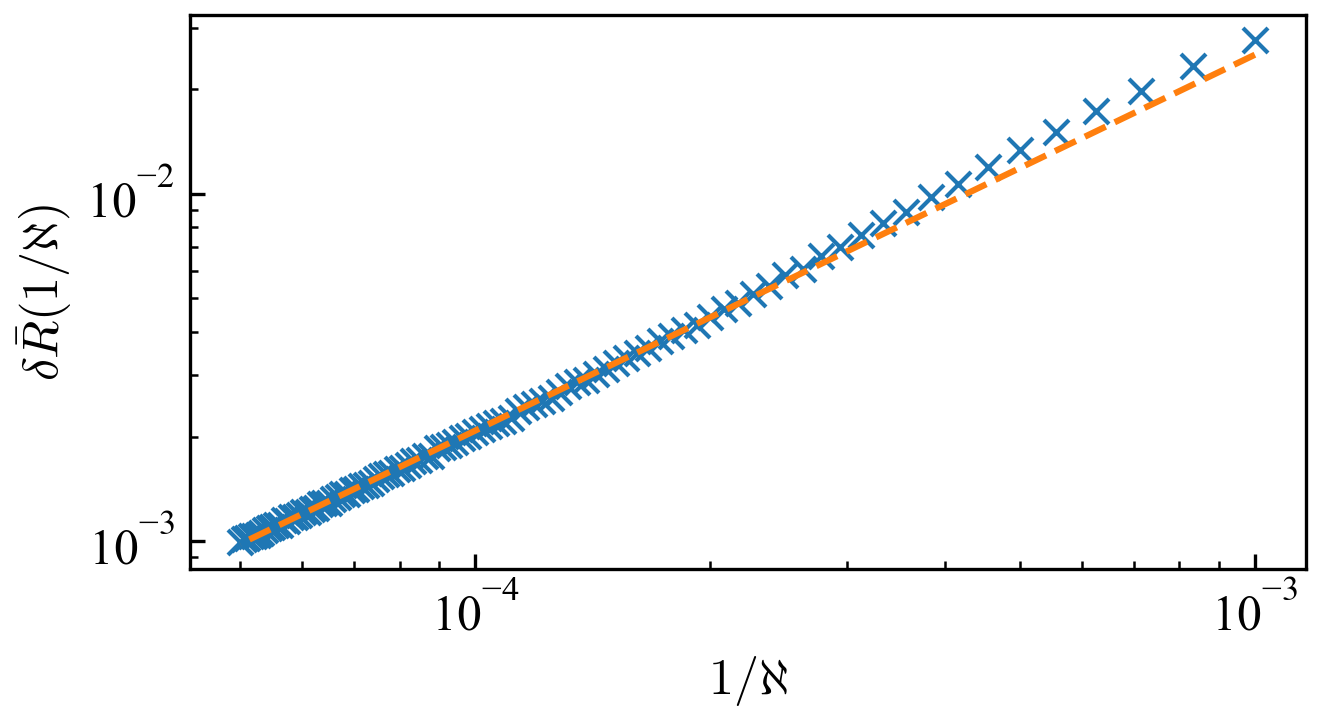}
	\caption{
		Relaxation rate of the period-averaged circular variance $\bar{R}(1/\aleph)$ (blue), together with the fitted power-law function (orange, dashed), according to the model given in Eq.~\eqref{eq: supp: power_law_scaling_relaxation_rates}.
	}
	\label{fig: supp: power_law_scaling_relaxation_rates_and_fit_1}
\end{figure}
\begin{figure}[t]
	\includegraphics[width=0.5\textwidth]{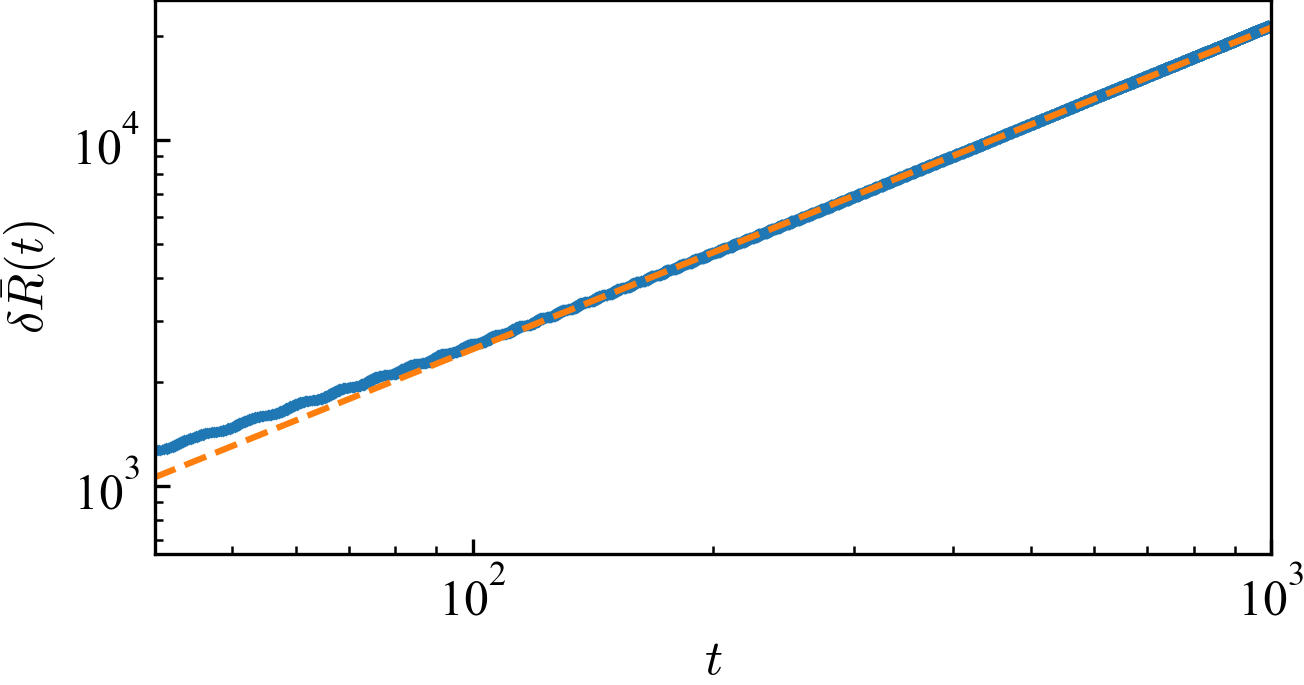}
	\caption{
		Relaxation rate of the period-averaged circular variance $\bar{R}(1/\aleph)$ (red), together with the fitted power-law function, according to the model given in Eq.~\eqref{eq: supp: power_law_scaling_relaxation_rates}.
		To exclude the short transient phase after initialization, we only consider times $t>100$ for the fit.
	}
	\label{fig: supp: power_law_scaling_relaxation_rates_and_fit_2}
\end{figure}

\begin{figure}[h!]
	\includegraphics[width=0.9\textwidth]{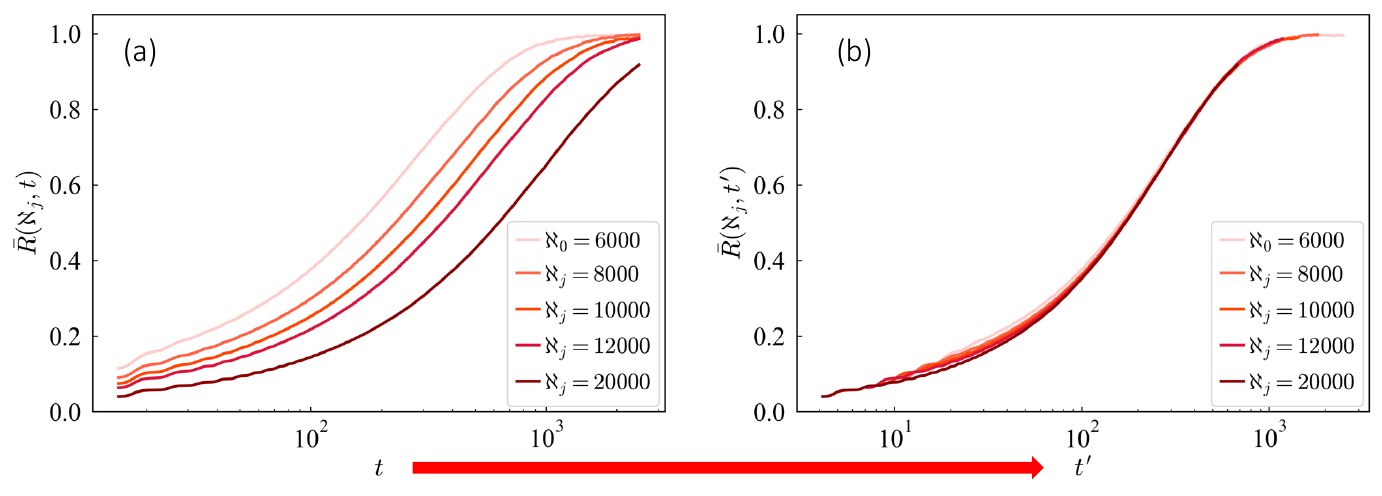}
	\caption{
		Collapse of $\bar{R}(\aleph_j,t)$ under rescaling of the time axis.
		(a) $\bar{R}(\aleph_j,t)$ for $5$ fixed values of $\aleph_j$. 
		(b) Collapsed data after rescaling transformation $t \to t^{\prime}$, see Eq.~\eqref{eq: supp: rescaling_transformations_collapse_part_1}.
	}
	\label{fig: supp: collapse_A}
\end{figure}

\begin{figure}[h!]
	\includegraphics[width=0.9\textwidth]{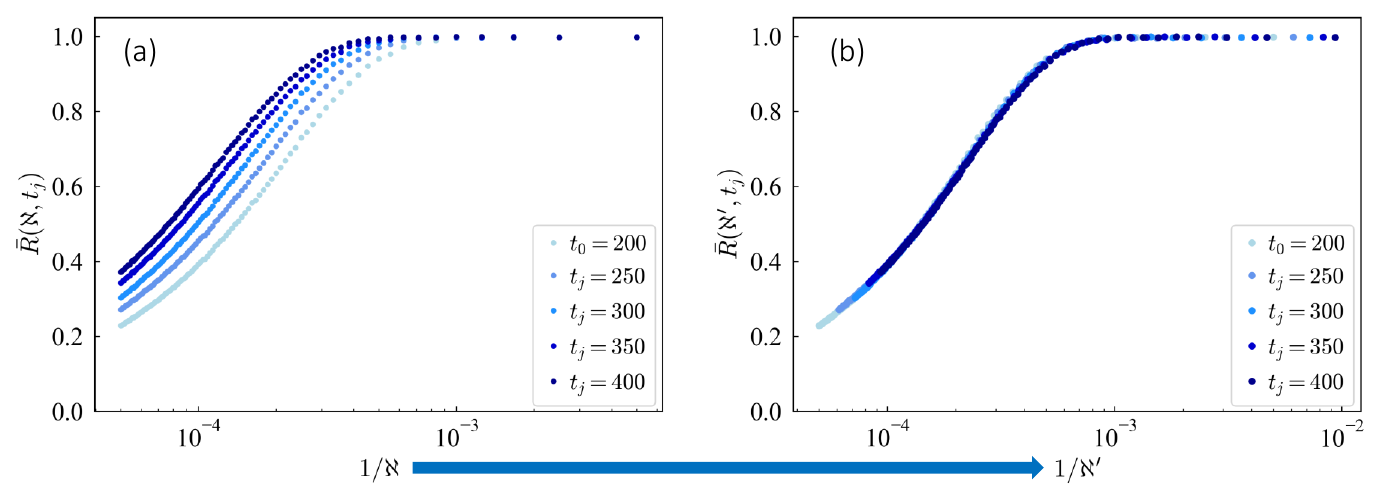}
	\caption{
		Collapse of $\bar{R}(\aleph,t_j)$ under rescaling of the scaling-parameter axis.
		(a) $\bar{R}(\aleph,t_j)$ for $5$ fixed values of $t_j$. 
		(b) Collapsed data after rescaling transformation $1/\aleph \to 1/\aleph^{\prime}$, see Eq.~\eqref{eq: supp: rescaling_transformations_collapse_part_1}.
	}
	\label{fig: supp: collapse_B}
\end{figure}

The scale invariant behavior along the two individual directions is further illustrated in Figure~\ref{fig: supp: collapse_A}, where we show a data collapse of $\bar{R}(\aleph_j,t)$ for $5$ fixed values $\aleph_j$, and in Figure~\ref{fig: supp: collapse_B}, where we show a data collapse of $\bar{R}(\aleph,t_j)$ for $5$ fixed values $t_j$.
This data collapse is achieved under the following rescaling transformations, reflecting the power-law behavior of both quantities:
\begin{equation}
	\begin{aligned}
		t^{\prime} &= (\aleph_j/\aleph_0)^{d_1} t
		\\
		(1/\aleph^{\prime}) &= (t_j/t_0)^{d_2} (1/\aleph) \;.
	\end{aligned}
	\label{eq: supp: rescaling_transformations_collapse_part_1}
\end{equation}
%
%

%
The scale invariance demonstrated above was found along the two individual directions $t$ and $\aleph$.
In a second step, we make a link between these two dimensions using the scaling ansatz
\begin{equation}
	t(\aleph) = \alpha \aleph^{\beta} \;.
	\label{eq: supp: time_scaling_relation}
\end{equation}
The relaxation rate as a function of the scaling parameter, $\delta_{\bar{R}}(\aleph^{-1})$, and as a function of time transformed according to Eq.~\eqref{eq: supp: time_scaling_relation}, $\delta_{\bar{R}}(t(\aleph))$, become equal if $\alpha=(c_1/c_2)^{1/d_2}$ and $\beta=d_1/d_2$, which can directly be seen from Eq.~\eqref{eq: supp: power_law_scaling_relaxation_rates}.
This relation reveals universal behavior of the system's relaxation dynamics with both, time and system size.
We demonstrate this universality by a complete collapse of the period-averaged circular variance $\bar{R}$ when taking the two collapsed curves shown in Fig.~\ref{fig: supp: collapse_A}(b) and Fig.~\ref{fig: supp: collapse_B}(b)---visualizing scale invariance along the two individual directions---and subsequent rescaling one of the axis employing Eq.~\eqref{eq: supp: time_scaling_relation}.
In particular, we rescale the time axis in Fig.~\ref{fig: supp: collapse_A}(b) according to
\begin{equation}
	t^{\prime} \to k_1 (t^{\prime})^{k_2}
\end{equation}
with $k_1 = \frac{c_2 t_0^{d_2}}{c_1 \aleph_0^{d_1}}\alpha^{1/\beta}$ and $k_2 = -1/\beta$,
which leads to the collapse of Fig.~\ref{fig: supp: collapse_A}(b) and Fig.~\ref{fig: supp: collapse_B}(b), which is shown in Fig.~(5) in the main text.


\section{Robustness of limit torus dynamics to single-photon loss and thermal noise}
\label{sec: appendix robustness}

To validate the physical relevance of our findings, we examine the robustness of the LT dynamics against realistic sources of dissipation and noise, namely single-photon loss and thermal fluctuations. 
These effects are unavoidable in experimental platforms due to system-environment coupling.
Throughout the main text, we considered an idealized model neglecting these effects in order to isolate and study the role of quantum fluctuations arising from engineered gain and two-photon processes. 
Here, we assume the gain and two-photon loss channels couple to engineered zero-temperature reservoirs. This assumption is experimentally justified, as two-photon processes are typically realized via parametric driving or reservoir engineering schemes involving nonlinear media or auxiliary lossy modes that are effectively immune to thermal occupation.
We therefore include Lindblad terms describing single-photon loss and thermal excitation through coupling to a thermal reservoir, while neglecting thermal effects on the other channels.

The Lindblad master equation including the single-photon loss with loss rate $\kappa_k$ and including thermal fluctuations with mean thermal occupation number $n_{th}$ reads
\begin{equation}
	\dfrac{d\hat{\rho}}{dt} =
	\dfrac{1}{i\hbar}\comm{\hat{\mathcal{H}}}{\hat{\rho}}
	+\sum_{k=1,2}
	\gamma_{k}\mathcal{D}[\hat{a}_k^\dagger]\hat{\rho}
	+\kappa_{k}(1+n_{th})\mathcal{D}[\hat{a}_k]\hat{\rho}
	+\kappa_{k}n_{th}\mathcal{D}[\hat{a}^\dagger_k]\hat{\rho}
	+\eta_k\mathcal{D}[\hat{a}_k^2]\hat{\rho}\;.
	\label{eq: supp: Master equation with single photon loss}
\end{equation}
The single-photon loss rate $\kappa_k$ depends on the specific experimental implementation of the system.
The mean thermal occupation number of the bath follows the Bose-Einstein distribution:
\begin{equation}
	n_{th} = \frac{1}{e^{\hbar\omega_k/(k_B T)}-1} \;.
\end{equation}
The Langevin equations corresponding to Eq.~\eqref{eq: supp: Master equation with single photon loss} are given by:
\begin{equation}
	\begin{aligned}
		i \frac{d\tilde{\alpha}_{1}}{dt}
		&=
		\left[
		\omega_1 
		+ i\frac{\gamma_1 -\kappa_1}{2} 
		+\left(
		\tilde{U}_1
		-i \tilde{\eta}_1
		\right)
		\left(\abs{\tilde{\alpha}_1}^2
		-\frac{1}{\aleph}\right)
		\right]
		\tilde{\alpha}_1
		\\
		&-2\tilde{J}\tilde{\alpha}_1^*\tilde{\alpha}_2
		+\sqrt{\dfrac{\gamma_1+\kappa_1(1+2n_{th})}{2\aleph}}\chi_{1}(t)
		+\sqrt{\dfrac{2\tilde{\eta}_1\abs{\tilde{\alpha}_1}^2}{\aleph}}\xi_{1}(t)
		\\
		i\frac{d\tilde{\alpha}_2}{dt}
		&=
		\left[\omega_2 + i\frac{\gamma_2 -\kappa_2}{2} +\left(\tilde{U}_2 - i \tilde{\eta}_2\right)\left(\abs{\tilde{\alpha}_2}^2-\frac{1}{\aleph}\right)\right]\tilde{\alpha}_2
		\\
		&-\tilde{J}\tilde{\alpha}_1^2
		+\sqrt{\dfrac{\gamma_2+\kappa_2(1+2n_{th})}{2\aleph}}\chi_{2}(t)
		+\sqrt{\dfrac{2\tilde{\eta}_2\abs{\tilde{\alpha}_2}^2}{\aleph}}\xi_{2}(t)\;.
	\end{aligned}
	\label{eq: supp: TWA Langevin with single photon loss}
\end{equation}
Note that the thermal occupation $\bar{n}_{th}$ only enters in the diffusion, while the drift term corresponding to the single-photon pump is effectively reduced by the rate of the single-photon loss, $\gamma_{k,\text{drift}}^{\text{eff}} = \gamma_k-\kappa_k$, but does not depend on temperature.
\subsubsection{Zero temperature ($T=0$)}
\begin{figure}[h]
	\centering
	\includegraphics[width=0.94\linewidth]{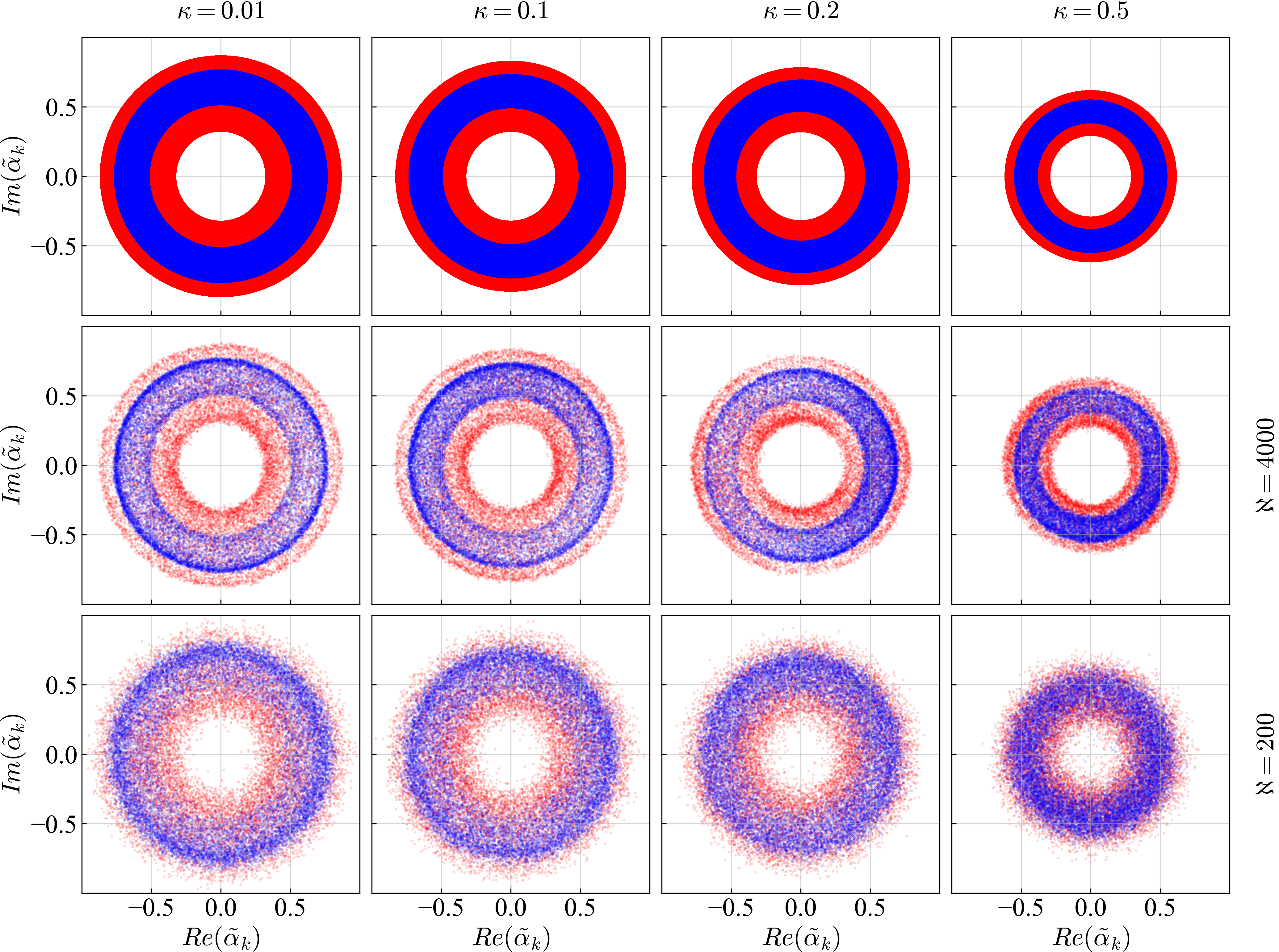}
	\caption{
		Impact of single-photon loss on the torus diameter and robustness of the torus structure.
		Upper row: GPE solutions for cavity 1 (red) and cavity 2 (blue). For increasing single-photon loss, the diameter of the torus is decreased.
		Second and third row:
		Wigner function reconstructions based on TWA fields $\alpha_1$ (red) and $\alpha_2$ (blue), shown at $t=4000$. In the second row, $\aleph=4000$, in the third row, $\aleph=200$. The torus shape remains robust for all values of $\kappa$.
	}
	\label{fig: supp: single_photon_loss_Wigner_function_reconstructions}
\end{figure}

We first focus on the case of zero temperature, studying the impact of the single-photon loss.
In Figure~\ref{fig: supp: single_photon_loss_Wigner_function_reconstructions} we show the GPE solutions and Wigner reconstructions at large times ($t=4000$), for different values of the single-photon loss rate $\kappa$ and the scaling parameter $\aleph$.
As explained above, the single-photon loss counteracts the incoherent single-photon pump, effectively reducing the pump rate. This effective change of the system parameter leads to a change of the diameter of the torus already at GPE level. This can be seen in the first row of Fig.~\ref{fig: supp: single_photon_loss_Wigner_function_reconstructions}.
In addition, the single-photon loss leads to increased diffusion, see Eq.~\eqref{eq: supp: TWA Langevin with single photon loss}. For all considered values of $\kappa$ this increased noise does not destruct the torus structure, see second and third row of Fig.~\ref{fig: supp: single_photon_loss_Wigner_function_reconstructions}.

We note again that the observations made here are system-dependent as the single-photon loss effectively changes the pump rate. In the following, we also take thermal fluctuations into account, which realistically occur in experiments and do not alter the effective drift, thus giving a system-independent picture of the impact of classical (thermal) noise.

\subsubsection{Finite temperature ($T\neq0$)}

To get a realistic estimate of the thermal fluctuations, we consider the example of a setup in a trapped-ion platform presented in the main text and detailed in Section~\ref{sec: appendix experimental realization}.
Here, typical trap frequencies are of the order $f_{\text{trap}}\approx1$MHz.
For our analysis, we focus on temperatures at or above the Doppler-cooling limit, which in a trapped ion system is typically $T_{\text{D}}\approx 0.5$mK, determined by the linewidth of the cooling transition, $T_{\text{D}}\approx \hbar\Gamma/(2k_B)$.
These values correspond to a thermal occupation at the Doppler cooling limit of $n_{th}\approx 9.93$.

\begin{figure}[H]
	\centering
	\includegraphics[width=0.82\linewidth]{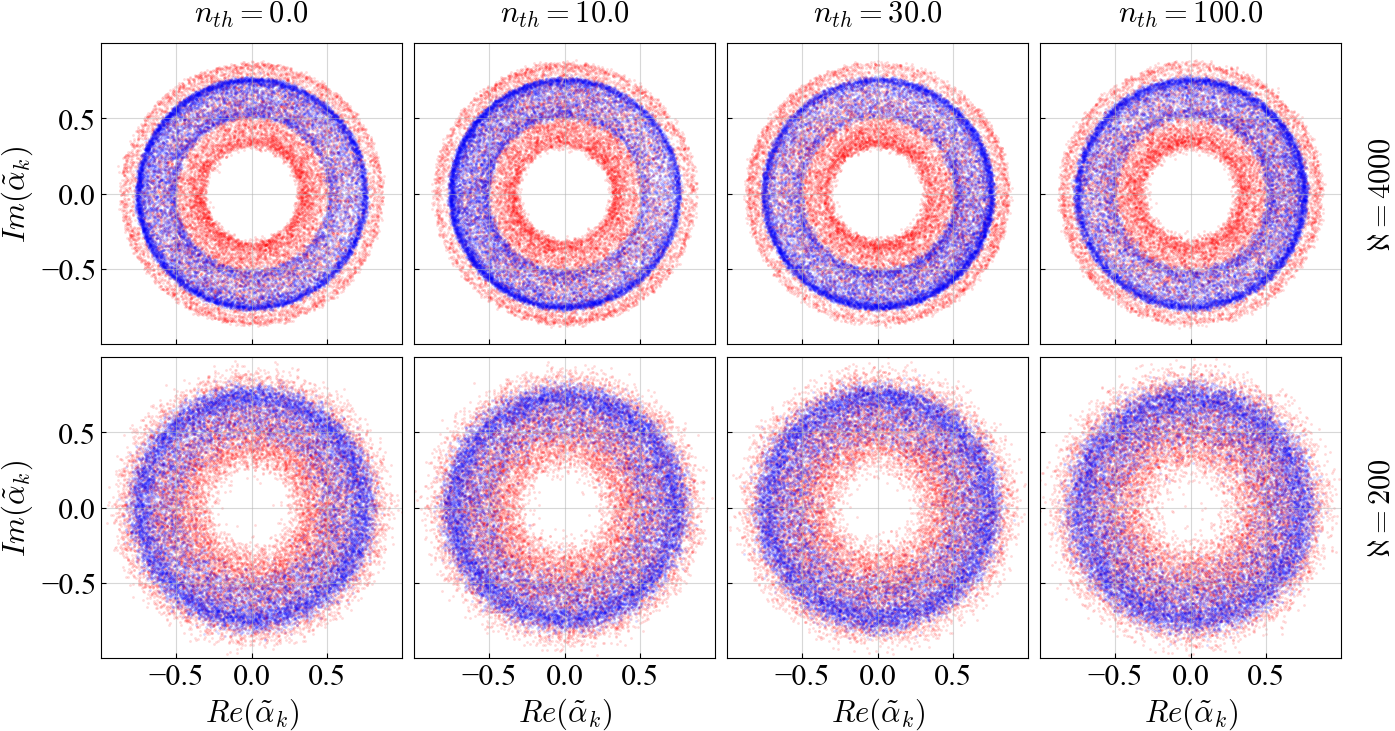}
	\caption{Wigner reconstructions at $t=2000$, with $\kappa=0.01$.}
	\label{fig: supp: wigner_reconstructions_gammaSys=0.01_t=2000}
\end{figure}

\begin{figure}[H]
	\centering
	\includegraphics[width=0.82\linewidth]{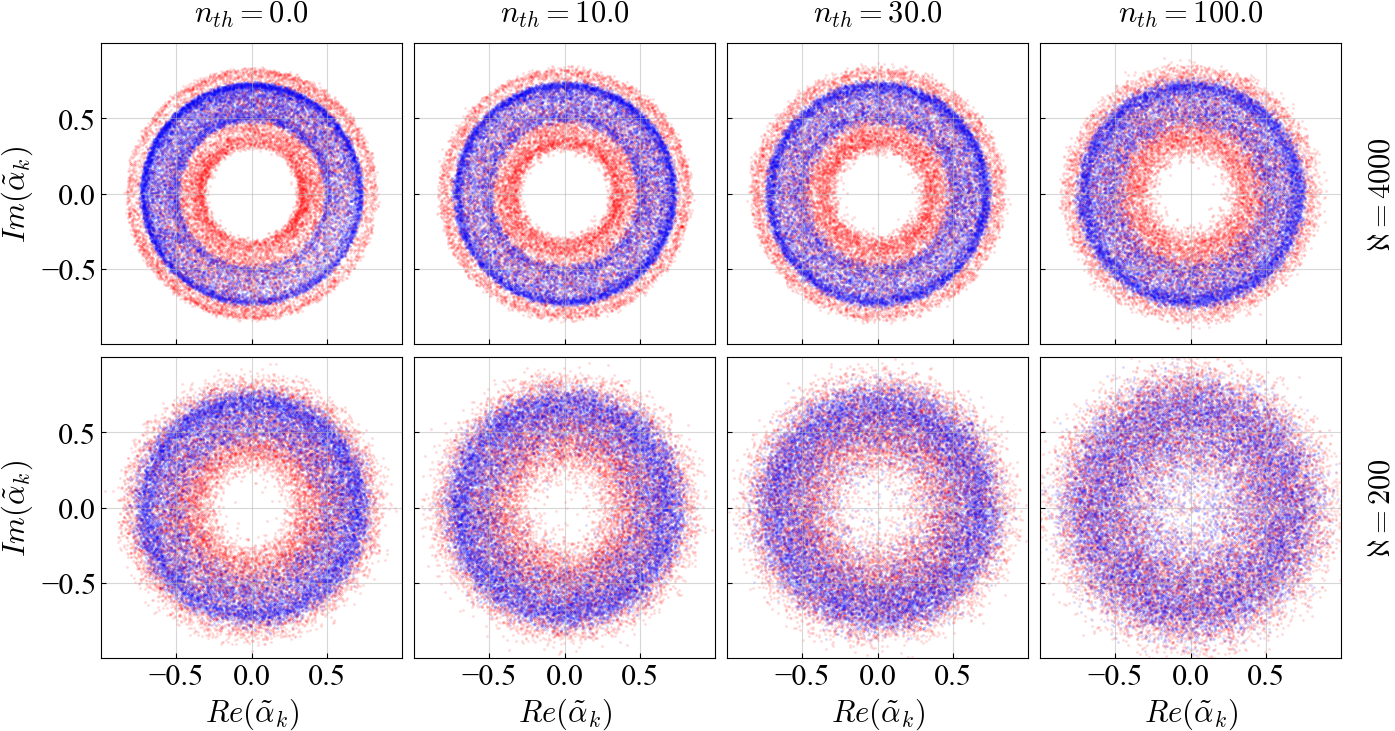}
	\caption{Wigner reconstructions at $t=2000$, with $\kappa=0.1$.}
	\label{fig: supp: wigner_reconstructions_gammaSys=0.1_t=2000}
\end{figure}

\begin{figure}[H]
	\centering
	\includegraphics[width=0.82\linewidth]{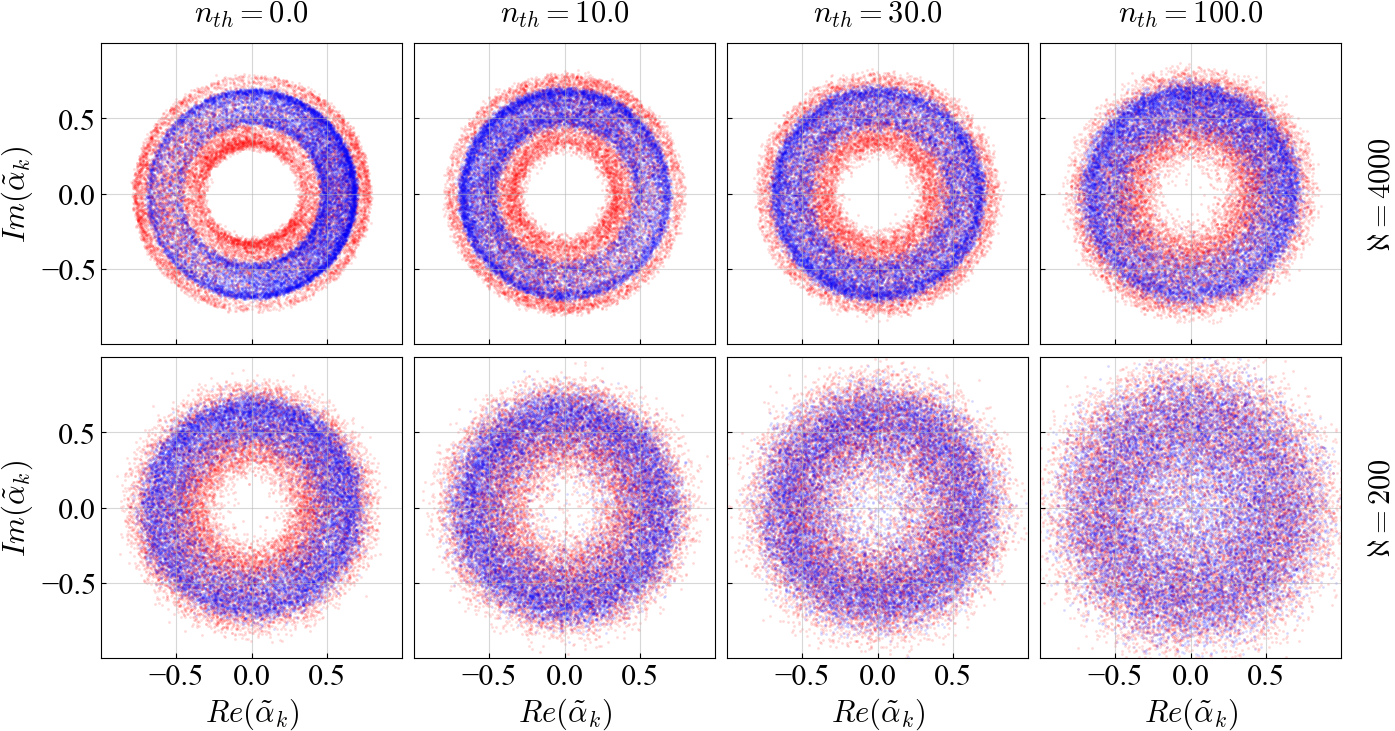}
	\caption{Wigner reconstructions at $t=2000$, with $\kappa=0.2$.}
	\label{fig: supp: wigner_reconstructions_gammaSys=0.2_t=2000}
\end{figure}

\begin{figure}[H]
	\centering
	\includegraphics[width=0.8\linewidth]{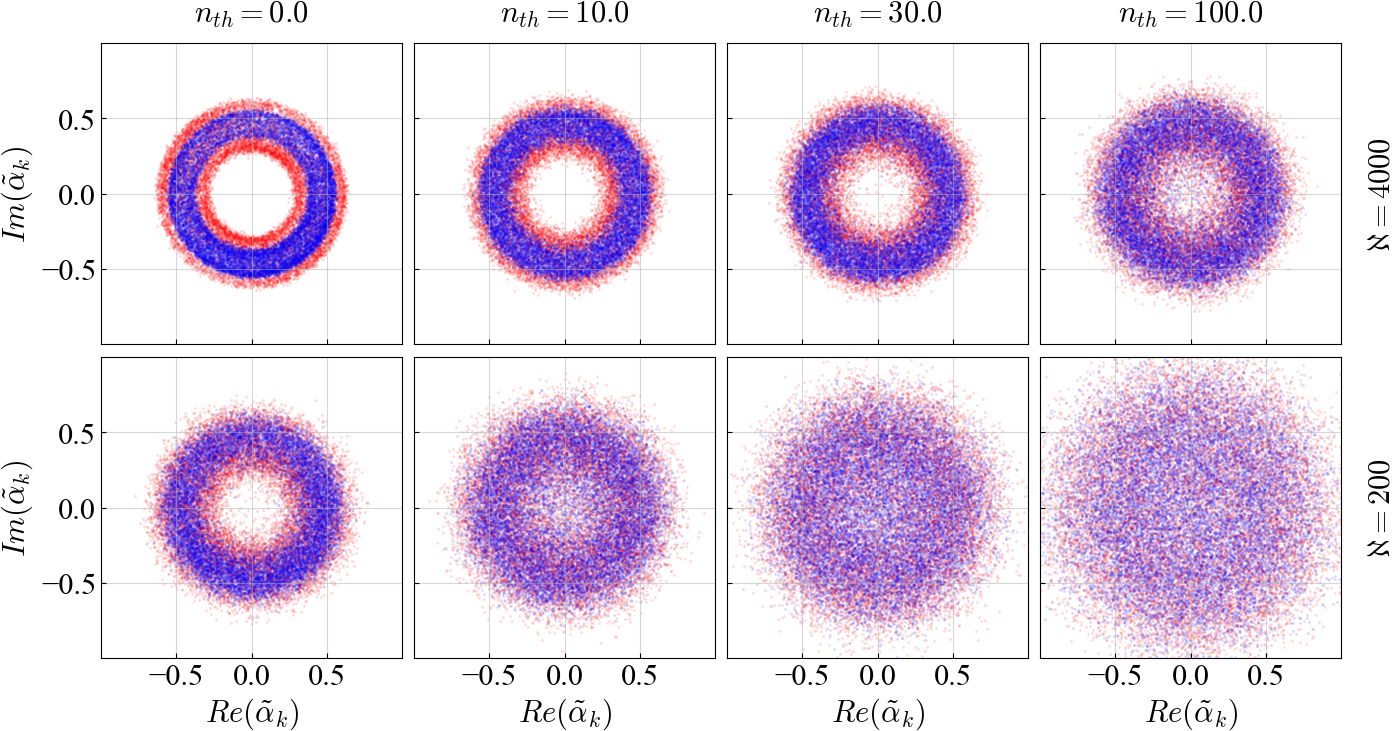}
	\caption{Wigner reconstructions at $t=2000$, with $\kappa=0.5$.}
	\label{fig: supp: wigner_reconstructions_gammaSys=0.5_t=2000}
\end{figure}

We examine the impact of thermal noise for each of the four values of the setup-dependent single-photon loss rate that we considered for $T=0$ in the previous subsection: $\kappa_1=\kappa_2=\kappa$, $\kappa\in\{0.01,0.1,0.2,0.5\}$.
For each $\kappa$, we simulate the time-evolution of the system for thermal occupation numbers $n_{th}\in[0, 100]$.
In Figures~\ref{fig: supp: wigner_reconstructions_gammaSys=0.01_t=2000}-\ref{fig: supp: wigner_reconstructions_gammaSys=0.5_t=2000} we show the resulting Wigner reconstructions in the steady state (late times $t=2000$), for different thermal occupations $n_{th}$ and scaling parameters $\aleph$. Each of the panels \ref{fig: supp: wigner_reconstructions_gammaSys=0.01_t=2000}-\ref{fig: supp: wigner_reconstructions_gammaSys=0.5_t=2000} is obtained for one of the values of $\kappa$.
We can see that for low mode occupations ($\aleph=200$) the impact of thermal fluctuations is more significant than for higher occupation ($\aleph=4000$). This is because the ratio of thermal noise and occupation is increased if the mode occupation is small.
For small $\kappa$, the torus structure is still preserved even for small mode occupations and high temperature.
For large $\kappa$, the torus is destroyed if the temperature of the system is too high. We note that for the largest considered value, $\kappa=0.5$, at the Doppler-cooling limit ($n_{th}=10.0$) the torus is still visible.
In general, we conclude that the structure of the torus is robust against experimentally realistic impact of single-photon loss and noise.


\begin{figure}[H]
	\centering
	\includegraphics[width=0.85\linewidth]{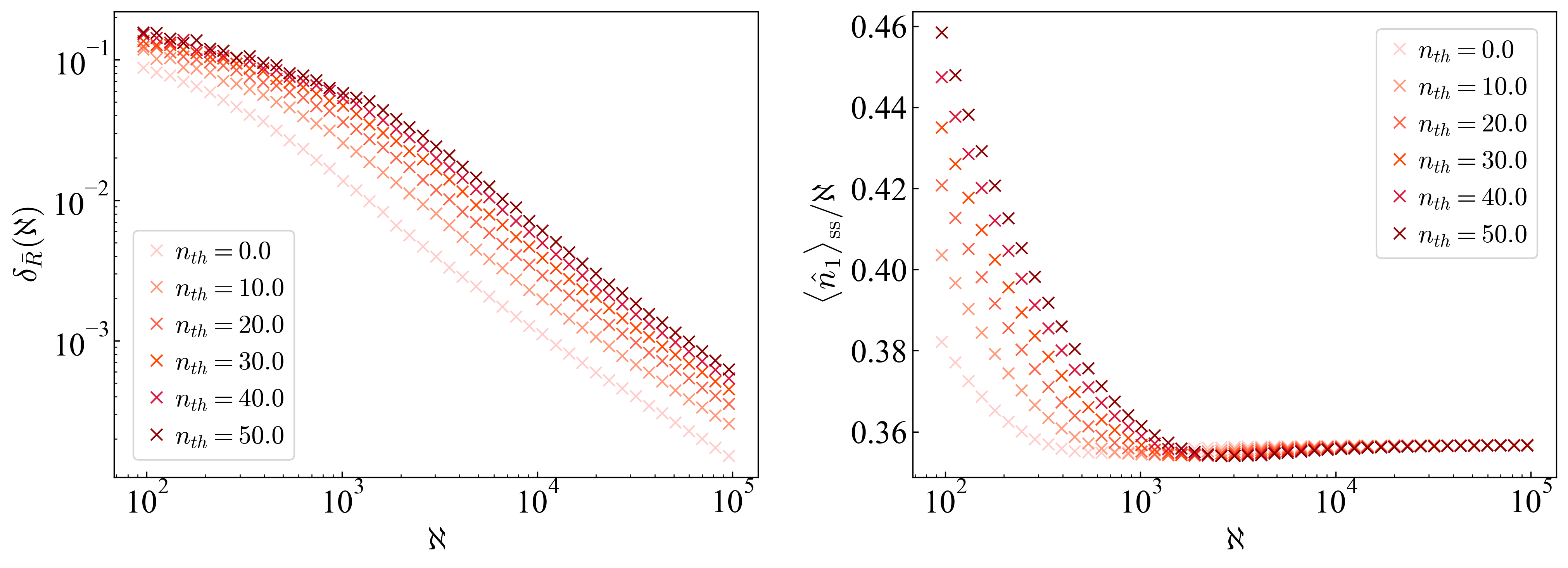}
	\caption{
		Effect of thermal noise on relaxation dynamics and steady-state occupation across the quantum-to-classical crossover.
		The shown results are for a fixed single-photon loss rate $\kappa = 0.1$.
		Left: Relaxation rate of the period-averaged circular variance $\delta_{\bar{R}}$ as a function of $\aleph$, for varying thermal photon numbers $n_{th}$.
		Right: Rescaled steady-state mean occupation $\langle \hat{n}_1 \rangle / \aleph$ of cavity 1.
		The plots highlight the transition from robust power-law scaling at large $\aleph$ to thermally modified dynamics at low $\aleph$, where the thermal population becomes comparable to the cavity occupation.
	}
	\label{fig: supp: dR_and_n1ss_vs_nth}
\end{figure}

To characterize the robustness of the scaling laws governing the melting of the torus versus single-photon loss and finite thermal noise, we further analyze the dynamical behavior of the system for a fixed loss rate $\kappa = 0.1$ across varying $\aleph$ and thermal occupations $n_{th}$. 
This analysis is based on the TWA given in Eq.~\eqref{eq: supp: TWA Langevin with single photon loss}. 
In Figure~\ref{fig: supp: dR_and_n1ss_vs_nth}, we show the impact of thermal noise on the dephasing dynamics and steady-state cavity population. The left panel presents the relaxation rate of the period-averaged circular variance $\delta_{\bar{R}}$, while the right panel shows the rescaled steady-state occupation $\langle \hat{n}_1 \rangle_{\text{ss}} / \aleph$. Both panels highlight the departure from the universal scaling behavior as thermal noise becomes significant in the small-$\aleph$ regime.

For $\aleph \gtrsim 2000$, we observe that increasing $n_{th}$ enhances the relaxation rate but leads to the same power-law scaling with $\aleph$, \textbf{with same scaling exponent}.
This behavior is consistent with Eq.~\eqref{eq: supp: TWA Langevin with single photon loss}, where thermal noise increases the diffusion rate but does not affect its scaling with $\aleph$, provided the system remains in the regime of approximately invariant rescaled population. 

In contrast, for $\aleph \lesssim 2000$, thermal noise qualitatively modifies the dynamics. As shown in the right panel of Fig.~\ref{fig: supp: dR_and_n1ss_vs_nth}, the rescaled steady-state occupation $\langle \hat{n}_1 \rangle_{\text{ss}}/\aleph$ becomes $\aleph$-dependent in this regime, with the exact dependence dominated by $n_{th}$. 
This indicates that in Eq.~\eqref{eq: supp: TWA Langevin with single photon loss}, the noise term proportional to $\sqrt{|\tilde{\alpha}_k|^2/\aleph}$ introduces a nontrivial $n_{th}$-dependent scaling. The consequence is visible in the left panel of Fig.~\ref{fig: supp: dR_and_n1ss_vs_nth}, where the different thermal occupations lead to different slopes of the relaxation rate. 
Notably, for small $\aleph$, the thermal population becomes comparable to the cavity photon number, signaling a breakdown of the scaling observed in the large-$\aleph$ limit. Although our main text focuses on the zero-temperature regime---appropriate to the quantum melting transition---the crossover to thermally modified scaling uncovered here may represent another axis of universality. 


\section{Experimental Realization}
\label{sec: appendix experimental realization}

In the main text, we propose an implementation of the system of coupled driven-dissipative Kerr cavities in a trapped ion setup.
As described, in this setup the two bosonic modes correspond to two motional modes of a trapped ion.
The implementation of the nonlinear interaction between these two modes, the two-phonon loss, and the incoherent single-phonon pump rely on coupling of the two motional modes to internal states of the ion.
This coupling can be realized via dipole interaction, mediated and controlled by external laser drives.
In this section, we show how suitable adjustment of the driving frequencies and phases leads to the effective interaction parameters and dissipation rates given in the main text.
%
\subsection{Dipole interaction}
First, we consider a single laser field $\vec{E}$ and a single internal electronic transition of the ion between two states $\ket*{g}$ and $\ket*{e}$, with transition frequency $\omega_{eg}$.
The dipole interaction is modeled by $\hat{V}=-\hat{d} \cdot \vec{E}$, with dipole moment operator $\hat{d}$, and electric field
\begin{equation}
	\vec{E}=\vec{E}_{0} \left(e^{i \vec{k}\vec{r}-i\omega_{d} t +\phi} + \mathrm{c.c.} \right)
	\;.
\end{equation}
Here, $\vec{E}_0$ denotes the field amplitude, $\vec{k}$ the wavevector, $\omega_{d}$ the driving frequency, and $\phi$ the phase of the electric field.
For a trapped ion with two motional modes $l\in\{1,2\}$ with frequencies $\omega_l$, the coupling Hamiltonian can be written as \cite{Leibfried_2003}
\begin{equation}
	\begin{aligned}
		\hat{V}(t)
		&=
		\sum_{l=1,2}
		\frac{\hbar\Omega}{2}
		\left(
		\hat{\sigma}_x
		e^{i[\eta_{\text{LD},l}(\hat{a}_l+\hat{a}_l^{\dagger})-\omega_{d}t +\phi]}
		+
		\mathrm{H.c.}
		\right) \;,
	\end{aligned}
	\label{eq: supp: coupling_hamiltonian_dipole_interaction}
\end{equation}
where $\Omega$ is the on-resonance Rabi frequency, $\eta_{\text{LD},l}= \abs{k} \cos(\theta_l)\sqrt{\hbar/(2\omega_l)}$ the Lamb-Dicke parameter, which depends on the angle $\theta_l$ between wavevector $\vec{k}$ and vibrational direction of mode $l$, and $\hat{\sigma}_x$ the Pauli-$X$ operator acting in the subspace of the electronic transition.
The two motional modes directly correspond to the two modes of the system Hamiltonian in the main text. Note that in this subsection, we use mode index \q{$l$} instead of \q{$k$} to avoid notational conflict with the wavevector $\vec{k}$.
In the following, we assume that the electric field is perfectly aligned with the direction of one mode $l$, i.e., $\cos(\theta_l)=1$ and $\cos(\theta_{l'\neq l})=0$.
In the interaction picture, and under RWA, the coupling Hamiltonian in Eq.~\eqref{eq: supp: coupling_hamiltonian_dipole_interaction} then takes the form
\begin{equation}
	\begin{aligned}
		\hat{V}_{I}(t)
		&=
		\frac{\hbar\Omega}{2}
		\left(\hat{\sigma}_+ e^{-i(\delta t -\phi)}
		e^{i\eta_{\text{LD},l}(\hat{a}_l e^{-i\omega_l t}+\hat{a}^{\dagger}_l e^{i\omega_l t})}
		+\mathrm{H.c.}
		\right) \;,
	\end{aligned}
\end{equation}
where the detuning between the driving frequency and the electronic transition is denoted as $\delta=\omega_{d}-\omega_{eg}$.
Assuming small Lamb-Dicke parameter $\eta_{\text{LD},l}$, the interaction Hamiltonian can be expanded to second order,
\begin{equation}
	\hat{V}_{I}(t)
	\approx
	\frac{\hbar\Omega}{2}
	\hat{\sigma}_{+}
	\left(
	1
	+i\eta_{\text{LD},l}
	(
	\hat{a}_l e^{-i\omega_l t}+\hat{a}^{\dagger}_l e^{i\omega_l t}
	)
	-\frac{\eta_{\text{LD},l}^2}{2}(
	\hat{a}_l\hat{a}_l e^{-2i\omega_l t}
	+\hat{a}^{\dagger}_l\hat{a}^{\dagger}_l e^{2i\omega_l t}
	+\hat{a}^{\dagger}_l\hat{a}_l
	+\hat{a}_l\hat{a}^{\dagger}_l
	)
	\right)
	e^{-i\delta t+i\phi}
	+ \mathrm{H.c.}
	\label{eq: supp: interaction_second_order}
\end{equation}

The implementation of the nonlinear interaction and dissipative processes in Eq.~(2) in the main text requires driving with multiple laser fields $\vec{E}_i$, which allow for addressing the motional modes $l\in\{1,2\}$ individually and using multiple distinct driving frequencies $\omega_{d,i}$.
To suppress interference, cross-coupling effects and unwanted scattering, it can be beneficial to employ distinct internal electronic transitions $\ket{g}\leftrightarrow \ket{e_j}$ for the implementation of the different processes.
In this more general case, the interaction Hamiltonian is given by a sum of terms of the form of Eq.~\eqref{eq: supp: interaction_second_order}, each corresponding to one of the applied driving fields $\vec{E}_i$:
\begin{equation}
	\hat{V}_I(t)
	\approx
	\sum_{i,l}
	\frac{\hbar\Omega_i}{2}
	\hat{\sigma}_{+}^{(i)}
	\left(
	1
	+i\eta_{\text{LD},i,l}
	(
	\hat{a}_l e^{-i\omega_l t}+\hat{a}^{\dagger}_l e^{i\omega_l t}
	)
	-\frac{\eta_{\text{LD},i,l}^2}{2}(
	\hat{a}_l\hat{a}_l e^{-2i\omega_l t}
	+\hat{a}^{\dagger}_l\hat{a}^{\dagger}_l e^{2i\omega_l t}
	+\hat{a}^{\dagger}_l\hat{a}_l
	+\hat{a}_l\hat{a}^{\dagger}_l
	)
	\right)
	e^{-i\delta_i t+i\phi_i}
	+ \mathrm{H.c.}
	\label{eq: supp: interaction_hamiltonian_RWA_multicomponent_drive}
\end{equation}
where $\hat{\sigma}_{+}^{(i)}=\ket{e_j(i)}\bra{g}$ denotes the raising operator corresponding to the electronic transition that is addressed by driving field $\vec{E}_i$, the Lamb-Dicke parameters $\eta_{\text{LD},i,l}$ depend on the angle $\theta_{i,l}$ between the field $\vec{E}_i$ and motional direction of mode $l$,
and the detuning between driving frequency $\omega_{d,i}$ and frequency $\omega_{eg}^{j(i)}$ of the addressed electronic transitions $\ket{g}\leftrightarrow\ket{e_j(i)}$ is denoted by $\delta_i^{(j)}$.

\subsection{Nonlinear coupling}
\begin{figure}
	\centering
	\includegraphics[width=0.96\linewidth]{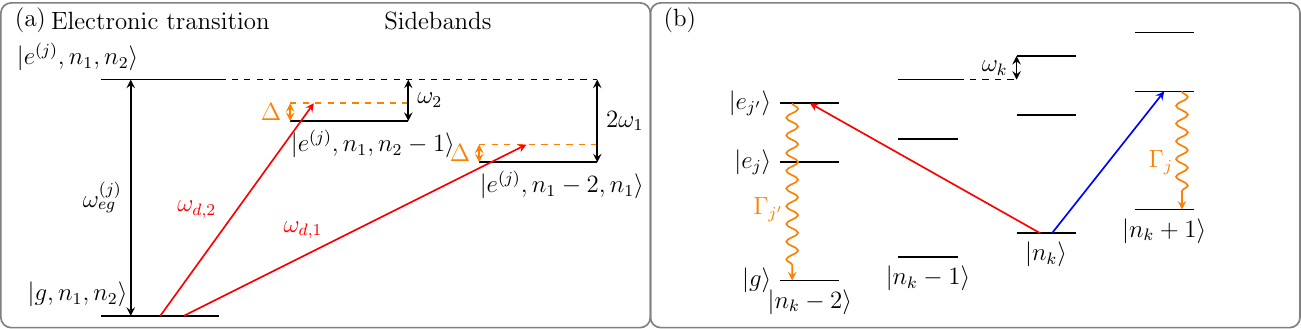}
	\caption{
		Driving schemes for the implementation of (a) nonlinear coupling of the modes, and (b) single-phonon pump and two-phonon loss.
	}
	\label{fig: supp: scheme_trapped_ions}
\end{figure}

The nonlinear interaction term
$\propto J (\hat{a}^{\dagger}_1\hat{a}^{\dagger}_1\hat{a}_2 + \hat{a}^{\dagger}_2\hat{a}_1\hat{a}_1)$ in the Hamiltonian, Eq.~(1) in the main text, involves two-phonon processes of the first mode and single-phonon processes of the second mode.
As described in the main text, these processes can be generated by driving with two laser fields $\vec{E}_i$, $i\in\{1,2\}$ with driving frequencies $\omega_{d,i}$.
The field $\vec{E}_1$ is predominantly aligned with mode $1$, and $\vec{E}_2$ is predominantly aligned with mode $2$, such that $\eta_{\text{LD},i,l}\approx 0$ for $i\neq l$, and we write $\eta_{\text{LD},i,l}\equiv\eta_{\text{LD},i}$ for $i=l$.
Both fields address the same electronic transition $\ket{g}\leftrightarrow\ket{e_j}$, such that $\hat{\sigma}_+^{(i)}=\ket{e_j}\bra{g}$.
The detunings $\delta_i^{(j)}=\omega_{d,i}-\omega_{eg}^{(j)}$ are adjusted to the first and second red sideband of the two modes, respectively, both shifted from the exact side-band transition by an off-resonance $\Delta$, such that
\begin{equation}
	\begin{aligned}
		\delta_1^{(j)} &= -2\omega_1 +\Delta\;,
		\\
		\delta_2^{(j)} &= -\omega_2 +\Delta\;.
	\end{aligned}
\end{equation}
This driving scheme is visualized in Figure~\ref{fig: supp: scheme_trapped_ions}(a).
Within the RWA, the resulting contributions to the interaction Hamiltonian arising from Eq.~\eqref{eq: supp: interaction_hamiltonian_RWA_multicomponent_drive} can be approximated by
\begin{equation}
	\hat{V}_{I}(t)
	\approx
	-\frac{\eta_{\text{LD},1}^2}{2}\frac{\hbar\Omega_1}{2}
	\ket{e_j}\bra{g}
	\hat{a}_1\hat{a}_1
	e^{-i\Delta t +i\phi_1}
	+
	i\eta_{\text{LD},2} \frac{\hbar\Omega_2}{2}
	\ket{e_j}\bra{g}
	\hat{a}_2
	e^{-i\Delta t +i\phi_2}
	+ \mathrm{H.c.} \;.
\end{equation}
The nonlinear coupling of the two modes arises as a second-order process of this interaction, $\propto \hat{V}_I(t)^2$.
More precisely, it appears at second order in the Magnus expansion of the time-evolution operator within the interaction picture:
\begin{equation}
	U_I(t) = \mathcal{T}\{e^{-i/\hbar  \int_0^t dt' V_I(t')}\}
	\approx
	1-\frac{i}{\hbar}\int_0^t dt' V_I(t')
	-\frac{1}{\hbar^2}\int_0^t dt_1\int_0^t dt_2 V_I(t_1)V_I(t_2) + ... \;.
\end{equation}
These second-order processes are dominant compared to the first-order contributions when operating with large off-resonances from the sideband-transitions, i.e., $\abs{\Delta} \gg \eta_{\text{LD},i}\Omega_i$.
In this case, and if the electronic part of the system is initially prepared in the ground state $\ket*{g}$, the excited state $\ket*{e_j}$ is only virtually populated via the second-order processes. We can therefore approximate the electronic part of the system to stay in the ground state, such that the first- and second-order contributions to the time evolution for the bosonic part of the system can be approximated as
\begin{equation}
	\begin{aligned}
		U_I^{(1)}(t) &\approx -\frac{i}{\hbar}\int_0^t dt' \bra{g}V_I(t')\ket{g}
		= 0 \;,
		\\
		U_I^{(2)}(t) &\approx
		-\frac{1}{\hbar^2}\int_0^t dt_1\int_0^t dt_2 \bra{g}V_I(t_1)V_I(t_2)\ket{g}
		\\
		&=
		-\int_0^t dt_1\int_0^t dt_2
		\Bigg[
		\frac{\eta_{\text{LD},1}^4\Omega_1^2}{16}
		\hat{a}^{\dagger}_1\hat{a}^{\dagger}_1\hat{a}_1\hat{a}_1 e^{-i\Delta (t_2-t_1)}
		-i \frac{\eta_{\text{LD},1}^2\eta_{\text{LD},2}\Omega_1\Omega_2}{8}
		\hat{a}^{\dagger}_1\hat{a}^{\dagger}_1 \hat{a}_2
		e^{-i\Delta(t_2-t_1) +i(\phi_2-\phi_1)}
		\\
		&\;\;\;\;\;\;\;\;\;\;\;\;\;\;\;\;\;\;\;\;\;\;\;\;\;\;\;\;\;\;\;\;\;\;
		+i \frac{\eta_{\text{LD},1}^2\eta_{\text{LD},2}\Omega_1\Omega_2}{8}
		\hat{a}^{\dagger}_2 \hat{a}_1\hat{a}_1
		e^{i \Delta(t_1-t_2) +i(\phi_1-\phi_2) }
		+ \frac{\eta_{\text{LD},2}^2\Omega_2^2}{4}
		\hat{a}^{\dagger}_2\hat{a}_2
		e^{i\Delta (t_1-t_2) }
		\Bigg]\;.
	\end{aligned}
\end{equation}
Substituting $\tau\equiv t_1-t_2$, $s\equiv t_2$, and integrating over $\tau$ yields
\begin{equation}
	\begin{aligned}
		U_I^{(2)}(t)
		\approx
		-\int_0^t ds
		\Bigg[ &
		\frac{\eta_{\text{LD},1}^4\Omega_1^2}{16}
		\hat{a}^{\dagger}_1\hat{a}^{\dagger}_1\hat{a}_1\hat{a}_1
		\frac{e^{i\Delta(t-s)}-1}{i\Delta}
		-i \frac{\eta_{\text{LD},1}^2\eta_{\text{LD},2}\Omega_1\Omega_2}{8} e^{i(\phi_2-\phi_1)}
		\hat{a}^{\dagger}_1\hat{a}^{\dagger}_1 \hat{a}_2
		\frac{e^{-i\Delta (s-t)}-1}{i\Delta}
		\\
		&\;\;\;
		+i \frac{\eta_{\text{LD},1}^2\eta_{\text{LD},2}\Omega_1\Omega_2}{8} e^{i(\phi_1-\phi_2)}
		\hat{a}^{\dagger}_2\hat{a}_1\hat{a}_1
		\frac{e^{-i \Delta (s-t)} -1}{i \Delta}
		+ \frac{\eta_{\text{LD},2}^2\Omega_2^2}{4}
		\hat{a}^{\dagger}_2\hat{a}_2
		\frac{e^{i\Delta(t-s)}-1}{i\Delta}
		\Bigg]\;.
	\end{aligned}
\end{equation}
Averaged over time, only secular terms are relevant:
\begin{equation}
	\begin{aligned}
		U_I^{(2)}(t) \approx
		-i\int_0^t ds
		\Bigg[ &
		-\frac{\eta_{\text{LD},1}^4\Omega_1^2}{16\Delta}
		\hat{a}^{\dagger}_1\hat{a}^{\dagger}_1\hat{a}_1\hat{a}_1
		+i \frac{\eta_{\text{LD},1}^2\eta_{\text{LD},2}\Omega_1\Omega_2}{8\Delta} e^{i(\phi_2-\phi_1)} \hat{a}^{\dagger}_1\hat{a}^{\dagger}_1\hat{a}_2
		\\
		&
		-i \frac{\eta_{\text{LD},1}^2\eta_{\text{LD},2}\Omega_1\Omega_2}{8\Delta} e^{i(\phi_1-\phi_2)} \hat{a}^{\dagger}_2\hat{a}_1\hat{a}_1
		-\frac{\eta_{\text{LD},2}^2\Omega_2^2}{4\Delta} \hat{a}^{\dagger}_2\hat{a}_2
		\Bigg]\,.
	\end{aligned}
\end{equation}
Choosing the relative phase $\phi_1-\phi_2$ such that $-ie^{i(\phi_1-\phi_2)}/\Delta = 1/\abs{\Delta}$, we obtain the effective Hamiltonian governing the approximate time evolution $U_I(t) \approx e^{-\frac{i}{\hbar}H_{\text{eff}} t}$,
\begin{equation}
	\hat{H}_{\text{eff}}/\hbar
	=
	-\frac{\eta_{\text{LD},1}^4\Omega_1^2}{16\Delta} \hat{a}^{\dagger}_1\hat{a}^{\dagger}_1\hat{a}_1\hat{a}_1
	-\frac{\eta_{\text{LD},2}^2\Omega_2^2}{4\Delta} \hat{a}^{\dagger}_2\hat{a}_2
	+\frac{\eta_{\text{LD},1}^2\eta_{\text{LD},2}\Omega_1\Omega_2}{8\abs{\Delta}} \left(\hat{a}^{\dagger}_1\hat{a}^{\dagger}_1\hat{a}_2 + \hat{a}^{\dagger}_2\hat{a}_1\hat{a}_1 \right) \;.
\end{equation}
This effective Hamiltonian includes the desired nonlinear interaction term, with effective coupling parameter $J=\eta_{\text{LD},1}^2\eta_{\text{LD},2}\Omega_1\Omega_2/(8\abs{\Delta})$ as given in the main text.
In addition, the driving scheme gives rise to linear and nonlinear Stark shifts for the second and first mode, respectively.
These shifts can be accounted for when setting up the trapping potential.

\subsection{Incoherent single-phonon pump and two-phonon excitations}
As described in the main text, the dissipative processes of the system---namely, incoherent single-phonon pump and two-phonon loss---can be implemented via sideband heating and cooling, respectively.
The key idea of sideband heating or cooling is to drive an electronic transition of the ion whose excited state decays rapidly. This allows energy to be added to or removed from the motional states in a controlled manner.
Here, we show how driving the sidebands and adiabatically eliminating the internal excited state $\ket*{e_j}$ leads to effective dissipators in the master equation as given in Eqs.~(24, 25) in the main text.
For simplicity, we show the derivation of the effective incoherent single-phonon pump for a single mode $k$ of frequency $\omega_k$, which can be directly extended to the case of two simultaneously driven modes.

For the incoherent single-phonon pump, the electronic ground state with motional mode having occupation $n_k$ is driven from $\ket{g,n_k} \to \ket{e_j,n_k+1}$, and then rapidly decays to $\ket{g,n_k+1}$, effectively gaining one occupation of the motional mode.
This process can be realized by a single laser field $\vec{E}_i$, driving the transition $\ket{g}\leftrightarrow\ket{e_j}$, such that $\hat{\sigma}_+^{(i)}=\ket{e_j}\bra{g}$.
The field is aligned with one of the modes $k$, such that $\eta_{\text{LD},i,k'\neq k}\approx 0$, and we can write $\eta_{\text{LD},i,k}\equiv\eta_{\text{LD},k}$, $\Omega_i\equiv\Omega_k$, and $\phi_i\equiv\phi_k$.
The laser is tuned to the first blue sideband, i.e, the detuning is chosen to be $\delta_i^{(j)}=\omega_k$.
With this drive, and under the RWA, the relevant terms in the interaction Hamiltonian Eq.~\eqref{eq: supp: interaction_second_order} read
\begin{equation}
	\hat{V}^{\text{RWA}}_I \approx
	\frac{\eta_{\text{LD},k}\hbar\Omega_k}{2}
	\left(
	i\ket{e_j}\bra{g}\hat{a}_k^{\dagger} e^{i\phi_k}
	-i\hat{a}_k\ket{e_j}\bra{g} e^{-i\phi_k}
	\right)\;.
\end{equation}
We now consider the master equation
$\frac{d}{dt}\hat{\rho}_{I}=-\frac{i}{\hbar}\comm{\hat{V}_{I}}{\hat{\rho}_{I}} +\Gamma_j\mathcal{D}[\hat{\sigma}_{-}]$,
where $\Gamma_j$ denotes the spontaneous decay rate of the excited state $\ket{e_j}$.
Writing the density matrix as $\hat{\rho}_{I}=\sum_{ij}\ketbra{i}{j}\otimes\hat{\rho}_B^{ij}$, where $\hat{\rho}_B^{ij}$ acts on the bosonic sector of the system, and $i,j \in \{g,e\}$ denoting the electronic states, allows to write the master equation for each of the electronic sectors:
\begin{equation}
	\begin{aligned}
		\frac{d \hat{\rho}_B^{ee}}{dt}
		&=
		\frac{\eta_{\text{LD,k}}\Omega_k}{2}\left(e^{i\phi_k}\hat{a}_k^{\dagger}\hat{\rho}_B^{ge}+\hat{\rho}_B^{eg}\hat{a}_k e^{-i\phi_k}\right)
		-\Gamma_j \hat{\rho}_B^{ee}
		\\
		\frac{d\hat{\rho}_B^{eg}}{dt}
		&=
		\frac{\eta_{\text{LD,k}}\Omega_k}{2}e^{i\phi_k}\left(\hat{a}_k^{\dagger} \hat{\rho}_B^{gg}-\hat{\rho}_B^{ee}\hat{a}_k^{\dagger} \right)
		-\frac{\Gamma_j}{2}\hat{\rho}_B^{eg}
		\\
		\frac{d\hat{\rho}_B^{ge}}{dt}
		&=
		-\frac{\eta_{\text{LD,k}}\Omega_k}{2}e^{-i\phi_k}\left(\hat{a}_k \hat{\rho}_B^{ee} -\hat{\rho}_B^{gg} \hat{a}_k \right)
		-\frac{\Gamma_j}{2}\hat{\rho}_B^{ge}
		\\
		\frac{d\hat{\rho}_B^{gg}}{dt}
		&=
		-\frac{\eta_{\text{LD,k}}\Omega_k}{2}\left(\hat{\rho}_B^{ge} \hat{a}_k^{\dagger} e^{i\phi_k} +\hat{a}_k e^{-i\phi_k} \hat{\rho}_B^{eg} \right)
		+\Gamma_j \hat{\rho}_B^{ee} \;.
	\end{aligned}
	\label{eq: supp: master_equation_electronic_sector}
\end{equation}
If the excited states decays rapidly, i.e., $\Gamma_j\gg\eta_{\text{LD},k}\Omega_k$, the population of the electronic excited state and the coherences can be assumed to be in a quasi-steady state, $\dot{\rho}_{eg}\approx 0$, $\dot{\rho}_{ge}\approx 0$, $\dot{\rho}_{ee}\approx 0$, and $\rho_{gg}\gg\rho_{ee}$. The master equation in Eq.~\eqref{eq: supp: master_equation_electronic_sector} simplifies to
\begin{equation}
	\begin{aligned}
		\hat{\rho}_B^{eg}
		&\approx
		\frac{\eta_{\text{LD},k}\Omega_k}{\Gamma_j}e^{i\phi_k}\hat{a}_k^{\dagger}\hat{\rho}_B^{gg}\\
		\hat{\rho}_B^{ge}
		&\approx 
		\frac{\eta_{\text{LD},k}\Omega_k}{\Gamma_j}e^{-i\phi_k}\hat{\rho}_B^{gg}\hat{a}_k\\
		\hat{\rho}_B^{ee}
		&\approx
		\frac{\eta_{\text{LD},k}\Omega_k}{2\Gamma_j}\left(e^{i\phi_k}\hat{a}_k^{\dagger}\hat{\rho}_B^{ge}+\hat{\rho}_B^{eg}\hat{a}_k e^{-i\phi_k}\right)
		\approx \frac{\eta_{\text{LD},k}^2\Omega^2}{\Gamma_j^2} \hat{a}_k^{\dagger}\hat{\rho}_B^{gg}\hat{a}_k \;,
	\end{aligned}
\end{equation}
and, consequently,
\begin{equation}
	\begin{aligned}
		\frac{d\hat{\rho}_B^{gg}}{dt}
		&\approx
		\frac{\eta_{\text{LD},k}^2\Omega_k^2}{2\Gamma_j}\left(
		2\hat{a}_k^{\dagger}\hat{\rho}_B^{gg}\hat{a}_k
		-\hat{\rho}_B^{gg}\hat{a}_k\hat{a}_k^{\dagger} -\hat{a}_k\hat{a}_k^{\dagger}\hat{\rho}_B^{gg}
		\right) \;.
	\end{aligned}
	\label{eq: supp: master_eq_gg}
\end{equation}
Eq.~\eqref{eq: supp: master_eq_gg} has the form of a Lindblad dissipator $\mathcal{D}[\hat{a}_k^{\dagger}]\hat{\rho}$ with effective dissipation rate
\begin{equation}
	\gamma_{k,\text{eff}} = \frac{\eta_{\text{LD},k}^2\Omega_k^2}{\Gamma_j} \;.
\end{equation}

Two-phonon loss can be implemented by driving second red-detuned sidebands, i.e., using laser frequencies with detuning $\delta_k^{(i)} = -2\omega_k$, which effectively removes two motional quanta via the transitions $\ket{g,n_k} \to \ket{e_j, n_k-2} \to \ket{g, n_k-2}$. Adiabatic elimination can be performed analogously to the computation shown above. This leads to the effective decay rates for the two-phonon loss
\begin{equation}
	\eta_{k,\text{eff}} = \frac{\eta_{\text{LD},k}^4\Omega_k^2}{4\Gamma_j} \;.
\end{equation}

The driving schemes for both single-phonon pump and two-phonon loss are depicted in Fig.~\ref{fig: supp: scheme_trapped_ions}(b).
Note that these heating and cooling processes can be independently implemented using distinct drives and different internal states $\ket{e_j}$.
\newpage

\stopcontents[supplement]

\restoregeometry

\end{document}

%% file: preamble.tex
\usepackage{color}
\usepackage{dcolumn}
\usepackage{graphicx}

\usepackage{amsmath,amsfonts,amssymb,amsthm}
\usepackage{mathtools}

\usepackage[driver=pdftex]{geometry}
\usepackage{bbold}
\usepackage{dsfont}

\usepackage{soul}

\usepackage{fancyhdr}
\usepackage{multirow}
\usepackage{bm}
\usepackage{dsfont}
\usepackage{cancel}

\usepackage{xifthen}

\usepackage{titletoc}

\makeatletter

\usepackage{enumitem}
\usepackage{float}
\graphicspath{{"./graphs/"}{"./scheme/"}}

\usepackage{graphicx}
\usepackage[usenames,dvipsnames,dvipsnames,svgnames]{xcolor}
\usepackage[colorlinks,bookmarks=false,
linkcolor=blue,
urlcolor=blue,
citecolor=blue
]{hyperref}

\usepackage{xspace}
\newcommand{\ket}[1]{\ensuremath{|#1\rangle}\xspace}
\newcommand{\bra}[1]{\ensuremath{\langle #1|}\xspace}

\def\commut#1#2{[ #1,#2 ]}

\usepackage{array,calc}

\newcommand{\q}[1]{``#1"}
\usepackage{physics}

\usepackage{xifthen}

\def\op#1{\hat{#1}}
\renewcommand{\ao}[1][]{%
	\ifthenelse{\equal{#1}{}}{\ensuremath{\op{a}}}{\ensuremath{\op{#1}}}%
}

\newcommand{\co}[1][]{%
	\ifthenelse{\equal{#1}{}}{\ensuremath{{{}\op{a}^{\dagger}}}}{\ensuremath{{{}\op{#1}^{\dagger}}}}%
}

\makeatletter
\newcommand*{\transpose}{\bgroup\@transpose}
\newcommand*{\@transpose}[1][0]{\mathpalette\@@transpose{#1}\egroup}
\newcommand*{\@@transpose}[2]{\setbox0=\hbox{\m@th$#1\mkern-#2mu\intercal$}\raise\dp0\box0}
\makeatother